\documentclass[onecolumn,draftcls]{IEEEtran}
\usepackage{amsmath,amsfonts,amssymb}
\usepackage{graphicx,subcaption,color}
\usepackage{epstopdf}
\usepackage{tabls,stackrel,cite}
\usepackage{notation}

\usepackage{tikz,pgfplots}
\usetikzlibrary{plotmarks}
\usetikzlibrary{patterns}
\pgfplotsset{compat=1.10}
\usepgfplotslibrary{fillbetween}

\newcommand{\refE}[1]   {(\ref{eqn:#1})}
\newcommand{\refS}[1]   {Section~\ref{sec:#1}}
\newcommand{\refF}[1]   {Fig.~\ref{fig:#1}}
\newcommand{\refFig}[1] {Figure~\ref{fig:#1}}
\IEEEoverridecommandlockouts

\begin{document}
\sloppy

\title{\huge Error Probability Bounds for {G}aussian Channels\\ under Maximal and Average Power Constraints}
\author{\IEEEauthorblockN{Gonzalo Vazquez-Vilar}
\thanks{G. Vazquez-Vilar is with the Universidad Carlos III de Madrid, Madrid, Spain, and with the Gregorio Mara\~n\'on Health Research Institute, Madrid, Spain. This work has been funded in part by the European Research Council (ERC) under grant 714161, and by the Spanish Ministry of Economy and Competitiveness under grant \mbox{TEC2016-78434-C3}  (AEI/FEDER,~EU).}
\thanks{This work was presented in part at the 2019 IEEE International Symposium on Information Theory~\cite{isit2019}, Paris, France, and at the 2020 International Z\"urich Seminar on Communication and Information~\cite{izs2020}, Z\"urich, Switzerland.}}

%% Create the title:
\maketitle

\begin{abstract}
This paper studies the performance of block coding on an additive white Gaussian noise channel under different power limitations at the transmitter. New lower bounds are presented for the minimum error probability of codes satisfying maximal and average power constraints. These bounds are tighter than previous results in the finite blocklength regime, and yield a better understanding on the structure of good codes under an average power limitation. Evaluation of these bounds for short and moderate blocklengths is also discussed.
\end{abstract}


\section{Introduction}\label{sec:Intro}
We consider the problem of transmitting equiprobable messages over several uses of an additive white Gaussian noise (AWGN) channel. We consider different power restrictions at the transmitter: (i) equal power constraint (all the codewords in the transmission code have equal energy); (ii) maximal power constraint (the energy of all the codewords is below a certain threshold); and, (iii) average power constraint (while some codewords may violate the threshold, the power constraint is satisfied in average).
Given its practical importance, the AWGN channel under a power limitation has been widely studied in the literature.

In his seminal 1948 work~\cite{Shannon48}, Shannon established the capacity of this channel, which characterizes the highest transmission rate under which reliable communication is possible with arbitrarily long codewords. A more refined asymptotic analysis follows from the study of the reliability function, which characterizes the exponential dependence between the error probability and the length of the codewords for a certain transmission rate.
For the power-constrained AWGN channel, Shannon obtained the reliability function for rates close to the channel capacity~\cite{Shannon59}. Both the capacity~\cite{Shannon48} and the reliability function~\cite{Shannon59} of the AWGN channel do not depend on the specific power restriction considered at the transmitter.
We conclude that equal, maximal and average power constraints can be cast as asymptotically equivalent.\footnote{Note however that some asymptotic differences still exits. For instance, the strong-converse error exponent (relevant for rates above capacity) under equal and maximal power constraint is strictly positive, while it is equal to zero under average-power constraint~\cite[Sec. 4.3]{PolThesis}.}
While the focus of the work by Shannon \cite{Shannon59} was on the reliability function, he also obtained both upper and lower bounds on the error probability of the best codebook with a certain blocklength $n$. The proof of these bounds is based on geometric arguments applied to codewords lying on the surface of an $n$-dimensional sphere~\cite[Eq.~(20)]{Shannon59} (i.e., satisfying an equal power constraint) and it is then extended to maximal and average power limitations~\cite[Sec.~XIII]{Shannon59}. 

An alternative proof technique in the derivation of converse bounds to the error probability is based on hypothesis testing.
The channel coding error probability can be related to that of a surrogate binary hypothesis test between the distribution induced by the codebook and an auxiliary distribution carefully chosen~\cite{tit16a}. An application of this technique was used in~\cite{Shannon67I} to obtain the sphere-packing bound on the channel coding reliability function for general channels (see also \cite{Haroutunian68, sason2008, altug2014, Nakiboglu19-SP} for alternative derivations and refinements). To obtain the sphere-packing exponent, the hypothesis testing technique needs to be applied with an specific auxiliary distribution, denoted as \textit{exponent-achieving output distribution} (analogously to the \textit{capacity-achieving output distribution} that follows from the channel capacity analysis).

Also using the hypothesis testing technique, Polyanskiy {\em et al.} obtained a fundamental lower bound to the error probability in the finite blocklength regime~\cite[Th. 27]{Pol10}. This result is usually referred to as \textit{meta-converse} since several converse bounds in the literature can be recovered from  it. The standard application of the meta-converse bound for a specific channel requires the choice of the auxiliary distribution used in the hypothesis test. 
To obtain more intuition in the structure of this auxiliary distribution, Polyanskiy analyzed in~\cite{Pol13} the properties of the solution to the minimax optimization problem in~\cite[Th. 27]{Pol10}. Exploiting the existing symmetries in the AWGN channel with an equal power constraint, \cite[Sec.~VI]{Pol13} shows that, for a certain non-product auxiliary distribution, the meta-converse bound coincides with Shannon lower bound~\cite[Eq.~(20)]{Shannon59}. Therefore, Shannon lower bound is still the tightest finite-length bound for the AWGN channel with an equal power constraint and this bound is often used as a benchmark for practical codes (see, e.g., \cite{lazic98, dolinar1998, Via99, shi2007, PolThesis, vazquez2016multiclass}).
While the choice for the auxiliary distribution in \cite[Sec.~VI]{Pol13} yields the tightest meta-converse bound, the resulting expression is still difficult to evaluate. For an auxiliary distribution equal to the capacity achieving output distribution, the meta-converse particularizes to \cite[Th. 41]{Pol10}. This bound is slightly weaker than Shannon's \cite[Eq.~(20)]{Shannon59} and can be extended to maximal and average power constraints using the techniques in~\cite[Sec.~XIII]{Shannon59} (see also \cite[Lem. 39]{Pol10}).
%The minimum error probability of codes satisfying an equal power constraint can be related to that of codes satisfying maximal and average power limitations~(see \cite[Sec.~XIII]{Shannon59} and also \cite[Lem. 39]{Pol10}). These results can be used to extend the Shannon'59 lower bound, \cite[eq. (20)]{Shannon59}, and PPV'10 lower bound, \cite[Th. 41]{Pol10}, beyond the equal power constraint assumed in their derivation. While the impact of this extension  becomes negligible in the asymptotic regime, it can be relevant at finite blocklengths. 

In this work, we complement the existing results with direct lower bounds on the error probability of codes for the AWGN channel under maximal and average power limitations at the transmitter. In particular, the main contributions in this article are the following:
\begin{enumerate}
\item We provide an exhaustive characterization of the error probability of a binary hypothesis test between two Gaussian distributions. The error probability of this test corresponds to the meta-converse bound for an equal power constraint and an auxiliary independent and identically distributed (i.i.d.) zero-mean Gaussian distribution (not necessarily capacity achieving).
\item Using this characterization, we optimize the meta-converse bound over input distributions satisfying maximal and average power constraints. The resulting hypothesis testing lower bound holds directly under a maximal power limitation. For an average power limitation, we obtain that the hypothesis testing bound holds directly if the codebook size is below a certain threshold and that it requires a simple transformation above this threshold.
\item We propose a saddlepoint expansion to estimate the error probability of a hypothesis test between two i.i.d. Gaussian distributions. This expansion yields a simple expression that can be used to evaluate \cite[Th. 41]{Pol10} and the bounds for  maximal and average power constraints presented in this work.
\item We provide several numerical examples that compare the new bounds with previous results in the literature. We show that considering an exponent-achieving auxiliary distribution under equal, maximal and average power constraints yields tighter bounds in general. 
\end{enumerate}

Given the difficulty of computing \cite[eq. (20)]{Shannon59} (see, e.g.,~\cite{Slepian63, Via99, Val04, Sas08}), the bounds proposed are not only tighter (for  maximal and average power constraints) but also simpler to evaluate than the original lower bound by Shannon. While the results obtained are specific for the AWGN channel, the techniques used in this work can in principle be extended to other scenarios requiring the optimization of the meta-converse bound over input distributions.

The organization of the manuscript is as follows. \refS{model} presents the system model and a formal definition of the power constraints. This section also introduces the meta-converse bound that will be used in the remainder of the article. \refS{equal} compares Shannon lower bound with the meta-converse for the AWGN channel with an equal power constraint. This section provides a geometric interpretation of \cite[Th. 41]{Pol10} analogous to that presented in \cite{Shannon59}. Sections \ref{sec:maximal} and \ref{sec:average} introduce new bounds for  maximal and average power constraints, respectively. The evaluation of the proposed bounds is studied in \refS{computation}. \refS{numerical} presents a numerical comparison of the bounds with previous results  and studies the effect of considering capacity and exponent achieving auxiliary distributions.
Finally, \refS{discussion} concludes the article discussing the main contributions of this work.

\section{System Model and Preliminaries}\label{sec:model}
%%%%%%%%%%%%%%%%%%%%%%%%%%%%%%%%%%%%%%%%%%%%%%%%%%%

We consider the problem of transmitting $M$ equiprobable messages over $n$ uses of an AWGN channel with noise power~$\sigma^2$. Specifically,  we consider a channel $W \triangleq \Pyx$ which, for an input $\x=(x_1,x_2,\ldots,x_n)\in\Xc$ and output $\y=(y_1,y_2,\ldots,y_n)\in\Yc$, with $\Xc=\Yc=\RR^n$, has a probability density function (pdf)
\begin{align}\label{eqn:Gaussian-channel}
  w(\y|\x) = \prod_{i=1}^{n} \varphi_{x_i,\sigma}(y_i),
\end{align}
where $\varphi_{\mu,\sigma}(\cdot)$ denotes the pdf of the Gaussian distribution,
\begin{align}
  \varphi_{\mu,\sigma}(y)\triangleq\frac{1}{\sqrt{2 \pi}\sigma} e^{-\frac{(y-\mu)^2}{2\sigma^2}}.\label{eqn:Gaussian-pdf}
\end{align}

In our communications system, the source generates a message $v\in\{1,\ldots,M\}$ randomly with equal probability. This message is then mapped by the encoder to a codeword $\cc_v$ using a codebook~$\Cc \triangleq \bigl\{\cc_1, \ldots,\cc_M \bigr\}$, and $\x = \cc_v$ is transmitted over the channel. Then, based on the channel output $\y$, the decoder guesses the transmitted message $\hat{v}\in\{1,\ldots,M\}$. In the following we shall assume that maximum likelihood (ML) decoding is used at the receiver.\footnote{Since the ML decoder minimizes the error probability, lower bounds to ML decoding error probability also apply to other decoding schemes.}
We define the average error probability of a codebook $\Cc$ as
\begin{align}
  \Pe(\Cc) \triangleq  \Pr \{\hat{V} \neq V\},
\end{align}
where the underlying probability is induced by the chain of source, encoder, channel, and ML decoder.%
\footnote{All the results in this article are derived under the average error probability formalism.
For the maximal error probability, defined as 
$\eps_{\max}(\Cc) \triangleq \max_{v\in\{1,\ldots,M\}} \Pr \{\hat{V} \neq V\,|\,V=v\}$,
it holds that $\eps_{\max}(\Cc) \geq \Pe(\Cc)$ and lower bounds on $\Pe(\Cc)$
also apply in this case.}

\subsection{Power constrained codebooks}
The focus of this work is on obtaining lower bounds to the error probability $\Pe(\Cc)$ for codebooks $\Cc \triangleq \bigl\{\cc_1, \ldots,\cc_M \bigr\}$ satisfying the following power constraints.
\begin{enumerate}
\item Equal power constraint:
\begin{align}
\Fc_{\text{e}}(n,M,\Upsilon) \triangleq
  \Bigl\{ \Cc \;\big|\; \|\cc_i\|^2 = n \Upsilon,\quad i=1,\ldots,M \Bigr\}.
\end{align}
\item Maximal power constraint:
\begin{align}
\Fc_{\text{m}}(n,M,\Upsilon) \triangleq \Bigl\{ \Cc \;\big|\; \|\cc_i\|^2 \leq n\Upsilon,\quad i=1,\ldots,M \Bigr\}.
\end{align}
\item Average power constraint:
\begin{align}
\Fc_{\text{a}}(n,M,\Upsilon) \triangleq \Bigl\{ \Cc \;\big|\; \tfrac{1}{M}\sum\nolimits_{i=1}^M\|\cc_i\|^2 \leq n\Upsilon \Bigr\}.
\end{align}
\end{enumerate}

Clearly, $\Fc_{\text{e}} \subset \Fc_{\text{m}} \subset \Fc_{\text{a}}$. Then, any lower bound on the error probability of an average power constrained codebook will also hold in the maximal power constraint setting, and any bound for a maximal power constraint also holds under an equal power constraint.
The next result relates the minimum error probability in the three scenarios considered via simple inequalities.
For fixed $n,M,\Upsilon$, we define the minimum error probability under a power constraint $i \in \{\text{e}, \text{m}, \text{a}\}$ as
\begin{align}
  \epsilon_i^{\star}(n,M,\Upsilon) \triangleq \min_{\Cc \in \Fc_{i}(n,M,\Upsilon)} \Pe(\Cc).
\end{align}
\begin{lemma}[{\hspace{-.1mm}\cite[Sec.~XIII]{Shannon59}, \cite[Lemma~65]{PolThesis}}]
\label{lem:relations}
For any $n,M,\Upsilon>0$, and $0<s<1$, the following inequalities hold:
\begin{align}
  \quad \epsilon_{\text{e}}^{\star}(n,M,\Upsilon)
  \ &\geq \ \epsilon_{\text{m}}^{\star}(n,M,\Upsilon)
  \ \geq \epsilon_{\text{e}}^{\star}(n+1,M,\Upsilon),
  \label{eqn:relations-1}\\
  \epsilon_{\text{m}}^{\star}(n,M,\Upsilon)
  \ &\geq \ \epsilon_{\text{a}}^{\star}(n,M,\Upsilon)
  \ \geq \ s\epsilon_{\text{m}}^{\star}\Bigl(n, sM,\tfrac{\Upsilon}{1-s}\Bigr).
  \label{eqn:relations-2}
\end{align}
\end{lemma}
\begin{remark}
The relations \refE{relations-1}-\refE{relations-2} were first proposed by Shannon in \cite[Sec.~XIII]{Shannon59}.
Nevertheless, there is a typo in the last equation of \cite[Sec.~XIII]{Shannon59}, which has been corrected in \refE{relations-2}.
In \cite[Sec.~XIII]{Shannon59}, Shannon states that ``The probability of error for the new code [satisfying the maximal power constraint] cannot exceed $1/{\alpha}$ times that of the original code  [satisfying the average power constraint]'' (brackets added). While this reasoning is right, using his notation, this statement traslates to $P_{\text{e opt}}' \leq \frac{1}{\alpha} P_{\text{e opt}}''$ and hence $P_{\text{e opt}}'' \geq \alpha P_{\text{e opt}}'$, which does not coincide with the last equation of \cite[Sec.~XIII]{Shannon59}.
These relations were rederived in \cite[Lemma~65]{PolThesis}, where the statement of the bound corresponding to \refE{relations-2} is correct.
%For the maximal error probability formalism the relations between the different power constraints were derived in \cite[Lem. 39]{Pol10} (see also \cite[Lemma~65]{PolThesis}).
\end{remark}

The relations from Lemma~\ref{lem:relations} show that lower and upper bounds on the error probability under a given power constraint can be adapted to other settings via simple transformations. If we focus on converse bounds, the analysis under an  equal power constraint is usually simpler. However, since the maximal power constraint and average power constraint are more relevant in practice, the transformations from Lemma~\ref{lem:relations} are often used to adapt the results derived under an equal power constraint.
While the loss incurred by using these transformations becomes negligible in the asymptotic regime, it can have a relevant impact at finite blocklengths. In Sections \ref{sec:maximal} and \ref{sec:average} we will prove direct lower bounds for  maximal and average power constraints, without resorting to the transformations from Lemma~\ref{lem:relations}.

\subsection{Meta-converse bound}
\label{sec:metaconverse-bound}

In \cite{Pol10}, Polyanskiy {\em et al.} proved that the error probability of a binary hypothesis test with certain parameters can be used to lower bound the error probability $\Pe(\Cc)$ for a certain channel $W$. In particular, \cite[Th. 27]{Pol10} shows that
\begin{align}
  \Pe(\Cc)
&\geq \infp_{P\in\Pc} \sup_{Q} \left\{
                 \alpha_{\frac{1}{M}} \bigl(PW, P \times Q \bigr)\right\},\label{eqn:metaconverse}
\end{align}
where $\Pc$ is the set of distributions over the input alphabet $\Xc$ satisfying a certain constraint, $Q$ is an auxiliary distribution over the output alphabet $\Yc$ which is not allowed to depend on the input $\x$, and where $\alpha_{\beta}\left(PW, P \times Q\right)$ denotes the minimum type-I error for a maximum type-II error $\beta\in[0,1]$ in a binary hypothesis testing problem between the distributions $PW$ and $P \times Q$. Formally, for two distributions $A$ and $B$ defined over an alphabet $\Zc$, the minimum type-I error for a maximum type-II error $\beta\in[0,1]$ is given by
\begin{align}
\alpha_{\beta}(A, B)
  \triangleq \inf_{\substack{0\leq T \leq 1:\\ \Ex_{B}[T(Z)] \leq \beta}} \Bigl\{ 1- \Ex_{A}[T(Z)] \Bigr\},\label{eqn:bht-alpha}
\end{align}
where $T:\Zc\to[0,1]$ and $\Ex_{P}[\cdot]$ denotes the expectation operator with respect to the random variable $Z\sim P$.

The bound \refE{metaconverse} is usually referred to as the \textit{meta-converse bound} since several converse bounds in the literature can be recovered from it via relaxation~\cite{Pol10}.

The results in this work are based on the following inequality chain, which always holds
\begin{align}
  \infp_{P\in\Pc} \sup_{Q} \left\{
                 \alpha_{\frac{1}{M}} \bigl(PW, P \times Q \bigr)\right\} &\geq  \sup_{Q} \infp_{P\in\Pc} \left\{
                 \alpha_{\frac{1}{M}} \bigl(PW, P \times Q \bigr)\right\}\label{eqn:metaconverse-maxmin}\\
               &\geq  \infp_{P\in\Pc} \left\{
                 \alpha_{\frac{1}{M}} \bigl(PW, P \times Q \bigr)\right\}.\label{eqn:metaconverse-fixedQ}
\end{align}
Here, the first step follows from the max-min inequality, and the second is the result of fixing the auxiliary distribution $Q$. 

The properties of the exact minimax solution to the optimizations in the left-hand side of \refE{metaconverse-maxmin} are studied in~\cite{Pol13}.
Under mild assumptions, \refE{metaconverse-maxmin} holds with equality and the saddle point property holds~\cite[Sec.~V]{Pol13}. Therefore, in practice it is possible to fix the auxiliary distribution $Q$ in  \refE{metaconverse} and still obtain tight lower bounds. However, the minimization needs to be carried out over all the input probability distributions $P$ (non necessarily product) satisfying the constraint $P\in \Pc$.
In the following sections we consider the optimization of the meta-converse bound over input distributions for the AWGN channel under equal, maximal and average power constraints and a certain auxiliary distribution $Q$.

\section{Lower Bounds for Equal Power Constraints}\label{sec:equal}
%%%%%%%%%%%%%%%%%%%%%%%%%%%%%%%%%%%%%%%%%%%%%%%%%%%

In this section we briefly discuss the results from \cite{Shannon59}, \cite{Pol10} and \cite{Pol13}.  The bounds presented here apply for codes $\Cc \in \Fc_{\text{e}}(n,M,\Upsilon)$ satisfying an equal power constraint, and they will be relevant in the sequel.

\subsection{Shannon cone-packing bound} \label{sec:equal-shannon}

Let $\theta$ be the half-angle of a $n$-dimensional cone with vertex at the origin and with axis going through the vector $\x= (1,\ldots,1)$. We let $\Phi_n(\theta, \bar\sigma^2)$ denote the probability that such vector be moved outside this cone by effect of the i.i.d. Gaussian noise with variance $\bar\sigma^2$ in each dimension.
\begin{theorem}[{\hspace{-.1mm}\cite[Eq. (20)]{Shannon59}}]
Let $\theta_{n,M}$ denote the half-angle of a cone with solid angle equal to $\Omega_n/M$, where $\Omega_n$ is the surface of the $n$-dimensional hypersphere. Then, the error probability of an equal-power constrained code satisfies
\begin{align} \label{eqn:shannon_lower_bound}
\epsilon_{\text{e}}^{\star}(n,M,\Upsilon) \geq \Phi_n\biggl(\theta_{n,M},\frac{\sigma^2}{\Upsilon} \biggr).
\end{align}\label{thm:shannon_lower_bound}
\end{theorem}

The derivation of this bound follows from deforming the optimal decoding regions, which for codewords lying on the surface of an sphere correspond to pyramids, to cones of the same volume (see \cite[Fig.~1]{Shannon59}) and analyzing the resulting error probability.
Given this geometric interpretation, Theorem~\ref{thm:shannon_lower_bound} is often referred to as cone-packing bound. While the resulting expression is conceptually simple and accurate for low SNRs and relatively short codes~\cite{Sason06},  it is difficult to evaluate. Approximate and exact computation of this bound is treated, e.g., in~\cite{Val04, Sas08}.

\subsection{Meta-converse bound for the AWGN channel}
\label{sec:equal-metaconverse}

We consider now the meta-converse bound \refE{metaconverse} for the equal power constrained AWGN channel.
Solving the minimax optimization in the right-hand side of \refE{metaconverse}, this bound recovers \refE{shannon_lower_bound}~\cite[Sec. VI.F]{Pol13}.
Indeed, for an equal power constraint the codewords $\cc_i$ are restricted to lie on the surface of a sphere of squared radius $n \Upsilon$. The optimal decoder does not depend on the norm of the received sequence and only on its direction, and it can operate over the equivalent channel from $\tilde\x = \frac{\x}{\sqrt{n\Upsilon}} \in\tilde\Xc $ to $\tilde{\y} = \frac{\y}{\|\y\|}\in\tilde\Yc$, where $\tilde\Xc=\tilde\Yc=\SS^{n-1}$ correspond to the $(n-1)$-dimensional sphere centered at the origin and with unit radius.
Applying the meta-converse bound \refE{metaconverse} to the random map $W = P_{\tilde\Y|\tilde\X}$, we obtain that both the optimizing $P$ and $Q$ correspond to the uniform distributions on $\SS^{n-1}$~\cite[Sec. VI.F]{Pol13}.
Mapping this result to the original channel $W = P_{\Y|\X}$, it shows that the tightest bound that can be obtained from the meta-converse \refE{metaconverse} coincides with Shannon cone-packing bound \refE{shannon_lower_bound}. As it occurs with \refE{shannon_lower_bound}, the resulting expression is difficult to evaluate.

As discussed in \refS{metaconverse-bound}, the meta-converse bound \refE{metaconverse} can be weakened by fixing the auxiliary distribution $Q$.
For any input distribution lying on the surface of a $n$-dimensional hyper-sphere of squared radius $n\Upsilon$ (equal power constraint) and an auxiliary distribution $Q$ that is invariant under rotations around the origin, it holds that~\cite[Lem.~29]{Pol10}
\begin{align}\label{eqn:mc-spherical-symmetry}
\alpha_{\frac{1}{M}} \bigl(PW, P \times Q \bigr)
 = \alpha_{\frac{1}{M}} \bigl( \varphi^n_{\sqrt{\Upsilon},\sigma}, Q \bigr).
\end{align}

In \cite[Sec.~III.J.2]{Pol10}, Polyanskiy {\em et al.} fixed the auxiliary distribution $Q$ to be an i.i.d. Gaussian distribution with zero-mean and variance $\theta^2$, with pdf
\begin{align}\label{eqn:Q-theta-def}
  q(\y) = \prod_{i=1}^{n} \varphi_{0,\theta}(y_i),
\end{align}
to obtain the following result.
\begin{theorem}[{\hspace{-.1mm}\cite[Th. 41]{Pol10}}]
%Let $\Cc \in \Fc_{\text{e}}(n,M,\Upsilon)$ be a length-$n$ code of cardinality $M$ satisfying an equal power constraint  $\Upsilon$.
Let $\theta^2=\Upsilon+\sigma^2$. The error probability of an equal-power constrained code satisfies
\begin{align} \label{eqn:PPV_lower_bound}
  \epsilon_{\text{e}}^{\star}(n,M,\Upsilon) \geq \alpha_{\frac{1}{M}} \bigl( \varphi^n_{\sqrt{\Upsilon},\sigma}, \varphi^n_{0,\theta} \bigr).
\end{align}\label{thm:PPV_lower_bound}
\end{theorem}

This expression admits a parametric form involving two Marcum-$Q$ functions (see Proposition~\ref{prop:alpha-beta-marcumQ} in Appendix~\ref{apx:f-beta-gamma}). However, for fixed rate $R\triangleq\frac{1}{n}\log_2 M$, the term $\frac{1}{M} = 2^{-n R}$ decreases exponentially with the blocklength and traditional series expansions of the Marcum-$Q$ function fail even for moderate values of $n$. 
Nevertheless, in contrast with the formulation in \refE{shannon_lower_bound}, the distributions appearing in \refE{PPV_lower_bound} are i.i.d., and Laplace methods can be used to evaluate this bound (this point will be treated in  Section \ref{sec:computation}).

\subsection{Geometric interpretation of Theorem~\ref{thm:PPV_lower_bound}}

Shannon lower bound from Theorem~\ref{thm:shannon_lower_bound} corresponds to the probability that the additive Gaussian noise moves a given codeword out of the $n$-dimensional cone centered at the codeword that roughly covers $1/M$-th of the output space. We show next that the hypothesis-testing bound from Theorem~\ref{thm:PPV_lower_bound} admits an analogous geometric interpretation.

Let $\x = \bigl(\sqrt{\Upsilon},\ldots, \sqrt{\Upsilon} \bigr)$ and let $\theta>\sigma$.
For the hypothesis test on the right-hand side of \refE{PPV_lower_bound}, the condition
\begin{align}
\log 
\frac{\varphi_{\sqrt{\Upsilon},\sigma}^n(\y) }{\varphi_{0,\theta}^n(\y)} &= n\log \frac{\theta}{\sigma} + \frac{\|\y\|^2}{2 \theta^2} - \frac{\|\y-\x\|^2}{2 \sigma^2} = t
\label{eqn:metaconverse-rho-boundary-1}
\end{align}
defines the boundary of the decision region induced by the optimal Neyman-Pearson test for some $-\infty < t < \infty$. We next study the shape of this region. To this end, we first write
\begin{align}
\frac{\|\y\|^2}{2 \theta^2} -\frac{\|\y-\x\|^2}{2 \sigma^2} 
%\notag\\
%&= - \frac{\theta^2 \|\y-\x\|^2-\sigma^2 \|\y\|^2}{2 \sigma^2 \theta^2}\\
&= - \frac{\theta^2-\sigma^2}{2 \sigma^2 \theta^2}
     \bigl(\|\y\|^2 - 2a \langle\x,\y\rangle + a\|\x\|^2\bigr)\label{eqn:y2-yx2}\\
&=- \frac{\theta^2\!-\!\sigma^2}{2 \sigma^2 \theta^2}
     \bigl(\|\y-a\x\|^2 + (a\!-\! a^2) \|\x\|^2 \bigr),\label{eqn:metaconverse-rho-boundary-2}
\end{align}
where we defined $a = \frac{\theta^2}{\theta^2-\sigma^2}$, and where $\langle\x,\y\rangle = \x^T \y$ denotes the inner product between $\x$ and $\y$.

The boundary of the decision region induced by the optimal NP test, defined by \refE{metaconverse-rho-boundary-1} corresponds to \refE{metaconverse-rho-boundary-2} being equal to $t-n\log\frac{\theta}{\sigma}$. 
Using that $\|\x\|^2 = n\Upsilon$ and for $\theta^2=\Upsilon+\sigma^2$ from Theorem~\ref{thm:PPV_lower_bound}, it yields
\begin{align}
\biggl\|\y-\Bigl(1+\frac{\sigma^2}{\Upsilon}\Bigr)\x\biggr\|^2 = r,
\label{eqn:metaconverse-rho-boundary-3}
\end{align}
where $r = n\sigma^2 \bigl(1+\frac{\sigma^2}{\Upsilon}\bigr) \bigl(  1 - \frac{2t}{n} + \log \bigl(1+\frac{\Upsilon}{\sigma^2}\bigr)\bigr)$.

\begin{figure}[t]%
\begin{center}
\begin{subfigure}[b]{.4\linewidth}
\centering\includegraphics[width=.8\linewidth]{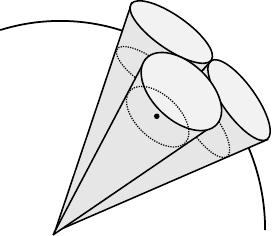}%
\caption{}
\end{subfigure}%
\begin{subfigure}[b]{.4\linewidth} 
\centering\includegraphics[width=.8\linewidth]{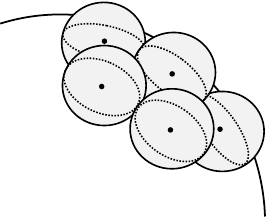}%
\caption{}
\end{subfigure}
\caption{Induced regions by (a) the Shannon cone-packing bound in~\refE{shannon_lower_bound}, and (b) the hypothesis-testing bound in \refE{PPV_lower_bound}, for codewords ($\bullet$) located on the shell of the sphere with squared radius $n\Upsilon$.}\label{fig:regions-bounds}
\end{center}
\end{figure}%
The region inside the boundary \refE{metaconverse-rho-boundary-3} corresponds to an $n$-dimensional sphere centered at $\bigl(1+\frac{\sigma^2}{\Upsilon}\bigr)\x$ with squared radius $r$. 
Then, we can describe the lower bound in Theorem~\ref{thm:PPV_lower_bound} as the probability that the additive Gaussian noise moves a given codeword $\x$ out of the $n$-dimensional sphere centered at $\bigl(1+\frac{\sigma^2}{\Upsilon}\bigr)\x$ that roughly covers $1/M$-th of the auxiliary measure $\varphi_{0,\theta}^n$.

The ``regions'' induced by Shannon lower bound in from Theorem~\ref{thm:shannon_lower_bound} correspond to cones, while those induced by the hypothesis-testing bound in Theorem~\ref{thm:PPV_lower_bound} correspond to spheres (see Fig.~\ref{fig:regions-bounds}).
Cones are close to the optimal ML decoding regions for codewords evenly distributed on surface of an $n$-dimensional sphere with squared radius $n\Upsilon$.\footnote{Indeed, in $n=2$ dimensions Shannon lower bound yields the exact error probability of an $M$-PSK constellation. See Section \ref{sec:constellations} for a numerical example.} On the other hand, ``spherical regions'' allow different configurations of the codewords. This fact suggests that the hypothesis-testing bound in Theorem~\ref{thm:PPV_lower_bound} may hold beyond the equal power constraint setting. This intuition is shown to be correct in the next sections.

\section{Lower Bounds for Maximal Power Constraints}\label{sec:maximal}
%%%%%%%%%%%%%%%%%%%%%%%%%%%%%%%%%%%%%%%%%%%%%%%%%%%

We consider now the family of codes satisfying a maximal power limitation, $\Cc\in \Fc_{\text{m}}(n,M,\Upsilon)$. As discussed in \refS{model}, Theorems \ref{thm:shannon_lower_bound} and \ref{thm:PPV_lower_bound} can be extended to the maximal power constraint via Lemma~\ref{lem:relations}. Indeed, the second inequality in \refE{relations-1} can be slightly tightened to
\begin{align}\label{eqn:equal-to-maximal}
  \epsilon_{\text{m}}^{\star}(n,M,\Upsilon) \geq \epsilon_{\text{e}}^{\star}\Bigl(n+1,M,\tfrac{n\Upsilon}{n+1}\Bigr).
\end{align}
The proof of Lemma~\ref{lem:relations} (see~\cite[Sec. XIII]{Shannon59} and \cite[Lem. 39]{Pol10}) is based on extending a maximal power constrained codebook of length $n$ by adding an extra coordinate. The energy of the new codewords of length $n+1$ is then normalized to $(n+1)\Upsilon$, so that the equal power constraint is satisfied. The proof of \refE{equal-to-maximal} follows the same lines, but normalizing the energy of the codewords to $n\Upsilon$, instead to $(n+1)\Upsilon$.
Applying \refE{equal-to-maximal} to Theorem~\ref{thm:shannon_lower_bound} we obtain the following result.
\begin{corollary}\label{cor:shannon_lower_bound}
Let $\theta_{n,M}$ denote the half-angle of a cone with solid angle equal to $\Omega_n/M$, where $\Omega_n$ is the surface of the $n$-dimensional hypersphere. Then, 
\begin{align} \label{eqn:cor_shannon_lower_bound}
   \epsilon_{\text{m}}^{\star}(n,M,\Upsilon) \geq \Phi_{n+1}\biggl(\theta_{n+1,M},\frac{(n+1)\sigma^2}{n\Upsilon} \biggr).
\end{align}
\end{corollary}

%\subsection{Direct lower bound under a maximal power constraint}
%\label{sec:maximal-A}

We now present an alternative lower bound to the error probability under maximal power constraint. To this end, we consider the weakening of the meta-converse in~\refE{metaconverse} obtained by fixing the auxiliary distribution $Q$ to be the zero-mean i.i.d. Gaussian distribution~\refE{Q-theta-def}. 

%%%%%%%%%%%%%%%%%%%%%%%%%%%%%%%%%%%%%%%%%%%%%%%%%%%%%%%%%%%%%%%%%%%%%%%%%%%%%%%%
\begin{theorem}[Converse, maximal power constraint]\label{thm:maximal_lower_bound}
%Let $\Cc \in \Fc_{\text{m}}(n,M,\Upsilon)$ be a length-$n$ code of cardinality $M$ satisfying the maximal power constraint $\Upsilon$.
Let $\theta \geq \sigma$, $n\geq 1$. Then,
\begin{align} \label{eqn:maximal_lower_bound}
  \epsilon_{\text{m}}^{\star}(n,M,\Upsilon) \geq \alpha_{\frac{1}{M}} \bigl( \varphi^n_{\sqrt{\Upsilon},\sigma} , \varphi^n_{0,\theta} \bigr).
\end{align}
\end{theorem}
%%%%%%%%%%%%%%%%%%%%%%%%%%%%%%%%%%%%%%%%%%%%%%%%%%%%%%%%%%%%%%%%%%%%%%%%%%%%%%%%
\begin{IEEEproof}
See Section~\ref{sec:maximal_lower_bound}.
\end{IEEEproof}

Setting  $\theta^2=\Upsilon+\sigma^2$ in \refE{maximal_lower_bound}, we recover the bound from Theorem~\ref{thm:PPV_lower_bound}. We conclude that this lower bound also holds for maximal power constraints and not only for the restriction of equal power codewords. This is not the case however for the Shannon cone-packing bound from Theorem~\ref{thm:shannon_lower_bound} as we show with the following example.

We consider the problem of transmitting $M=16$ codewords over an additive Gaussian noise channel with $n=2$ dimensions. For $n=2$, Shannon cone-packing bound (SCPB) from Theorem~\ref{thm:shannon_lower_bound}  coincides with the ML decoding error probability of a $M$-PSK constellation $\Cc_{M\text{-PSK}}$ satisfying the equal power constraint $\Upsilon$ (as $2$-dimensional cones are precisely the ML decoding regions of the $M$-PSK constellation). For instance, for a $2$-dimensional Gaussian channel with a signal-to-noise ratio (SNR) $\frac{\Upsilon}{\sigma^2}=10$ and $M=16$ codewords, we obtain SCPB $=\Pe(\Cc_{16\text{-PSK}}) \approx 0.38$. Let now define a code $\Cc_{M\text{-APSK}}$ composed by the points of an $(M-1)$-PSK constellation and an additional codeword located at $\x = (0,0)$. While this code satisfies the maximal error constraint $\Upsilon$, its error probability violates SCPB for sufficiently large $M$.
Indeed, for the previous example, the modified codebook attains $\Pe(\Cc_{16\text{-APSK}}) \approx 0.34 < 0.38 \approx$ SCPB. We conclude that  in general Theorem~\ref{thm:shannon_lower_bound} holds only under an equal power constraint.

Evaluation of Corollary~\ref{cor:shannon_lower_bound} and Theorem~\ref{thm:maximal_lower_bound} with $\theta^2=\Upsilon+\sigma^2$ yields Cor.~1 $\approx 0.08$ and Th.~3 $\approx 0.15$, respectively. We can see that the direct lower bound from Theorem~\ref{thm:maximal_lower_bound} is tighter than that from Corollary~\ref{cor:shannon_lower_bound}. For a more detailed discussion comparing the bounds under different power constraints, see~\refS{numerical}.

\subsection{Proof of Theorem~\ref{thm:maximal_lower_bound}}
\label{sec:maximal_lower_bound}

We consider the set of input distributions $\X\sim P$ satisfying the maximal power constraint
\begin{align}
  \Pc_{\text{m}}(\Upsilon) \triangleq \Bigl\{ P \;\Big|\; \Pr\bigl[ \|\X\|^2 \leq n\Upsilon \bigr] =1 \Bigr\}.
\end{align}
Then, the meta-converse bound~\refE{metaconverse}
for some fixed $Q$ becomes
\begin{align}
\epsilon_{\text{m}}^{\star}(n,M,\Upsilon)
&\geq \inf_{P\in\Pc_{\text{m}}(\Upsilon)} \left\{\alpha_{\frac{1}{M}} \bigl(P W, P \times Q \bigr)\right\}.\label{eqn:metaconverse-maximal}
\end{align}
In order to make the minimization over $P$ tractable we shall use the following result.

\begin{lemma}\label{lem:alpha-split}%[{\hspace{-.1mm}\cite[Lem. 25]{Pol13}}]
Let $\bigl\{P_{\lambda}\bigr\}$ be a family of probability measures defined over the input alphabet $\Xc$, 
parametrized by $\lambda\in\RR$. Assume that the distributions $P_{\lambda}$ have pairwise disjoint supports
and that there exists a probability distribution $S$ over the parameter $\lambda$ such that
$P = \int P_{\lambda} S(\textrm{d} \lambda)$.
Then, the hypothesis testing error trade-off function satisfies 
\begin{align}\label{eqn:alpha-split}
&\alpha_{\beta} \bigl(PW, P\times Q\bigr) = \min_{\substack{\{\beta_{\lambda}\}:\\\beta=\int \beta_{\lambda} S(\textrm{d} \lambda) }} \int \alpha_{\beta_{\lambda}} \bigl(P_{\lambda} W, P_{\lambda} \times Q \bigr) S(\textrm{d} \lambda).
\end{align}
\end{lemma}
\begin{IEEEproof}
This lemma is analogous to the second part of \cite[Lem. 25]{Pol13}.
Since here we require the family $P_{\lambda}$ to be parametrized by a continuous $\lambda$, for completeness, we include the proof here.

First, we observe that $\alpha_{\beta} \bigl(PW, P\times Q\bigr)$ is a jointly convex function on $(\beta,P)$ \cite[Thm.~6]{Pol13}.
Let $\bigl\{\beta_{\lambda}\bigr\}$ and $\bigl\{P_{\lambda}\bigr\}$ satisfy
$\beta = \Ex_S[\beta_{\lambda}]=\int \beta_{\lambda} S(\textrm{d} \lambda)$ and $P=\Ex_S[P_{\lambda}]=\int P_{\lambda} S(\textrm{d} \lambda)$.
Then, using Jensen's inequality it follows that
\begin{align}
\Ex_S\bigl[ \alpha_{\beta_{\lambda}} \bigl(P_{\lambda} W, P_{\lambda} \times Q \bigr) \bigr] 
&\geq \alpha_{\Ex_S[\beta_{\lambda}]} \bigl(\Ex_S[P_{\lambda}] W,\, \Ex_S[P_{\lambda}] \times Q \bigr)
= \alpha_{\beta} \bigl(P W, P \times Q \bigr). \label{eqn:alpha-split-1}
\end{align}
We conclude that the right-hand side of \refE{alpha-split} is an upper bound on
$\alpha_{\beta} \bigl(P W, P \times Q \bigr)$.

To prove the identity \refE{alpha-split}, it remains to show that there exists $\bigl\{\beta_{\lambda}\bigr\}$ such that 
$\beta = \int \beta_{\lambda} S(\textrm{d} \lambda)$ and such that \refE{alpha-split-1} holds with equality.
We consider the Neyman-Pearson test for the original testing problem
$\alpha_{\beta} \bigl(P W, P \times Q \bigr)$, which is given by
\begin{align}\label{eqn:alpha-split-T}
  T(\x,\y) = \openone\biggl[\log\frac{W(\y|\x)}{Q(\y)}>t'\biggr] + c \openone\biggl[\log\frac{W(\y|\x)}{Q(\y)} = t'\biggr]
\end{align}
for some $t'\geq 0$ and $c\in[0,1]$
such that $\beta = \int T(\x,\y) Q(\textrm{d} \y) P(\textrm{d} \x)$.
We apply this test to the testing problem between $P_{\lambda} W$ and $P_{\lambda} \times Q$
and obtain a type-I and type-II error probabilities
\begin{align}
  \epsilon_{1}(\lambda) &= 1 -\int T(\x,\y) W(\textrm{d} \y|\x) P_{\lambda}(\textrm{d} \x),\\
  \epsilon_{2}(\lambda) &= \int T(\x,\y) Q(\textrm{d} \y) P_{\lambda}(\textrm{d} \x).
\end{align}
For the choice $\beta_{\lambda} = \epsilon_{2}(\lambda)$, 
the test \refE{alpha-split-T} is precisely the Neyman-Pearson test
of the problem 
$\alpha_{\beta_{\lambda}} \bigl(P_{\lambda} W, P_{\lambda} \times Q\bigr)$.
Therefore, 
\begin{align}
\Ex_S\bigl[ \alpha_{\beta_{\lambda}} \bigl(P_{\lambda} W, P_{\lambda} \times Q \bigr) \bigr] 
&= \int \epsilon_{1}(\lambda) S(\textrm{d} \lambda)\\
&= 1 - \int T(\x,\y) W(\textrm{d} \y|\x) P_{\lambda}(\textrm{d} \x) S(\textrm{d} \lambda)\\
&= 1 - \int T(\x,\y) W(\textrm{d} \y|\x) P(\textrm{d} \x)\\
&= \alpha_{\beta} \bigl(P W, P \times Q \bigr).
\end{align}
Similarly, we can show that $\Ex_S\bigl[\beta_{\lambda}\bigr] = \Ex_S\bigl[\epsilon_{2}(\lambda)\bigr] = \beta$.
We conclude that this choice of $\bigl\{\beta_{\lambda}\bigr\}$ yields equality in \refE{alpha-split-1}.
Given the bound \refE{alpha-split-1}, it also attains the minimum in \refE{alpha-split} and the result follows.
\end{IEEEproof}

Lemma \ref{lem:alpha-split} asserts that it is possible to express a binary hypothesis test as a convex combination of disjoint sub-tests provided that the type-II error is optimally distributed among them. 
For any $\gamma\geq 0$, we define the input set $\Sc_{\gamma} \triangleq \bigl\{ \x \,|\, \|\x\|^2=n\gamma \bigr\}$. In words, the set $\Sc_{\gamma}$ corresponds to the spherical shell centered at the origin that contains all input sequences with energy $n\gamma$. Note that, whenever $\gamma_1\neq\gamma_2$, the sets $\Sc_{\gamma_1}$ and $\Sc_{\gamma_2}$ are disjoint.
We now decompose the input distribution~$P$ based on the parameter $\gamma = \|\x\|^2/n$. To this end, we define the distribution $S(\gamma) \triangleq \Pr\{\X \in \Sc_{\gamma}\}$, and we let $P_{\gamma}$ be a distribution defined over $\Xc$ satisfying $P_{\gamma}(\x) = 0$ for any $\x\notin\Sc_{\gamma}$, and
\begin{align}\label{eqn:P_decomposition}
  P(\x) = \int P_{\gamma}(\x) S(\textrm{d}\gamma).
\end{align}
This condition implies that $P_{\gamma}(\x) = \frac{P(\x)}{S(\gamma)} \openone[\x \in \Sc_{\gamma}]$ for $S(\gamma)>0$ and where $\openone[\cdot]$ denotes the indicator function.
When $S(\gamma)=0$, then $P_{\gamma}$ can be an arbitrary distribution  such that $P_{\gamma}(\x) = 0$ for any $\x\notin\Sc_{\gamma}$.

Given \refE{P_decomposition} and since the measures $P_{\gamma}$ have disjoint supports for different values of $\gamma$, the conditions in Lemma~\ref{lem:alpha-split} hold for $\lambda \leftrightarrow \gamma$. Then, using \refE{alpha-split}, we obtain that the right-hand side of \refE{metaconverse-maximal} with $Q$ given by \refE{Q-theta-def} becomes
\begin{align}
\inf_{P\in\Pc_{\text{m}}(\Upsilon)} \left\{\alpha_{\frac{1}{M}} \bigl(P W, P \times Q \bigr)\right\}
&= \inf_{\substack{\{S,\beta_{\gamma}\}:\,\gamma\leq\Upsilon,\\ \int \beta_{\gamma} S(\textrm{d}\gamma)  = \frac{1}{M}}} \left\{ \int \alpha_{\beta_{\gamma}} \bigl(P_{\gamma} W, P_{\gamma} \times Q \bigr) S(\textrm{d}\gamma) \right\}
\label{eqn:metaconverse-split-1}\\
&= \inf_{\substack{\{S,\beta_{\gamma}\}:\,\gamma\leq\Upsilon,\\ \int \beta_{\gamma} S(\textrm{d}\gamma)   = \frac{1}{M} }} \left\{ \int
\alpha_{\beta_{\gamma}} \bigl(\varphi_{\sqrt{\gamma},\sigma}^n, \varphi_{0,\theta}^n \bigr) S(\textrm{d}\gamma) \right\},
\label{eqn:metaconverse-split-2}
\end{align}
where \refE{metaconverse-split-2} follows given the spherical
symmetry of each of the sub-tests in \refE{metaconverse-split-1}, since $\x=(\sqrt{\gamma},\ldots,\sqrt{\gamma})\in\Sc_{\gamma}$.

We transformed the original optimization over the $n$-dimensional distribution $P$ in the left-hand side of \refE{metaconverse-split-1} into an optimization over a one-dimensional distribution $S$ and auxiliary function $\beta_{\gamma}$ in the right-hand side of \refE{metaconverse-split-2}. To obtain the lower bound in the theorem, we make use of the following properties of the function $\alpha_{\beta}\bigl(\varphi_{\sqrt{\gamma},\sigma}^n, \varphi_{0,\theta}^n \bigr)$.
%%%%%%%%%%%%%%%%%%%%%%%%%%%%%%%%%%%%%%%%%%%%%%%%%%%%%%%%%%%%%%%%%%%%%%%%%%%%%%%%
\begin{lemma}\label{lem:alpha-decreasing-gamma}
Let $0 < \sigma \leq \theta$, with $\sigma,\theta\in\RR$ and $n \geq 1$. Then, the function 
\begin{align}\label{eqn:f-def}
f(\beta,\gamma) \triangleq \alpha_{\beta} \bigl(\varphi_{\sqrt{\gamma},\sigma}^n, \varphi_{0,\theta}^n \bigr)
\end{align}
is non-increasing in~$\gamma$ for any fixed $\beta\in[0,1]$, and convex non-increasing in~$\beta$ for any fixed $\gamma > 0$.
\end{lemma}
%%%%%%%%%%%%%%%%%%%%%%%%%%%%%%%%%%%%%%%%%%%%%%%%%%%%%%%%%%%%%%%%%%%%%%%%%%%%%%%%
\begin{IEEEproof}
 The minimum type-I error $\alpha$ is a non-increasing convex function of the type-II error~$\beta$ (see, \textit{e.g.},~\cite[Sec. I]{Pol13}).
Since $f(\beta,\gamma)$ characterizes the trade-off between the type-I and type-II errors of a hypothesis test, for fixed $\gamma\geq 0$, $f(\beta,\gamma)$ is non-increasing and convex in $\beta\in[0,1]$.

To characterize the behavior of $f(\beta,\gamma)$ with respect to $\gamma$, in Appendix~\ref{apx:f-beta-gamma} we show that $f(\beta,\gamma)$ is differentiable and obtain the derivative of $f(\beta,\gamma)$ with respect to $\gamma$. In particular, it follows from \refE{partial-f-g} that
\begin{align}
\frac{\partial f(\beta,\gamma)}{\partial \gamma}
&= - \frac{n}{2\delta} \biggl(\frac{t\delta}{\sigma^2\sqrt{n \gamma}}\biggr)^{\frac{n}{2}} e^{-\frac{1}{2}\left( \frac{n\gamma\sigma^2}{\delta^2} + \frac{t^2}{\sigma^2}\right)} I_{\frac{n}{2}}\biggl(\frac{t\sqrt{n\gamma}}{\delta}\biggr), \label{eqn:partial-f-g-bis}
\end{align}
where $\delta = \theta^2-\sigma^2$, $t$ satisfies $\beta(\gamma,t) = \beta$  for $\beta(\gamma,t)$ defined in \refE{beta-marcumQ} and $I_{m}(\cdot)$ denotes the $m$-th order modified Bessel function of the first kind.
For any $\gamma\geq 0$ and $\beta\in[0,1]$, the parameter $t$ that follows from the identity $\beta(\gamma,t) = \beta$ is non-negative. Then, using that $e^{-x/2}\geq 0$ and since $x\geq 0$ implies $I_{m}(x)\geq 0$, we conclude that \refE{partial-f-g-bis} is non-positive for any $\delta = \theta^2 -\sigma^2 > 0$. As a result, the function $f(\beta,\gamma) = \alpha_{\beta} \bigl(\varphi_{\sqrt{\gamma},\sigma}^n, \varphi_{0,\theta}^n \bigr)$ is non-increasing in $\gamma$ for any fixed value of $\beta$, provided that the conditions of the lemma hold.
\end{IEEEproof}

According to Lemma~\ref{lem:alpha-decreasing-gamma}, for any $0 \leq \gamma\leq \Upsilon$, it holds that $\alpha_{\beta} \bigl(\varphi_{\sqrt{\gamma},\sigma}^n, \varphi_{0,\theta}^n \bigr) = f(\beta,\gamma) \geq f(\beta,\Upsilon)$. As any maximal power constrained input distribution $P\in\Pc_{\text{m}}(\Upsilon)$ satisfies $S(\gamma) = 0$ for $\gamma>\Upsilon$, it follows that
\begin{align}
\inf_{\substack{\{S,\beta_{\gamma}\}:\, \gamma\leq\Upsilon,\\ \int \beta_{\gamma} S(\textrm{d}\gamma)  = \frac{1}{M} }} \left\{ \int
f\bigl(\beta_{\gamma},\gamma\bigr) S(\textrm{d}\gamma) \right\}
%\notag\\&\qquad\qquad
&\geq \inf_{\substack{\{S,\beta_{\gamma}\}:\, \gamma\leq\Upsilon,\\ \int  \beta_{\gamma} S(\textrm{d}\gamma) = \frac{1}{M} }} \left\{ \int
f\bigl(\beta_{\gamma},\Upsilon\bigr) S(\textrm{d}\gamma) \right\}\\
&\geq f\bigl(\tfrac{1}{M},\Upsilon\bigr), \label{eqn:metaconverse-bound-opt}
\end{align}
where in \refE{metaconverse-bound-opt} we used that the function $f(\beta,\Upsilon)$ is convex with respect to $\beta$ (Lemma~\ref{lem:alpha-decreasing-gamma}); hence, by Jensen's inequality and using the constraint $ \int  \beta_{\gamma} S(\textrm{d}\gamma) = \frac{1}{M}$, we obtain
\begin{align}
\int f\bigl(\beta_{\gamma},\Upsilon\bigr) S(\textrm{d}\gamma) \geq
f\bigl(\textstyle{\int \beta_\gamma S(\textrm{d}\gamma)},\Upsilon\bigr) = f\bigl(\tfrac{1}{M},\Upsilon\bigr).
\end{align}

Then, using \refE{metaconverse-maximal}, \refE{metaconverse-split-2} and \refE{metaconverse-bound-opt}, since $f\bigl(\frac{1}{M},\Upsilon\bigr) = \alpha_{\frac{1}{M}} \bigl(\varphi_{\sqrt{\Upsilon},\sigma}^n, \varphi_{0,\theta}^n \bigr)$, the result follows.

\section{Lower Bounds for Average Power Constraints}\label{sec:average}
%%%%%%%%%%%%%%%%%%%%%%%%%%%%%%%%%%%%%%%%%%%%%%%%%%%

In this section we study lower bounds to the error probability of codes satisfying an average power limitation. To this end, we first introduce some concepts of convex analysis.
The Legendre-Fenchel (LF) transform of a function $g$ is
\begin{align}
g^{*}(b) = \max_{a\in\Ac} \bigl\{\langle a,b \rangle - g(a)\bigr\},
\label{eqn:LF-transform}
\end{align}
where $\Ac$ is the domain of the function $g$ and $\langle a,b \rangle$ denotes the interior product between $a$ and $b$.

The function $g^*$ is usually referred to as Fenchel's conjugate (or convex conjugate) of $g$. If $g$ is a convex function with closed domain, applying the LF transform twice recovers the original function, \textit{i.e.}, $g^{**} = g$. If $g$ is not convex, applying the LF transform twice returns the lower convex envelope of $g$, which is the largest lower semi-continuous convex function function majorized by $g$.

For our problem, for $f(\beta,\gamma)=\alpha_{\beta} \bigl(\varphi_{\sqrt{\gamma},\sigma}^n, \varphi_{0,\theta}^n \bigr)$ with domain $\beta\in[0,1]$ and $\gamma\geq 0$, we define
\begin{align}
\underline{f}(\beta,\gamma) &\triangleq f^{**}(\beta,\gamma),
\label{eqn:conv-f}
\end{align}
and note that $\underline{f}(\beta,\gamma)\leq f(\beta,\gamma)$. 
The lower convex envelope \refE{conv-f} is a lower bound to the error probability in the average power constraint setting, as the next result shows.

%%%%%%%%%%%%%%%%%%%%%%%%%%%%%%%%%%%%%%%%%%%%%%%%%%%%%%%%%%%%%%%%%
\begin{theorem}[Converse, average power constraint]\label{thm:average_lower_bound}
%Let $\Cc \in \Fc_{\text{a}}(n,M,\Upsilon)$ be a length-$n$ code of cardinality $M$ satisfying the average power constraint $\Upsilon$.
Let $\theta \geq \sigma$, $n\geq 1$. Then,
\begin{align} \label{eqn:average_lower_bound}
  \epsilon_{\text{a}}^{\star}(n,M,\Upsilon) \geq \underline{f}\bigl(\tfrac{1}{M},\Upsilon\bigr),
\end{align}
where $\underline{f}(\beta,\gamma)$ is the lower convex envelope \refE{conv-f} of $f(\beta,\gamma)=\alpha_{\beta} \bigl(\varphi_{\sqrt{\gamma},\sigma}^n, \varphi_{0,\theta}^n \bigr)$.
\end{theorem}
%%%%%%%%%%%%%%%%%%%%%%%%%%%%%%%%%%%%%%%%%%%%%%%%%%%%%%%%%%%%%%%%%%%%%%%%%%
\begin{IEEEproof}
We start by considering the general meta-converse bound in~\refE{metaconverse}
where $\Pc$ is the set of distributions satisfying an average power constraint, i.e., $\Pc = \Pc_{\text{a}}(\Upsilon)$ with 
\begin{align}
  \Pc_{\text{a}}(\Upsilon) \triangleq \Bigl\{ \X\sim \Px \;\Big|\; \Ex\bigl[ \|\X\|^2 \bigr] \leq n\Upsilon \Bigr\}.\label{eqn:Pa_def}
\end{align}

Proceeding analogously as in \refE{metaconverse-split-1}-\refE{metaconverse-split-2}, but with the average power constraint $\int \gamma S(\textrm{d}\gamma) \leq \Upsilon$ instead of the maximal power constraint, it follows that
\begin{align}
&\inf_{P\in\Pc_{\text{a}}(\Upsilon)} \left\{\alpha_{\frac{1}{M}} \bigl(P W, P \times Q \bigr)\right\}
%\notag\\
%&\qquad\quad= \inf_{\substack{\{S,\beta_{\gamma}\}:\\
%\int \gamma \diff P_{\gamma} = \Upsilon\\
%\int \beta_{\gamma} S(\textrm{d}\gamma) =\frac{1}{M}}} \left\{ \int \alpha_{\beta_{\gamma}} \bigl(P_{\gamma} W, P_{\gamma} \times Q \bigr) S(\textrm{d}\gamma) \right\}
%\label{eqn:metaconverse-split-1}\\
%&\qquad\quad
= \inf_{\substack{\{S,\beta_{\gamma}\}:\\
\int \gamma S(\textrm{d}\gamma) \leq \Upsilon\\
\int \beta_{\gamma} S(\textrm{d}\gamma)  = \frac{1}{M}}} \left\{ \int
\alpha_{\beta_{\gamma}} \bigl(\varphi_{\sqrt{\gamma},\sigma}^n, \varphi_{0,\theta}^n \bigr) S(\textrm{d}\gamma) \right\}.
\label{eqn:metaconverse-split-2a}
\end{align}

The function $f(\beta,\gamma) = \alpha_{\beta} \bigl(\varphi_{\sqrt{\gamma},\sigma}^n, \varphi_{0,\theta}^n \bigr)$
is non-increasing in~$\gamma$ for any fixed $\beta\in[0,1]$ (see Lemma~\ref{lem:alpha-decreasing-gamma}).
Therefore, 
\begin{align}
\inf_{\substack{\{S,\beta_{\gamma}\}:\\
\int \gamma S(\textrm{d}\gamma) \leq \Upsilon\\
\int \beta_{\gamma} S(\textrm{d}\gamma)  = \frac{1}{M}}} \left\{ \int
f\bigl(\beta_{\gamma},\gamma\bigr) S(\textrm{d}\gamma) \right\}
%\notag\\[-8pt]&\qquad\qquad\qquad
&\geq
\inf_{\substack{\{S,\beta_{\gamma}\}:\\
\int \gamma S(\textrm{d}\gamma) = \Upsilon\\
\int \beta_{\gamma} S(\textrm{d}\gamma)  = \frac{1}{M}}} \left\{ \int
f\bigl(\beta_{\gamma},\gamma\bigr) S(\textrm{d}\gamma) \right\}.
\label{eqn:metaconverse-split-3a} 
\end{align}
That is, the average power constraint can be assumed to hold with equality as this relaxation does not increase the bound.
Using \refE{metaconverse-split-3a} and since $f(\beta,\gamma)\geq\underline{f}(\beta,\gamma)$,
we lower-bound the right-hand side of \refE{metaconverse-split-2a} as
\begin{align}
\inf_{\substack{\{S,\beta_{\gamma}\}:\\
\int \gamma S(\textrm{d}\gamma) \leq \Upsilon\\
\int \beta_{\gamma} S(\textrm{d}\gamma)  = \frac{1}{M}}} \left\{ \int
f\bigl(\beta_{\gamma},\gamma\bigr) S(\textrm{d}\gamma) \right\}
%\notag\\[-8pt]&\qquad\qquad\qquad
&\geq
\inf_{\substack{\{S,\beta_{\gamma}\}:\\
\int \gamma S(\textrm{d}\gamma) = \Upsilon\\
\int \beta_{\gamma} S(\textrm{d}\gamma)  = \frac{1}{M}}} \left\{ \int
\underline{f}\bigl(\beta_{\gamma},\gamma\bigr) S(\textrm{d}\gamma) \right\}
\label{eqn:metaconverse-split-4a}\\
&\geq
\inf_{\substack{\{S,\beta_{\gamma}\}:\\
\int \gamma S(\textrm{d}\gamma) = \Upsilon\\
\int \beta_{\gamma} S(\textrm{d}\gamma)  = \frac{1}{M}}} 
\Bigl\{ \underline{f}\bigl(\textstyle\int\beta_{\gamma}S(\textrm{d}\gamma),\textstyle\int\gamma S(\textrm{d}\gamma)\bigr) \Bigr\}
\label{eqn:metaconverse-split-5a}\\
&=
\underline{f}\bigl(\tfrac{1}{M},\Upsilon\bigr),
\label{eqn:metaconverse-split-6a}
\end{align}
where \refE{metaconverse-split-5a} follows from applying Jensen's inequality, and
\refE{metaconverse-split-6a} holds since, given the constraints $\int\beta_{\gamma}S(\textrm{d}\gamma)=\frac{1}{M}$ and
$\int\gamma S(\textrm{d}\gamma)=\Upsilon$, the objective of the optimization does not depend on $\{S,\beta_{\gamma}\}$. 
The bound \refE{conv-f} then follows from combining \refE{metaconverse},
\refE{metaconverse-split-2a} and the inequalities  \refE{metaconverse-split-4a}-\refE{metaconverse-split-6a}.
\end{IEEEproof}

The function $\underline{f}(\beta,\gamma)$ can be evaluated numerically by considering a $2$-dimensional grid on the parameters $(\beta,\gamma)$, computing $f(\beta,\gamma)$ over this grid, and obtaining
the corresponding convex envelope. Nevertheless, sometimes $\underline{f}(\beta,\gamma) = f(\beta,\gamma)$ and these steps can be avoided, as the next result shows.

\begin{lemma}\label{lem:envelope_equals_f}
Let $\sigma,\theta,\gamma>0$ and $n\geq 1$, be fixed parameters, and define $\delta \triangleq\theta^2-\sigma^2$. For $t\geq 0$, we define
\begin{align}
\!\xi_{1}(t) &\triangleq Q_{\frac{n}{2}}\left( \sqrt{n\gamma}\frac{\sigma}{\delta}, \frac{t}{\sigma} \right) -  Q_{\frac{n}{2}}\left( 0, \sqrt{\bigl(\tfrac{t^2}{\sigma^2}-n\gamma\tfrac{\theta^2}{\delta^2} \bigr)_{+}} \right),\\
\!\xi_{2}(t) &\triangleq \frac{\theta^n}{\sigma^n} e^{\frac{1}{2}\left(\frac{n\gamma}{\delta}-\frac{\delta t^2}{\sigma^2\theta^2}\right)}
   \biggl( Q_{\frac{n}{2}}\left( 0, \sqrt{\bigl(\tfrac{t^2}{\theta^2}-n\gamma\tfrac{\sigma^2}{\delta^2} \bigr)_{+}} \right)
   %\notag\\  &\qquad\qquad\qquad\qquad\qquad 
  -Q_{\frac{n}{2}}\left( \sqrt{n\gamma}\frac{\theta}{\delta}, \frac{t}{\theta} \right) \biggr),\\
\!\xi_{3}(t) &\triangleq \frac{n \gamma}{2\delta}\biggl(\frac{t\delta}{\sigma^2\sqrt{n\gamma}}\biggr)^{\frac{n}{2}} 
   e^{-\frac{1}{2}\left(n\gamma\frac{\sigma^2}{\delta^2}+\frac{t^2}{\sigma^2}\right)} I_{\frac{n}{2}}\Bigl(\sqrt{n\gamma} \frac{t}{\delta}\Bigr),\!
\end{align}
where $(a)_{+} = \max(0,a)$,  $Q_m(a,b)$ is the Marcum $Q$-function and $I_{m}(\cdot)$ denotes the $m$-th order modified Bessel function of the first kind.
Let $t_0$ be the solution to the implicit equation
\begin{align} \label{eqn:avmc_t1_cond}
  \xi_{1}(t_0) + \xi_{2}(t_0) + \xi_{3}(t_0) = 0.
\end{align}
Then, for any $\beta$ satisfying $\bigl(1- Q_{\frac{n}{2}}\bigl( \sqrt{n\gamma}{\theta}/{\delta},\, t_0/{\theta} \bigr) \bigr) \leq \beta\leq 1$, it holds that
\begin{align} \label{eqn:envelope_equals_f}
  \underline{f}(\beta,\gamma) = f(\beta,\gamma).
\end{align}
\end{lemma}
\begin{IEEEproof}
See Appendix~\ref{apx:envelope_equals_f}.
\end{IEEEproof}

Combining Theorem~\ref{thm:average_lower_bound} and Lemma~\ref{lem:envelope_equals_f} we obtain a simple lower bound on the error probability of any code satisfying an average-power constraint, provided that its cardinality is below a certain threshold.

\begin{corollary}\label{cor:average_metaconverse_bound}
Let $\sigma,\theta>0$ and $n\geq 1$, be fixed parameters, and $\delta = \theta^2-\sigma^2$. Let $t_0$ be the solution to the implicit equation \refE{avmc_t1_cond} with $\gamma=\Upsilon$ and define
\begin{align} \label{eqn:avmc_barM_def}
\bar{M}_n \triangleq \left(1- Q_{\frac{n}{2}}\biggl( \frac{\sqrt{n\Upsilon}{\theta}}{\delta},\, \frac{t_0}{\theta} \biggr)\right)^{-1}.
\end{align}
Then, for any code $\Cc \in \Fc_{\text{a}}(n,M,\Upsilon)$ with cardinality $M\leq\bar{M}_n$,
\begin{align} \label{eqn:cor_average_metaconverse_bound}
  \epsilon_{\text{a}}^{\star}(n,M,\Upsilon) \geq \alpha_{\frac{1}{M}} \bigl( \varphi^n_{\sqrt{\Upsilon},\sigma} , \varphi^n_{0,\theta} \bigr).
\end{align}
\end{corollary}

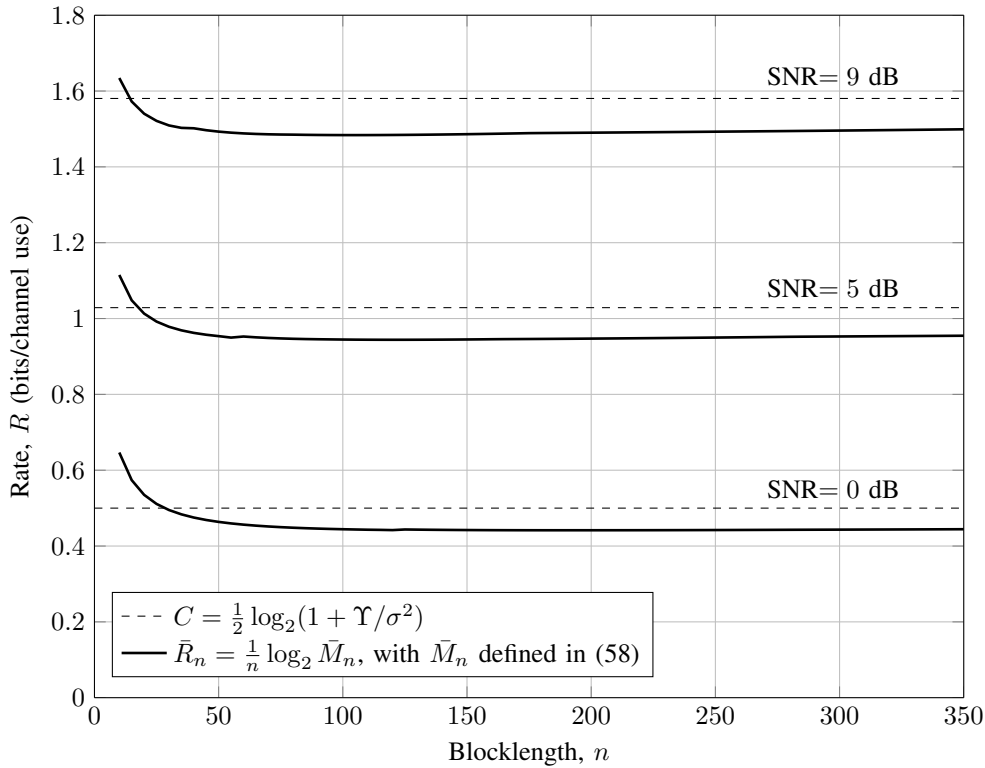
\begin{figure}[t]%
\centering% This file was created by matlab2tikz v0.6.0 running on MATLAB 8.2.
%Copyright (c) 2008--2014, Nico Schlömer <nico.schloemer@gmail.com>
%All rights reserved.
%Minimal pgfplots version: 1.3
%  http://www.mathworks.com/matlabcentral/fileexchange/22022-matlab2tikz
\begin{tikzpicture}

\begin{axis}[%
width=.7\linewidth,
height=.55\linewidth,
scale only axis,
unbounded coords=jump,
separate axis lines,
every outer x axis line/.append style={black},
every x tick label/.append style={font=\color{black}},
xmin=0,
xmax=350,
xlabel={Blocklength, $n$},
xmajorgrids,
every outer y axis line/.append style={black},
every y tick label/.append style={font=\color{black}},
ymin=0,
ymax=1.8,
ylabel={Rate, $R$ (bits/channel use)},
ymajorgrids,
legend style={at={(0.02,0.03)},anchor=south west,legend cell align=left,align=left,draw=black}
]
\addplot [color=black,dashed]
  table[row sep=crcr]{%
0	1.58040221195651\\
350	1.58040221195651\\
};
\addlegendentry{$C = \frac{1}{2}\log_2(1+\Upsilon/\sigma^2)$};

\addplot [color=black,solid,line width=1pt]
  table[row sep=crcr]{%
10	1.63436041501144\\
15	1.57234838807521\\
20	1.5402434082836\\
25	1.52128889892891\\
30	1.50924439741826\\
35	1.50285114269349\\
40	1.50175916566835\\
45	1.49664981187623\\
50	1.49293749175944\\
55	1.49022254360489\\
60	1.48824668525533\\
65	1.4868367005258\\
70	1.48587282848622\\
75	1.48527016444755\\
80	1.48494998291861\\
85	1.48446489852515\\
90	1.4841263094383\\
95	1.48390944566065\\
100	1.48379526797434\\
105	1.48376897642328\\
110	1.48381895435748\\
115	1.48393600848252\\
120	1.48411281375356\\
125	1.4843435023822\\
130	1.48462335576679\\
135	1.48494857094968\\
140	1.4853160817295\\
145	1.48572342032823\\
150	1.4861686094805\\
155	1.48665007757653\\
160	1.48716659144155\\
165	1.48771720272789\\
170	1.48830120490239\\
175	1.48891809854623\\
180	1.48919525820704\\
185	1.4894319944881\\
190	1.48967201264621\\
195	1.48991513671801\\
200	1.49016123801582\\
205	1.49041022708554\\
210	1.49066204712346\\
215	1.49091666857101\\
220	1.49117408466397\\
225	1.49143430775867\\
230	1.49169736629253\\
235	1.49196330226459\\
240	1.49223216914352\\
245	1.4925040301282\\
250	1.49277895670005\\
255	1.4930570274172\\
260	1.49333832690984\\
265	1.49362294504342\\
270	1.49391097622191\\
275	1.49420251880852\\
280	1.49449767464492\\
285	1.49479654865335\\
290	1.49509924850851\\
295	1.49540588436837\\
300	1.49571656865486\\
305	1.49603141587665\\
310	1.49635054248781\\
315	1.49667406677684\\
320	1.49700210878154\\
325	1.497334790226\\
330	1.49767223447631\\
335	1.49801456651242\\
340	1.49836191291372\\
345	1.49871440185647\\
350	1.49907216312133\\
};
\addlegendentry{$\bar{R}_n = \frac{1}{n}\log_2 \bar{M}_n$, with $\bar{M}_n$ defined in \refE{avmc_barM_def}};

\addplot [color=black,dashed,forget plot]
  table[row sep=crcr]{%
0	1.0286866043034\\
350	1.0286866043034\\
};
\addplot [color=black,solid,line width=1pt,forget plot]
  table[row sep=crcr]{%
10	1.1149149357214\\
15	1.04806390506688\\
20	1.01302623984698\\
25	0.991983412353077\\
30	0.978299366107811\\
35	0.968880177317102\\
40	0.962161300811329\\
45	0.957246724456531\\
50	0.953570901689023\\
55	0.949726447810113\\
60	0.952551133831914\\
65	0.950437883497042\\
70	0.948798132184642\\
75	0.947532344648106\\
80	0.946567834981187\\
85	0.945850506997717\\
90	0.94533943703603\\
95	0.944968891703713\\
100	0.944594550160006\\
105	0.944323597114843\\
110	0.944141160825606\\
115	0.944035341793648\\
120	0.943996540896525\\
125	0.944016959130737\\
130	0.944090220590137\\
135	0.944211085143552\\
140	0.944375227217016\\
145	0.944579063854191\\
150	0.944819619903969\\
155	0.945094421458165\\
160	0.945401410983693\\
165	0.945664074862339\\
170	0.945829765776423\\
175	0.946008386467375\\
180	0.946198964104328\\
185	0.946400675403031\\
190	0.94661282351382\\
195	0.946834818959818\\
200	0.947066163844537\\
205	0.947306438712348\\
210	0.94755529157223\\
215	0.947812428693884\\
220	0.948077606862459\\
225	0.94835062683883\\
230	0.948631327820338\\
235	0.948919582735075\\
240	0.94921529423323\\
245	0.949518391263479\\
250	0.949828826142112\\
255	0.950146572038512\\
260	0.950471620813635\\
265	0.95080398115866\\
270	0.951143676989689\\
275	0.95149074606151\\
280	0.951845238769283\\
285	0.952118804326488\\
290	0.952310438145064\\
295	0.952502341445976\\
300	0.952694557193319\\
305	0.952887130187145\\
310	0.953080106735897\\
315	0.953273534378481\\
320	0.953467461648827\\
325	0.953661937876816\\
330	0.953857013020427\\
335	0.954052737524637\\
340	0.954249162203319\\
345	0.9544463381409\\
350	0.954644316610984\\
};
\addplot [color=black,dashed,forget plot]
  table[row sep=crcr]{%
0	0.5\\
350	0.5\\
};
\addplot [color=black,solid,line width=1pt,forget plot]
  table[row sep=crcr]{%
10	0.646700637456979\\
15	0.573836439405738\\
20	0.535021593008207\\
25	0.51110295109314\\
30	0.495030463416759\\
35	0.483602987827676\\
40	0.47514457096989\\
45	0.468693462593751\\
50	0.463659609570658\\
55	0.45966186945254\\
60	0.456430595767643\\
65	0.453789837270967\\
70	0.451612446367545\\
75	0.449804511898956\\
80	0.448295800295142\\
85	0.44703288583313\\
90	0.445974473582111\\
95	0.445075990314143\\
100	0.444294777495486\\
105	0.443639414029623\\
110	0.443045837143952\\
115	0.442629014252243\\
120	0.442039516313615\\
125	0.443643442324062\\
130	0.443260944066015\\
135	0.442919417415931\\
140	0.442629437935429\\
145	0.442385218943499\\
150	0.442181837430447\\
155	0.442015084912696\\
160	0.44188134764945\\
165	0.441777509776434\\
170	0.441700874465278\\
175	0.441649099364409\\
180	0.441620143432311\\
185	0.441612222917045\\
190	0.441623774723369\\
195	0.441635466786989\\
200	0.441634897587121\\
205	0.44164673984028\\
210	0.441670070253088\\
215	0.441704070118809\\
220	0.441748012068394\\
225	0.441801248741745\\
230	0.441863203068414\\
235	0.441933359902439\\
240	0.442011258800662\\
245	0.44209648776996\\
250	0.442188677838305\\
255	0.442287498328377\\
260	0.442392652732355\\
265	0.442503875102469\\
270	0.442620926885388\\
275	0.442743594139622\\
280	0.442871685084208\\
285	0.442979985381006\\
290	0.443069618393244\\
295	0.44316202732823\\
300	0.443257092104558\\
305	0.443354704525319\\
310	0.443454767134794\\
315	0.443557192200625\\
320	0.443661900806112\\
325	0.443768822039559\\
330	0.443877892269144\\
335	0.443989054493416\\
340	0.444102257758678\\
345	0.444217456635692\\
350	0.444334610749053\\
};
\node at (rel axis cs:0.85,0.305){SNR$=0$ dB};
\node at (rel axis cs:0.85,0.6) {SNR$=5$ dB};
\node at (rel axis cs:0.85,0.91) {SNR$=9$ dB};

\end{axis}
\end{tikzpicture}%
%
\caption{Condition from Corollary~\ref{cor:average_metaconverse_bound} for AWGN channels with SNR of $0$ dB, $5$ dB and $9$ dB compared with the channel capacity $C$.}\label{fig:AWGN-avcond}
\end{figure}%

%Corollary~\ref{cor:average_metaconverse_bound} implies that the PPV'10 lower bound holds in the average power constraint setting if the cardinality of the codebook is sufficiently small. 
\refFig{AWGN-avcond} shows the condition $M\leq\bar{M}_n$ in Corollary~\ref{cor:average_metaconverse_bound} as an upper bound on the transmission rate of the system, given by $R = \frac{1}{n}\log_2 M$. In this example, we let $\theta^2=\Upsilon+\sigma^2$ and consider three different values of the signal-to-noise ratio, SNR=$10\log_{10}\frac{\Upsilon}{\sigma^2}$. 
The channel capacity $C=\frac{1}{2}\log_2\bigl(1+\frac{\Upsilon}{\sigma^2}\bigr)$ for each of the SNRs is also included for reference.
According to Corollary~\ref{cor:average_metaconverse_bound},
for any code satisfying an average power constraint $\Upsilon$ with blocklength $n$ and rate $R \leq \bar{R}_n$, the lower bound \refE{cor_average_metaconverse_bound} holds. 
Then, in \refF{AWGN-avcond} we can see that the condition in Corollary~\ref{cor:average_metaconverse_bound} holds except for rates very close to and above capacity (provided that the blocklegth $n$ is sufficiently large).
For rates above the curves $\bar{R}_n$ shown in the figure, \refE{cor_average_metaconverse_bound} does not apply and the refined bound from Theorem~\ref{thm:average_lower_bound} needs to be used instead.

This observation agrees with previous results in the literature. Indeed, the asymptotic analysis of the right-hand side of \refE{cor_average_metaconverse_bound} yields a strong converse behavior for rates above capacity~\cite[Th. 74]{PolThesis}. Nevertheless, \cite[Th. 77]{PolThesis} shows that under an average power constraint no strong converse holds, and therefore the bound \refE{cor_average_metaconverse_bound} cannot hold in general.

\subsection{Optimal input distribution}
\label{sec:mc-implicit-distribution}

A careful analysis of the derivation of the bounds in Theorem~\ref{thm:average_lower_bound} and Corollary~\ref{cor:average_metaconverse_bound} shows that they
are indeed tight in the sense that, for the auxiliary distribution $Q$ given in \refE{Q-theta-def}, they correspond to the tightest meta-converse bound
\begin{align}
\inf_{P\in\Pc_{\text{a}}(\Upsilon)} \left\{\alpha_{\frac{1}{M}} \bigl(P W, P \times Q \bigr)\right\} =  \underline{f}\bigl(\tfrac{1}{M},\Upsilon\bigr).
\label{eqn:mc-average-uf}
\end{align}
Moreover, the optimal input distribution in the left-hand side of \refE{mc-average-uf} is characterized by the convex envelope $ \underline{f}$.
To see this, note that the right-hand side of \refE{mc-average-uf} corresponds to the value of the convex envelope $\underline{f}$ at the point $\bigl(\tfrac{1}{M},\Upsilon\bigr)$.
Using Carath\'eodory's theorem, it follows that any point on $\underline{f}$ can be written as a convex combination of (at most) $4$ points in $f$.\footnote{For a $2$-dimensional function its epigraph is a $3$-dimensional set. Therefore, Carath\'eodory's theorem implies that at most $3+1$ points are needed to construct the convex hull of the epigraph, which corresponds to the convex envelope of the original function.}
Let us denote these $4$ points as $\bigl(\beta_i,\gamma_i\bigr)$, $i=1,\ldots,4$.
Then, for some $\lambda_i\geq 0$, $i=1,\ldots,4$, such that $\sum_{i=1}^4 \lambda_i=1$,
the following identities hold
\begin{align}
  \underline{f}\bigl(\tfrac{1}{M},\Upsilon\bigr) = \sum_{i=1}^4 \lambda_i f\bigl(\beta_i,\gamma_i\bigr),\quad
  \frac{1}{M} = \sum_{i=1}^4 \lambda_i \beta_i,\quad
  \Upsilon = \sum_{i=1}^4 \lambda_i \gamma_i.
\end{align}

Let $S$ be the probability distribution that sets its mass points at $\gamma_i$, $i=1,\ldots,4$, with probabilities $S(\gamma_i) = \lambda_i$. Let $\beta_{\gamma_i} = \beta_i$, $i=1,\ldots,4$. This choice of $\{S,\beta_{\gamma}\}$ satisfies the constraints of the left-hand side of \refE{metaconverse-split-4a}. Moreover, for this choice of $\{S,\beta_{\gamma}\}$ the left-hand side of \refE{metaconverse-split-4a} becomes $\sum_{i=1}^4 \lambda_i f\bigl(\beta_i,\gamma_i\bigr) = \underline{f}\bigl(\tfrac{1}{M},\Upsilon\bigr)$ and therefore the inequality chain in \refE{metaconverse-split-4a}-\refE{metaconverse-split-6a} holds with equality.
Also, increasing the power limit $\Upsilon$ yields a strictly lower error probability, and therefore  \refE{metaconverse-split-3a} holds with equality.
Then, using \refE{metaconverse-split-2a} we conclude that the identity \refE{mc-average-uf} holds.

To characterize the input distribution minimizing the left-hand side of \refE{mc-average-uf}, we recall
\refE{P_decomposition}. We conclude that the input distribution $P$ optimizing the left-hand side of \refE{metaconverse-split-2a} concentrates its mass on (at most) 4 spherical shells with squared radius $n\gamma_i$, $i=1,\ldots,4$. The probability of each of these shells is precisely $S(\gamma_i) = \lambda_i$ and it is uniformly distributed over the surface of each of the shells~\cite[Sec. VI.F]{Pol13}.

\begin{figure}[t]
\centering\input{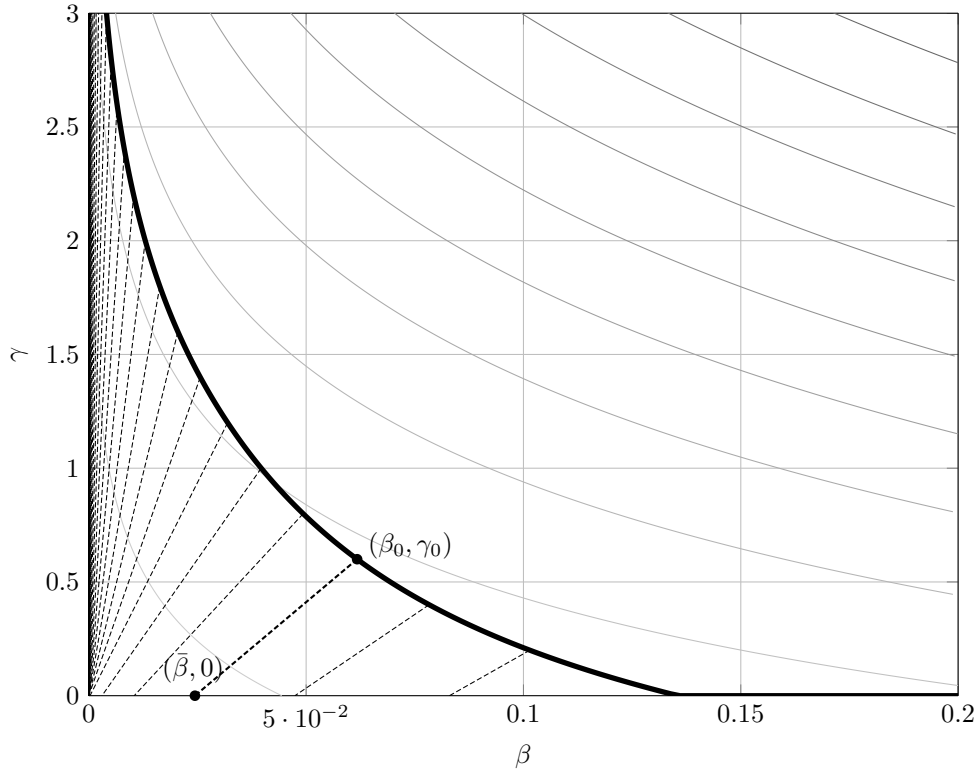}%
\caption{In gray, the level curves of $f(\beta,\gamma)$ for $n=6$, $\sigma^2=1$, $\theta^2=2$. The bold line corresponds to the boundary \refE{chull-boundary}. The dashed lines correspond to the convex combinations that yield the convex envelope $\underline{f}(\beta,\gamma)$ below the boundary.}\label{fig:chull-n6-s1-t2}%
\end{figure}%

The results above hold for a general function $f(\beta,\gamma)$ (provided that it is strictly decreasing in $\gamma$). The structure of the convex envelope of the function $f(\beta,\gamma)=\alpha_{\beta} \bigl(\varphi_{\sqrt{\gamma},\sigma}^n, \varphi_{0,\theta}^n \bigr)$ is implicit in the proof of Lemma~\ref{lem:envelope_equals_f} in Appendix~\ref{apx:envelope_equals_f}.
Let $(\beta_0, \gamma_0)$ denote the boundary described in Lemma~\ref{lem:envelope_equals_f}, i.e., the points $(\beta_0, \gamma_0)$ satisfying
\begin{align}
  \beta_0 = 1- Q_{\frac{n}{2}}\bigl( \sqrt{n\gamma_0}{\theta}/{\delta},\, t_0/{\theta} \bigr) \label{eqn:chull-boundary}. 
\end{align}
where $t_0$ is the solution of \refE{avmc_t1_cond} for $\gamma=\gamma_0$. We consider the two regions:

\subsubsection{Above the boundary \refE{chull-boundary}} In this region $\gamma \geq 0$ and $\bigl(1- Q_{\frac{n}{2}}\bigl( \sqrt{n\gamma}{\theta}/{\delta},\, t_0/{\theta} \bigr) \bigr) \leq \beta\leq 1$ with $t_0$ the solution of \refE{avmc_t1_cond}.
It follows that $\underline{f} = f$ and therefore $\underline{f}$ is the convex combination of a single point of $f$.

\subsubsection{Below the boundary \refE{chull-boundary}}  For $\gamma \geq 0$ and $0 \leq \beta < \bigl(1- Q_{\frac{n}{2}}\bigl( \sqrt{n\gamma}{\theta}/{\delta},\, t_0/{\theta} \bigr) \bigr)$, $\underline{f}$ and $f$ do not coincide.
Instead, the convex envelope $\underline{f}$ evaluated at $(\beta,\gamma)$ corresponds to a convex combination of the function $f$ at the points $(\beta_0,\gamma_0)$ and $(\bar\beta,\bar{\rho})$, where $(\beta_0,\gamma_0)$ satisfies \refE{chull-boundary}, $\bar{\rho}=0$ and $\bar\beta = 1- Q_{\frac{n}{2}}\bigl(0,\, \bar{t}_{\star}/{\theta} \bigr)$ for $\bar{t}_{\star}$ in \refE{first-order-opt-bart}.

An example of this property is shown in \refF{chull-n6-s1-t2}. This figure shows in gray the level lines of $f\bigl(\beta,\gamma\bigr)$; the bold line corresponds to the boundary \refE{chull-boundary}; and with dashed lines we show the convex combinations between  $(\beta_0,\gamma_0\bigr)$ and $(\bar\beta,0\bigr)$ for different values of $\gamma_0$.
Above the boundary, the convex envelope $\underline{f}$ coincides with $f$, and therefore the input distribution $P$ optimizing the left-hand side of \refE{mc-average-uf} is uniform over a spherical shell centered at the origin and with squared radius $n\Upsilon$.
Below the boundary, the convex envelope $\underline{f}$ corresponds to the convex combination of two points of $f$ and the optimal input distribution $P$ corresponds to a mass point at the origin ($\bar{\rho}=0$) and a spherical shell centered at the origin and with squared radius $n\gamma_0$, $\gamma_0 \geq \Upsilon$.
The probability mass of these two components of $P$ depends on the parameters of the system. %While this distribution does not describe how the codewords of a good code are distributed over the space, it suggest that several codewords could be concentrated at $(0,0)$.

%Numerical evaluation of the condition from Corollary~\ref{cor:average_metaconverse_bound} shows that the bound \refE{cor_average_metaconverse_bound} holds for practical SNR ranges and error probabilities of interest. A particular example for which the difference between the lower-bounds obtained from $f(\beta,\gamma)$ and from $\underline{f}(\beta,\gamma)$ becomes relevant will be presented in Section~\ref{sec:constellations}.

%%%%%%%%%%%%%%%%%%%%%%%%%%%%%%%%%%%%%%%%%%%%%%%%%%%%%%%%%
\section{Computation of $f(\beta,\gamma)=\alpha_{\beta} \bigl(\varphi_{\sqrt{\gamma},\sigma}^n, \varphi_{0,\theta}^n \bigr)$}
\label{sec:computation}

In the previous sections we showed that both the function $f(\beta,\gamma)$ and its convex envelope $\underline{f}(\beta,\gamma)$ yield lower bounds to the error probability for Gaussian channels under different power constraints. In this section we provide several tools that can be used in the numerical evaluation of these functions.

\subsection{Exact computation of $f(\beta,\gamma)$}

Proposition~\ref{prop:alpha-beta-marcumQ} in Appendix~\ref{apx:f-beta-gamma} provides a parametric formulation of the function $f(\beta,\gamma)$. A non-parametric expression for $f(\beta,\gamma)$ can be obtained by combining Proposition~\ref{prop:alpha-beta-marcumQ} and \cite[Lem.~1]{tit19qp}.
\begin{theorem}[Non-parametric formulation]\label{thm:alpha-beta-opt}
Let $\sigma,\theta>0$ and $n\geq 1$, be fixed parameters. 
Then, it holds that
\begin{align}
f(\beta,\gamma)=\max_{t\geq 0} \Biggl\{ Q_{\frac{n}{2}}\left(\sqrt{n\gamma}\frac{\sigma}{\delta},\frac{t}{\sigma} \right)
 + \frac{\theta^n}{\sigma^n} e^{\frac{1}{2}\left(\frac{n \gamma}{\delta}-\frac{\delta t^2}{\sigma^2\theta^2}\right)}\left( 1 - \beta -Q_{\frac{n}{2}}\left(\sqrt{n\gamma}\frac{\theta}{\delta},\frac{t}{\theta} \right) \right)\Biggr\}.\label{eqn:alpha-beta-opt}
\end{align}
\end{theorem}
\begin{IEEEproof}
According to \cite[Lem.~1]{tit19qp}, we obtain the following alternative expression for $\alpha_{\beta}\bigl(\varphi_{\sqrt{\gamma},\sigma}^n, \varphi_{0,\theta}^n \bigr)$,
\begin{align}
\alpha_{\beta}\bigl(\varphi_{\sqrt{\gamma},\sigma}^n, \varphi_{0,\theta}^n \bigr)
= \max_{t'} \Bigl\{ \Pr\bigl[  j(\Y_{0}) \leq t' \bigr]
\!+\!e^{t'}\bigl( \Pr\bigl[  j(\Y_{1}) > t'\bigr]
          \!-\! \beta \bigr)\!\Bigr\},\hspace{-1mm}\label{eqn:HT-im-formulation}
\end{align}
where $\Y_0 \sim \varphi_{\sqrt{\gamma},\sigma}^n$ and $\Y_1 \sim \varphi_{0,\theta}^n$
and where $j(\y)$ denotes the log-likelihood ratio
\begin{align}
  j(\y) &= \log \frac{\varphi_{\sqrt{\gamma},\sigma}^n(\y)}{\varphi_{0,\theta}^n (\y)}\\
        &=  n \log\frac{\theta}{\sigma} - \frac{1}{2} \sum_{i=1}^{n} \frac{\theta^2(y_i-\sqrt{\gamma})^2-\sigma^2 y_i^2}{\sigma^2 \theta^2}.
\end{align}
Following analogous steps as in the proof of Proposition~\ref{prop:alpha-beta-marcumQ} in Appendix~\ref{apx:f-beta-gamma}, we obtain that 
\begin{align}
\Pr\bigl[  j(\Y_{0}) \leq t' \bigr] &= Q_{\frac{n}{2}}\left(\sqrt{n\gamma}\frac{\sigma}{\delta},\frac{t}{\sigma} \right),\label{eqn:PjY0-marcumQ}\\
\Pr\bigl[  j(\Y_{1}) > t'\bigr] &= 1-Q_{\frac{n}{2}}\left(\sqrt{n\gamma}\frac{\theta}{\delta},\frac{t}{\theta} \right),\label{eqn:PjY1-marcumQ}
\end{align}
where $t'\leftrightarrow t$ are related according to \refE{change-of-variable-t}, c.f., $e^{t'} = \frac{\theta^n}{\sigma^n} e^{\frac{1}{2}\left(\frac{n \gamma}{\delta}-\frac{\delta t^2}{\sigma^2\theta^2}\right)}$. Then, the result follows from \refE{HT-im-formulation}, \refE{PjY0-marcumQ} and \refE{PjY1-marcumQ} via the change of variable $t'\leftrightarrow t$.
\end{IEEEproof}

The formulation in Theorem~\ref{thm:alpha-beta-opt} allows to obtain simple lower bounds on $f(\beta,\gamma)$ by fixing the value of $t$ in \refE{alpha-beta-opt}, and Verd\'u-Han-type lower bounds by using that $Q_{\frac{n}{2}}\bigl(\sqrt{n\gamma}\frac{\theta}{\delta},\frac{t}{\theta} \bigr) \leq 1$, hence
 \begin{align}
f(\beta,\gamma) \geq \max_{t\geq 0} \Biggl\{ Q_{\frac{n}{2}}\left(\sqrt{n\gamma}\frac{\sigma}{\delta},\frac{t}{\sigma} \right)
 - \frac{\theta^n}{\sigma^n} e^{\frac{1}{2}\left(\frac{n \gamma}{\delta}-\frac{\delta t^2}{\sigma^2\theta^2}\right)}\beta\Biggr\}.
\end{align}

In order to evaluate \refE{alpha-beta-opt} in Theorem~\ref{thm:alpha-beta-opt} we need to solve a maximization over the scalar parameter $t\geq 0$ with an objective involving two Marcum-$Q$ functions. The computation of the bounds from Theorems \ref{thm:PPV_lower_bound}, \ref{thm:maximal_lower_bound} and \ref{thm:average_lower_bound} for a fixed rate $R\triangleq\frac{1}{n}\log_2 M$, implies that the parameter $\beta = 2^{-n R}$ decreases exponentially with the blocklength $n$. Then, traditional Taylor series expansions of the Marcum-$Q$ function fail to achieve the required precision even for moderate values of $n$. In this regime, the following expansion yields an accurate approximation of $f(\beta,\gamma)$.

\subsection{Saddlepoint expansion of $f(\beta,\gamma)$}
\label{sec:saddlepoint}

We define the information density
\begin{align}
  j(y) \triangleq  \log \frac{\varphi_{\sqrt{\gamma},\sigma}(y)}{\varphi_{0,\theta} (y)} =  \log\frac{\theta}{\sigma}
  - \frac{1}{2} \frac{\theta^2(y-\sqrt{\gamma})^2-\sigma^2 y^2}{\sigma^2 \theta^2}, \label{eqn:j1-def}
\end{align}
and we consider the cumulant generating function of the random variable $j(Y)$, $Y \sim \varphi_{0,\theta}$, given by
\begin{align}
  \kappa_{\gamma}(s)
  &\triangleq \log \int_{-\infty}^{\infty}
\frac{\varphi_{\sqrt{\gamma},\sigma}(y)^s }{\varphi_{0,\theta}(y)^{s-1}} \diff y
%\label{eqn:kappa-Gaussian}\\&
= {\gamma} \frac{s(s-1)}{2\eta(s)}+\log\frac{\theta^{s}\sigma^{1-s}}{\sqrt{\eta(s)}},\label{eqn:d0kappa-Gaussian}
\end{align}
where we defined $\eta(s) \triangleq s\theta^2 +(1-s) \sigma^2$.
The first three derivatives of \refE{d0kappa-Gaussian} with respect to $s$ are, respectively,
\begin{align}
  \kappa_{\gamma}'(s) &= {\gamma} \frac{s^2 \theta^2 - (1-s)^2 \sigma^2}{2\eta(s)^2} -\frac{\theta^2 - \sigma^2}{2 \eta(s)} + \log\frac{\theta}{\sigma}, \label{eqn:d1kappa-Gaussian}\\
  \kappa_{\gamma}''(s) &= \gamma \frac{\theta^2\sigma^2}{\eta(s)^3}
  + \frac{(\theta^2 - \sigma^2)^2}{2\eta(s)^2},\label{eqn:d2kappa-Gaussian}\\
  \kappa_{\gamma}'''(s) &= - \left( \gamma \frac{3\theta^2\sigma^2 (\theta^2 - \sigma^2)}{\eta(s)^4}
  + \frac{(\theta^2 - \sigma^2)^3}{2\eta(s)^3}\right).\label{eqn:d3kappa-Gaussian}
\end{align}

\begin{theorem}[Saddlepoint expansion]\label{thm:alpha-beta-sp-formulation}
Let $\sigma,\theta>0$ and $n\geq 1$, be fixed parameters. Then, 
\begin{align}
f(\beta,\gamma) &= \max_{s} \Bigl\{ \bigl(a(s,\gamma)+b(s,\gamma)\bigr) e^{n(\kappa_{\gamma}(s)+(1-s)\kappa_{\gamma}'(s))}
%\notag\\&\qquad\;\;\;\; 
+ \openone[s>1] + \bigl(\openone[s<0] - \beta\bigr) e^{n\kappa_{\gamma}'(s)}\!\Bigr\},\label{eqn:HT-sp-formulation}
\end{align}
where $\openone[\cdot]$ denotes the indicator function and, for  $\lambda_{a}(s) \triangleq |1-s|\sqrt{n \kappa_{\gamma}''(s)}$ and $\lambda_{b}(s) \triangleq |s|\sqrt{n \kappa_{\gamma}''(s)}$,
\begin{align}
a(s,\gamma) &=  \sgn(1-s) \left(\Psi\bigl(\lambda_a(s)\bigr) + \frac{n (s-1)^3}{6} \left(\frac{\lambda_a(s)^{-1}-\lambda_a(s)^{-3}}{\sqrt{2\pi}} - \Psi\bigl(\lambda_a(s)\bigr)\right) \kappa_{\gamma}'''(s)\right) + o\bigl(n^{-\frac{1}{2}}\bigr),\label{eqn:sp-an}\\
b(s,\gamma) &=  \sgn(s) \left(\Psi\bigl(\lambda_b(s)\bigr) + \frac{n s^3}{6} \left(\frac{\lambda_b(s)^{-1}-\lambda_b(s)^{-3}}{\sqrt{2\pi}} - \Psi\bigl(\lambda_b(s)\bigr)\right) \kappa_{\gamma}'''(s)\right) + o\bigl(n^{-\frac{1}{2}}\bigr).\label{eqn:sp-bn}
\end{align}
Here, $\sgn(\cdot)$ denotes the sign function, defined as $\sgn(s)=-1$ for $s<0$ and $\sgn(s)=1$ otherwise; the function $\Psi(\lambda)$ is defined as $\Psi(\lambda) \triangleq  \Qsf(|\lambda|) e^{\frac{\lambda^2}{2}}$ with $\Qsf(\cdot)$ the Gaussian $\Qsf$-function; and $o\bigl(g(n)\bigr)$ summarizes the terms that approach zero faster than $g(n)$, \textit{i.e.}, $\lim_{n\to \infty}\frac{o(g(n))}{g(n)}=0$.
\end{theorem}
\begin{IEEEproof}
The proof follows the lines of \cite[Th. 2]{isit18} with a more refined expansion of $a(s,\gamma)$ and $b(s,\gamma)$.

We consider a sequence of i.i.d. non-lattice real-valued random variables with positive variance, $\{Z_{\ell}\}_{\ell=1}^n$, and we define their mean as
\begin{align}
  \bar{Z}_n \triangleq \frac{1}{n}\sum_{\ell=0}^n Z_{\ell}.
\end{align}
Let $\kappa_{Z}(s) \triangleq \log \Ex\bigl[e^{s Z_\ell}\bigr]$ denote the cumulant generating function of $Z_{\ell}$, and 
let $\kappa_{Z}'(s)$, $\kappa_{Z}''(s)$ and $\kappa_{Z}'''(s)$ denote its 1st, 2nd and 3rd derivatives with respect to $s$, respectively.
We apply the tilting from \cite[Sec. XVI.7]{Feller71}
to the random variable $\bar{Z}_n$ and then use the
expansion \cite[Sec. XVI.4, Th. 1]{Feller71}. We obtain the following result  (see \cite[Prop.~1, Part~1]{twc2020} for detailed derivation).
If there exists $s$ in the region of convergence of $\kappa_{Z}(s)$ such that $\kappa_{Z}'(s) = t$, with $t \geq \Ex\bigl[\bar{Z}_n\bigr]$,
the tail probability $\Pr\bigl[\bar{Z}_n \geq t\bigr]$ satisfies
\begin{align}
\Pr\bigl[\bar{Z}_n \geq t\bigr]
&=  e^{n(\kappa_{Z}(s)-s \kappa_{Z}'(s))} \left(\Psi(\lambda_{Z,s}) + \frac{n s^3}{6} \left(\frac{\lambda_{Z,s}^{-1}-\lambda_{Z,s}^{-3}}{\sqrt{2\pi}} - \Psi(\lambda_{Z,s})\right) \kappa_{Z}'''(s) + o\bigl(n^{-\frac{1}{2}}\bigr)\right),\label{eqn:sp-expansion-1}
\end{align}
where $\lambda_{Z,s} \triangleq |s|\sqrt{n \kappa_{Z}''(s)}$ and the error term $o\bigl(n^{-\frac{1}{2}}\bigr)$ holds uniformly in $s$.

The expansion \refE{sp-expansion-1} requires that the threshold $t$ is above the average value $\Ex\bigl[\bar{Z}_n\bigr] = \Ex\bigl[{Z}_{\ell}\bigr]$. That is, the expansion \refE{sp-expansion-1} is only accurate for evaluating the tail of the probability distribution. Whenever $t < \Ex\bigl[\bar{Z}_n\bigr]$, we use that
\begin{align}
  \Pr\bigl[\bar{Z}_n \geq t\bigr] 
    &= 1 - \Pr\bigl[\bar{Z}_n < t\bigr]
    \label{eqn:PZn-1}\\
    &= 1 - \Pr\bigl[-\bar{Z}_n > -t\bigr].
    \label{eqn:PZn-2}
\end{align}
For non-lattice distributions, the expansion
\refE{sp-expansion-1} coincides with that of
$\Pr\bigl[\bar{Z}_n > t\bigr]$ and therefore, 
it can be used to estimate $\Pr\bigl[-\bar{Z}_n > -t\bigr]$. Moreover, given the mapping $\kappa_{Z}'(s) = t$, it can be checked that the condition $t \geq \Ex\bigl[\bar{Z}_n\bigr]$ corresponds to $s\geq 0$. Similarly, for $t < \Ex\bigl[\bar{Z}_n\bigr]$ and the mapping $\kappa_{Z}'(s) = t$, we obtain $s<0$.

We now apply this expansion to the probability terms
$\Pr\bigl[  j(\Y_{0}) \leq t' \bigr]$ and $\Pr\bigl[  j(\Y_{1}) > t'\bigr]$ appearing in \refE{HT-im-formulation}.
To this end, we consider the random variables $Z_0 = -j(Y_0)$, $Y_0 \sim \varphi_{\sqrt{\gamma},\sigma}$, and $Z_1 = j(Y_1)$, $Y_1 \sim \varphi_{0,\theta}$, with $j(y)$ defined in \refE{j1-def}. The cumulant generating functions of the random variables $Z_0$ and $Z_1$ are, respectively, given by
\begin{align}
  \kappa_{Z_0}(s) &=\log \Ex\left[\left(
  \frac{\varphi_{\sqrt{\gamma},\sigma}(Y_0)}{\varphi_{0,\theta}(Y_0)}\right)^{-s}\right] = \kappa_{\gamma}(1-s),\label{eqn:HT-sp-kappa0}\\
  \kappa_{Z_1}(s) &=\log \Ex\left[\left(\frac{\varphi_{\sqrt{\gamma},\sigma}(Y_1)}{\varphi_{0,\theta}(Y_1)}\right)^s\right] = \kappa_{\gamma}(s),\label{eqn:HT-sp-kappa1}
\end{align}
for $\kappa_{\gamma}(s)$ defined in \refE{d0kappa-Gaussian}. The fact that the cumulant generating functions $\kappa_{Z_0}(s)$ and $\kappa_{Z_1}(s)$ are shifted and mirrored versions of each other will allow us to simplify the resulting expression.

Now, we use \refE{sp-expansion-1} for $\bar{Z}_{0,n} \triangleq -\frac{1}{n} j(\Y_{0}) = -\frac{1}{n} \sum_{\ell=1}^n j(Y_{0,\ell})$, and apply \refE{PZn-1}-\refE{PZn-2} whenever $t < \Ex\bigl[\bar{Z}_{0,n}\bigr]$, to obtain
\begin{align}
\Pr\bigl[\bar{Z}_{0,n} \geq t\bigr]
&=  \openone[\bar{s}< 0] + \sgn(\bar{s}) e^{n(\kappa_{Z_0}(\bar{s})-\bar{s} \kappa_{Z_0}'(\bar{s}))} \notag\\
&\qquad\qquad\times\left(\Psi(\lambda_{Z_0,\bar{s}}) + \frac{n \bar{s}^3}{6} \left(\frac{\lambda_{Z_0,\bar{s}}^{-1}-\lambda_{Z_0,\bar{s}}^{-3}}{\sqrt{2\pi}} - \Psi(\lambda_{Z_0,\bar{s}})\right) \kappa_{Z_0}'''(\bar{s}) + o\bigl(n^{-\frac{1}{2}}\bigr)\right),\label{eqn:sp-expansion-2}
\end{align}
where the value of $\bar{s}$ satisfies $\kappa_{Z_0}'(\bar{s})=t$.

We consider the change of variable $\bar{s} \leftrightarrow s$ with $s = 1-\bar{s} \Leftrightarrow \bar{s} = 1-s$ and use the identity \refE{HT-sp-kappa0}.
For this change of variables, from \refE{HT-sp-kappa0} we obtain the identities
\begin{align}
  \kappa_{Z_0}(\bar{s}) &=  \kappa_{\gamma}(1-\bar{s}) = \kappa_{\gamma}(s),\label{eqn:HT-sp-kappa0-0}\\
  \kappa_{Z_0}'(\bar{s}) &=  -\kappa_{\gamma}'(1-\bar{s}) = -\kappa_{\gamma}'(s),\label{eqn:HT-sp-kappa0-1}\\
  \kappa_{Z_0}''(\bar{s}) &=  \kappa_{\gamma}''(1-\bar{s}) = \kappa_{\gamma}''(s),\label{eqn:HT-sp-kappa0-2}\\
  \kappa_{Z_0}'''(\bar{s}) &=  -\kappa_{\gamma}'''(1-\bar{s}) = -\kappa_{\gamma}'''(s).\label{eqn:HT-sp-kappa0-3}
\end{align}
Then, using \refE{HT-sp-kappa0-0}-\refE{HT-sp-kappa0-3} in \refE{sp-expansion-2}, via the change of variable $\bar{s} \leftrightarrow s$ with $\bar{s} = 1-s$, we obtain that
\begin{align}
\Pr\bigl[  j(\Y_{0}) \leq t' \bigr] 
    = \openone[s>1] + a(s,\gamma) e^{n(\kappa_{\gamma}(s)+(1-s)\kappa_{\gamma}'(s))}\label{eqn:Pj0-sp}
\end{align}
where $s$ satisfies $\kappa_{\gamma}'(s)=t'/n$.

Proceeding analogously for $\bar{Z}_{1,n} = \frac{1}{n} j(\Y_{1}) = \frac{1}{n} \sum_{\ell=1}^n j(Y_{1,\ell})$, using \refE{sp-expansion-1} and \refE{HT-sp-kappa1}, we obtain
\begin{align}
\Pr\bigl[ j(\Y_{1}) > t' \bigr] 
    = \openone[s<0] + b(s,\gamma) e^{n(\kappa_{\gamma}(s)-s\kappa_{\gamma}'(s))}\label{eqn:Pj1-sp}
\end{align}
where $s$ satisfies again $\kappa_{\gamma}'(s)=t'/n$.

We replace \refE{Pj0-sp} and \refE{Pj1-sp} in \refE{HT-im-formulation} and change the optimization variable from $t'$ to $s$ acording to the relation $t'= n\kappa_{\gamma}'(s)$.
Noting that $e^{t'}$ in \refE{HT-im-formulation} becomes $e^{n\kappa_{\gamma}'(s)}$, then \refE{HT-im-formulation} becomes
\begin{align}
\alpha_{\beta}\bigl(\varphi_{\sqrt{\gamma},\sigma}^n, \varphi_{0,\theta}^n \bigr)
= \max_{s} \Bigl\{&\openone[s>1] + a(s,\gamma) e^{n(\kappa_{\gamma}(s)+(1-s)\kappa_{\gamma}'(s))}
\notag\\&
+ e^{n\kappa_{\gamma}'(s)} \bigl( \openone[s<0] + b(s,\gamma) e^{n(\kappa_{\gamma}(s)-s\kappa_{\gamma}'(s))}
           -  \beta \bigr)\Bigr\}.
\label{eqn:HT-sp-formulation-1}
\end{align}
The result then follows from \refE{HT-sp-formulation-1} by reorganizing terms.
\end{IEEEproof}

The refined expressions from \refE{sp-an}-\refE{sp-bn} are needed to obtain an error term $o\bigl(n^{-\frac{1}{2}}\bigr)$ that is uniform on~$s$. For practical purposes, however, the function $f(\beta,\gamma)$ can be approximated using \refE{HT-sp-formulation} from Theorem \ref{thm:alpha-beta-sp-formulation} with $a(s,\gamma)$ and $b(s,\gamma)$ replaced by the simpler expressions
\begin{align}
  \hat{a}(s,\gamma) &\triangleq \sgn(1-s) \Psi\bigl(\lambda_a(s)\bigr),\qquad
  \hat{b}(s,\gamma) \triangleq \sgn(s) \Psi\bigl(\lambda_b(s)\bigr),
\end{align}
respectively. This approximation yields accurate results for blocklengths as short as $n=20$ (see \cite{isit18} for details),
and still we obtain an approximation error of order $o\bigl(n^{-\frac{1}{2}}\bigr)$ for values of $s$ satisfying $0 < s_0 \leq s \leq s_1 < 1$.

\subsection{Exponent-achieving output distribution}
\label{sec:exponent-achieving}

Often, the variance of the auxiliary distribution in Theorems \ref{thm:PPV_lower_bound}, \ref{thm:maximal_lower_bound} and
\ref{thm:average_lower_bound} is chosen to be the variance of the capacity-achieving output distribution, $\theta^2=\Upsilon+\sigma^2$. While this choice of $\theta^2$ is adequate for rates approaching the capacity of the channel, it does not attain the sphere-packing exponent in general~\cite{Shannon67I}.
An auxiliary distribution that yields the right exponential behavior in these bounds is the so-called exponent-achieving output distribution.\footnote{If we restrict the auxiliary output distribution to be product, the exponent-achieving output distribution is unique in the sense that the resulting hypothesis-testing bound attains the sphere-packing exponent~\cite{Shannon67I}. Nevertheless, using other non-product auxiliary distributions can yield the right exponential behavior. One example is the optimal distribution in the meta-converse for the AWGN channel under an equal power constraint discussed in Section~\ref{sec:equal-metaconverse}, which recovers Shannon cone-packing bound with the sphere-packing exponent.}

The exponent-achieving output distribution for the power-constrained AWGN channel is precisely the zero-mean i.i.d. Gaussian distribution defined in \refE{Q-theta-def} with the variance given by $\theta^2 = \tilde\theta_s^2$ where~\cite[Sec. 6.2]{Nakiboglu19-Augustin}
\begin{align}\label{eqn:theta_s-def}
  \tilde\theta_s^2 = \sigma^2 + \frac{\gamma}{2} - \frac{\sigma^2}{2 s} + \sqrt{\biggl(\frac{\gamma}{2}-\frac{\sigma^2}{2 s}\biggr)^2 + \gamma\sigma^2}.
\end{align}
Here, $\gamma=\Upsilon$ is the power constraint and the parameter $s$ is the result of optimizing the sphere-packing exponent for a given transmission rate. Specifically, the sphere-packing exponent can be computed as~\cite{Nakiboglu19-SP}
\begin{align}\label{eqn:Esp-def}
  E_{\text{sp}}(R)
  &\triangleq \sup_{s\in(0,1)} \biggl\{ \frac{1-s}{s} \bigl( C_{s,W,\Upsilon} - R \bigr) \biggr\},
\end{align}
where $R$ denotes the transmission rate $R = \frac{1}{n}\log{M}$ in nats/channel use and $C_{s,W,\Upsilon}$ is the so-called Augustin capacity~\cite{Nakiboglu19-Augustin}, which for the power-constrained AWGN channel is given by~\cite[Eq.~(126)]{Nakiboglu19-Augustin}
\begin{align}\label{eqn:CAug}
  C_{s,W,\Upsilon} = 
\begin{cases}
\frac{s\Upsilon}{2\tilde\eta(s)} + \frac{1}{s-1}\log\frac{\tilde\theta_s^s \sigma^{1-s}}{\sqrt{\tilde\eta(s)}}, & s\geq0, s\neq 1,\\
\frac{1}{2}\log\left(1+\frac{\Upsilon}{\sigma^2} \right), & s= 1,
\end{cases}
\end{align}
where $\tilde\eta(s) \triangleq s\tilde\theta_s^2 + (1-s) \sigma^2$.

For transmission rates close to the channel capacity, the optimal value of $s$ in \refE{Esp-def} tends to $1$ as $n\to\infty$ and therefore $\theta_s^2 \to \theta_1^2 = \Upsilon +\sigma^2$. Hence, $\varphi_{0,\theta}^n(\y)$ becomes the capacity-achieving output distribution used in \cite[Th. 41]{Pol10}.

In principle, we can solve \refE{Esp-def} to determine the asymptotic optimal value of $s$, and then use it in the variance \refE{theta_s-def}. Nevertheless, the saddlepoint expansion from Theorem~\ref{thm:alpha-beta-sp-formulation} allows to introduce a dependence of $\theta^2$ with the auxiliary parameter~$s$ without incurring an extra computational cost, as the next result shows. 

For $\tilde\eta(s) = s\tilde\theta_s^2 + (1-s) \sigma^2$ we define
\begin{align}
  \tilde\kappa_{\gamma}(s)
  &\triangleq {\gamma} \frac{s(s-1)}{2\tilde\eta(s)}+\log\frac{\tilde\theta_s^{s}\sigma^{1-s}}{\sqrt{\tilde\eta(s)}},\label{eqn:d0tkappa-Gaussian}\\
   \tilde\kappa_{\gamma}^{(1)}(s) &\triangleq {\gamma} \frac{s^2 \tilde\theta_s^2 - (1-s)^2 \sigma^2}{2\tilde\eta(s)^2} -\frac{\tilde\theta_s^2 - \sigma^2}{2 \tilde\eta(s)} + \log\frac{\tilde\theta_s}{\sigma},
   \label{eqn:d1tkappa-Gaussian}\\
  \tilde\kappa_{\gamma}^{(2)}(s) &\triangleq \gamma \frac{\tilde\theta_s^2\sigma^2}{\tilde\eta(s)^3}
  + \frac{(\tilde\theta_s^2 - \sigma^2)^2}{2\tilde\eta(s)^2},\label{eqn:d2tkappa-Gaussian}\\
  \tilde\kappa_{\gamma}^{(3)}(s) &\triangleq - \left( \gamma \frac{3\tilde\theta_s^2\sigma^2 (\tilde\theta_s^2 - \sigma^2)}{\tilde\eta(s)^4}
  + \frac{(\tilde\theta_s^2 - \sigma^2)^3}{2\tilde\eta(s)^3}\right).\label{eqn:d3tkappa-Gaussian}
\end{align}
The definitions \refE{d0tkappa-Gaussian}-\refE{d3tkappa-Gaussian} correspond to \refE{d0kappa-Gaussian}-\refE{d3kappa-Gaussian} after setting $\theta=\tilde\theta_s$.
Note that \refE{d1tkappa-Gaussian}-\refE{d3tkappa-Gaussian} do not coincide with the derivatives of \refE{d0tkappa-Gaussian} in general, since in obtaining \refE{d1kappa-Gaussian}-\refE{d3kappa-Gaussian} we have considered $\theta$ to be independent of $s$.

\begin{corollary}[Exponent-achieving saddlepoint expansion]\label{cor:alpha-beta-sp-exponent}
Let $\sigma>0$ and $n\geq 1$, be fixed parameters.
Then,
\begin{align}
\max_{\theta\geq\sigma} \Bigl\{
\alpha_{\beta} \bigl(\varphi_{\sqrt{\gamma},\sigma}^n, \varphi_{0,\theta}^n \bigr) \Bigr\} \geq \tilde{f}(\beta,\gamma),\label{eqn:alpha-beta-sp-exponent}
\end{align}
where the function $\tilde{f}(\beta,\gamma)$ is defined as
\begin{align}
\tilde{f}(\beta,\gamma) &\triangleq \max_{s} \Bigl\{ \bigl(\tilde{a}(s,\gamma) + \tilde{b}(s,\gamma)\bigr) e^{n(\tilde\kappa_{\gamma}(s)+(1-s)\tilde\kappa_{\gamma}^{(1)}(s))}
%\notag\\&\qquad\;\;\;\; 
+ \openone[s>1] + \bigl(\openone[s<0] - \beta\bigr) e^{n\tilde\kappa_{\gamma}^{(1)}(s)}\!\Bigr\},\label{eqn:HT-sp-exponent}\\
\tilde{a}(s,\gamma) &=  \sgn(1-s) \left(\Psi\bigl(\tilde\lambda_a(s)\bigr) + \frac{n (s-1)^3}{6} \left(\frac{\tilde\lambda_a(s)^{-1}-\tilde\lambda_a(s)^{-3}}{\sqrt{2\pi}} - \Psi\bigl(\tilde\lambda_a(s)\bigr)\right) \tilde\kappa_{\gamma}^{(3)}(s)\right) + o\bigl(n^{-\frac{1}{2}}\bigr),\label{eqn:sp-tan}\\
\tilde{b}(s,\gamma) &=  \sgn(s) \left(\Psi\bigl(\tilde\lambda_b(s)\bigr) + \frac{n s^3}{6} \left(\frac{\tilde\lambda_b(s)^{-1}-\tilde\lambda_b(s)^{-3}}{\sqrt{2\pi}} - \Psi\bigl(\tilde\lambda_b(s)\bigr)\right) \tilde\kappa_{\gamma}^{(3)}(s)\right) + o\bigl(n^{-\frac{1}{2}}\bigr).\label{eqn:sp-tbn}
\end{align}
with $\tilde\lambda_{a}(s) \triangleq |1-s|\sqrt{n \tilde\kappa_{\gamma}^{(2)}(s)}$ and $\tilde\lambda_{b}(s) \triangleq |s|\sqrt{n \tilde\kappa_{\gamma}^{(2)}(s)}$.
\end{corollary}
\begin{IEEEproof}
Using \refE{HT-sp-formulation-1}, it follows that
\begin{align}
\max_{\theta\geq\sigma}&\Bigl\{
\alpha_{\beta} \bigl(\varphi_{\sqrt{\gamma},\sigma}^n, \varphi_{0,\theta}^n \bigr) \Bigr\}
\notag\\
&= \max_{\theta\geq\sigma} \max_{s} \Bigl\{ \bigl(a(s,\gamma)+b(s,\gamma)\bigr) e^{n(\kappa_{\gamma}(s)+(1-s)\kappa_{\gamma}'(s))}
%\notag\\&\qquad\;\;\;\; 
+ \openone[s>1] + \bigl(\openone[s<0] - \beta\bigr) e^{n\kappa_{\gamma}'(s)}\!\Bigr\}\\
&= \max_{s} \max_{\theta\geq\sigma} \Bigl\{ \bigl(a(s,\gamma)+b(s,\gamma)\bigr) e^{n(\kappa_{\gamma}(s)+(1-s)\kappa_{\gamma}'(s))}
%\notag\\&\qquad\;\;\;\; 
+ \openone[s>1] + \bigl(\openone[s<0] - \beta\bigr) e^{n\kappa_{\gamma}'(s)}\!\Bigr\}.
\end{align}
We can fix a value $\theta\geq\sigma$ and obtain a lower bound to $\max_{\theta\geq\sigma}\bigl\{\alpha_{\beta} \bigl(\varphi_{\sqrt{\gamma},\sigma}^n, \varphi_{0,\theta}^n \bigr) \bigr\}$.
Moreover, as the maximization over $\theta$ is inside the maximization over $s$, the chosen value for $\theta$ may depend on $s$. Then, letting $\theta=\tilde\theta_s$ in the inner maximization we obtain the desired result.
\end{IEEEproof}

According to Corollary~\ref{cor:alpha-beta-sp-exponent}, we can use $\tilde{f}\bigl(\frac{1}{M},\gamma\bigr)$ instead of $f\bigl(\frac{1}{M},\gamma\bigr)$ in Theorems \ref{thm:PPV_lower_bound}, \ref{thm:maximal_lower_bound} and Corollary~\ref{cor:average_metaconverse_bound}.
In Theorem~\ref{thm:average_lower_bound} however, the convex hull needs to be evaluated for a fixed variance $\theta^2$, which can be the capacity-achieving $\theta^2 = \Upsilon+\sigma^2$, the exponent-achieving $\theta^2=\tilde\theta_s^2$ for the asymptotic value of $s$ optimizing~\refE{Esp-def},
or $\theta^2=\tilde\theta_s^2$ for the asymptotic value of $s$ optimizing \refE{HT-sp-exponent} the system parameters under consideration.

%If we fix $s = \frac{1}{1+\rho}$ for $\rho$ corresponding to the triplet $(\rho,Q,r)$ maximizing the exponent in \cite[Eq. (7.3.21)]{Gall68}, $\varphi_{0,\theta_s}^n(\y)$ corresponds to the exponent-achieving output distribution.
%Note that when the rate $R$ approaches capacity, $\rho\to 0$, \textit{i.e.}, $s\to 1$ and $\theta^2 \to \Upsilon +\sigma^2$. Hence, when the rate is close to capacity $\varphi_{0,\theta_s}^n(\y)$ corresponds to the capacity-achieving output distribution.

%%%%%%%%%%%%%%%%%%%%%%%%%%%%%%%%%%%%%%%%%%%%%%%%%%%%%%%

\section{Numerical Examples}
\label{sec:numerical}

\begin{figure}[t]%
\vspace{-0.3mm}
\centering% This file was created by matlab2tikz v0.6.0 running on MATLAB 8.2.
%Copyright (c) 2008--2014, Nico Schlömer <nico.schloemer@gmail.com>
%All rights reserved.
%Minimal pgfplots version: 1.3
%  http://www.mathworks.com/matlabcentral/fileexchange/22022-matlab2tikz
\begin{tikzpicture}

\begin{axis}[%
width=.8\linewidth,
height=.55\linewidth,
scale only axis,
separate axis lines,
every outer x axis line/.append style={black},
every x tick label/.append style={font=\color{black}},
xmajorgrids,
xmin=0,
xmax=100,
xlabel={Blocklength, $n$},
every outer y axis line/.append style={black},
every y tick label/.append style={font=\color{black}},
ymode=log,
ymajorgrids,
ymin=1e-3,
ymax=0.1,
yminorticks=true,
ylabel={Error probability, $\epsilon_{\text{e}}^{\star}(n,M,\Upsilon)$},
legend style={at={(0.991,0.995)},anchor=north east,legend cell align=left,align=left,draw=black}
]

%\addplot [color=black,solid,line width=0.5pt,mark=*,mark options={solid,scale=0.5}]
\addplot [color=gray,dash pattern=on 1pt off 1pt on 4pt off 1pt,line width=1pt]
  table[row sep=crcr]{%
2	0.455955733219089\\
4	0.298266272463783\\
6	0.232825092838634\\
8	0.193415958578272\\
10	0.165736837942085\\
12	0.144633459590168\\
14	0.127777344897316\\
16	0.113874111297367\\
18	0.102138894188286\\
20	0.0921030051648228\\
22	0.0833850896930203\\
24	0.0757472429348077\\
26	0.0690036349818882\\
28	0.0630122543302112\\
30	0.0576601816961487\\
32	0.0529241114413876\\
34	0.0485335066867251\\
36	0.044626002880964\\
38	0.041084384667068\\
40	0.0378673698078911\\
42	0.0349387914965699\\
44	0.0322663461099882\\
46	0.0298244606161055\\
48	0.0275888781188258\\
50	0.0255393991987647\\
52	0.0236586933600312\\
54	0.0219295372861573\\
56	0.0203388956804966\\
58	0.0188741201465949\\
60	0.0175236198359466\\
62	0.0161306229294946\\
64	0.0151432866727508\\
66	0.0140636838968213\\
68	0.0130802102210177\\
70	0.0121702407840231\\
72	0.0113276444927851\\
74	0.0105471173388172\\
76	0.00982304560155835\\
78	0.00915212561085251\\
80	0.00852945524331934\\
82	0.00795123423288897\\
84	0.00741413901444186\\
86	0.00691509404755351\\
88	0.00645125400557414\\
90	0.00601998715636451\\
92	0.00561885985041757\\
94	0.00524562204452637\\
96	0.00489819379827264\\
98	0.00457465268664452\\
100	0.00427322207647964\\
};
\addlegendentry{Achievability \cite[Eq.~(20)]{Shannon59}};

%\addplot [color=gray,dash pattern=on 2pt off 1pt on 2pt off 1pt,line width=0.8pt]
\addplot [color=black,solid,line width=0.8pt]
  table[row sep=crcr]{%
2	0.087049668551521\\
4	0.0674793315545587\\
6	0.062130523253862\\
8	0.0578319146837727\\
10	0.0538268509396664\\
12	0.0500468291688573\\
14	0.0464952069049285\\
16	0.0431758908824696\\
18	0.040085759455498\\
20	0.0372164965609607\\
22	0.0345563522632975\\
24	0.0320924509406799\\
26	0.0298115013696256\\
28	0.0277006657124931\\
30	0.0257467333164426\\
32	0.0239381426302476\\
34	0.0222632964548267\\
36	0.0207123649673819\\
38	0.0192749961995984\\
40	0.0179430155551334\\
42	0.0167082769958404\\
44	0.0155625662953534\\
46	0.0144997482922194\\
48	0.0135129859435369\\
50	0.0125966182361117\\
52	0.0117456919440054\\
54	0.0109545208311098\\
56	0.0102192120097656\\
58	0.00953555459380961\\
60	0.00889950073579302\\
62	0.00830754822595631\\
64	0.00775658986084426\\
66	0.0072438892205652\\
68	0.00676610664093838\\
70	0.0063210799890924\\
72	0.00590638452507471\\
74	0.00551993601935682\\
76	0.00515933857943288\\
78	0.00482345309146015\\
80	0.0045100976071068\\
82	0.00421765853056049\\
84	0.00394473173477974\\
86	0.00368999621374883\\
88	0.00345221138932129\\
90	0.00323021406396209\\
92	0.00302291512646593\\
94	0.00282929609979286\\
96	0.00264840560435701\\
98	0.00247935579645121\\
100	0.0023213188296439\\
};
\addlegendentry{Theorem~\ref{thm:shannon_lower_bound}, Eq.~\refE{shannon_lower_bound}};

\addplot [color=black,solid,line width=1.6pt]
  table[row sep=crcr]{%
2	0.0227771513207305\\
4	0.0315379363092323\\
6	0.0344147744112234\\
8	0.0349399399283127\\
10	0.0343427624661513\\
12	0.0331597680731154\\
14	0.0316880575946233\\
16	0.0300785587518243\\
18	0.0284146434301356\\
20	0.0267649208162544\\
22	0.0251667933961467\\
24	0.0236006674686066\\
26	0.0221399543879728\\
28	0.0207397922855153\\
30	0.0194117509291246\\
32	0.0181575407124722\\
34	0.0169683367623379\\
36	0.0158894542139695\\
38	0.0148535460669512\\
40	0.013885645616868\\
42	0.0129881678765352\\
44	0.0121276102118403\\
46	0.0113466357780985\\
48	0.0106138753770168\\
50	0.00990224068088397\\
52	0.00928279292319438\\
54	0.00867478448219586\\
56	0.00809238338130142\\
58	0.00759104829677458\\
60	0.00710475778067897\\
62	0.00662786081068722\\
64	0.00620143439840396\\
66	0.00581605513294296\\
68	0.00544574520395657\\
70	0.0050903590842115\\
72	0.00475394689436381\\
74	0.00445472354208347\\
76	0.00417832907126781\\
78	0.00391715108654566\\
80	0.0036668854765047\\
82	0.00342525869468506\\
84	0.00320211695908252\\
86	0.00299835129662343\\
88	0.00281397240606852\\
90	0.00264453620162069\\
92	0.00248011736075191\\
94	0.00232390118887678\\
96	0.00217431729234844\\
98	0.00203111518089692\\
100	0.00190169878648021\\
};
\addlegendentry{Th.~\ref{thm:PPV_lower_bound} with $\theta^2=\tilde\theta_s^2$ in \refE{theta_s-def}};

\addplot [color=black,dotted,mark=o,mark options={solid, scale=0.7}]
  table[row sep=crcr]{%
2	0.0236137190343139\\
4	0.0314635805919035\\
6	0.0341845220883097\\
8	0.0346713875762002\\
10	0.0340556168587276\\
12	0.0328753860359481\\
14	0.0314147946520058\\
16	0.0298126535020477\\
18	0.0281666191638222\\
20	0.0265211623100416\\
22	0.0249337627333844\\
24	0.023386841659996\\
26	0.0219211548346289\\
28	0.0205375278655629\\
30	0.0192032256656708\\
32	0.0179804367403764\\
34	0.0168176663845406\\
36	0.0157244389310973\\
38	0.0147008552663084\\
40	0.0137408148709859\\
42	0.0128338854184869\\
44	0.0119934799463356\\
46	0.0112240521912968\\
48	0.0104794404193837\\
50	0.00980413805759572\\
52	0.00916345754502867\\
54	0.00856899533582087\\
56	0.00800743490598469\\
58	0.0074965834228491\\
60	0.00698757601597191\\
62	0.0065582545426727\\
64	0.00612969441824277\\
66	0.00572308090408807\\
68	0.00537048530637074\\
70	0.00502210178742826\\
72	0.00468206485539744\\
74	0.00439767215101845\\
76	0.0041212917710025\\
78	0.00385179965556015\\
80	0.00359127964982777\\
82	0.00337636262872645\\
84	0.00316760171205492\\
86	0.00296493029400256\\
88	0.00276953356111984\\
90	0.00258540440720445\\
92	0.00243280188189541\\
94	0.00228439713581649\\
96	0.00214098978351462\\
98	0.00200315881587902\\
100	0.00187130143366536\\
};
\addlegendentry{Th.~\ref{thm:PPV_lower_bound} with $\theta^2=\Upsilon+\sigma^2$};

\end{axis}
\end{tikzpicture}%
%
\caption{Upper and lower bounds to the channel coding error probability for an AWGN channel under an equal power constraint. System parameters: $\text{SNR}=10$~dB, rate $R = 1.5$ bits/channel use.}\vspace{-2mm}\label{fig:equal-AWGN-Pevsn-snr10dB}
\end{figure}
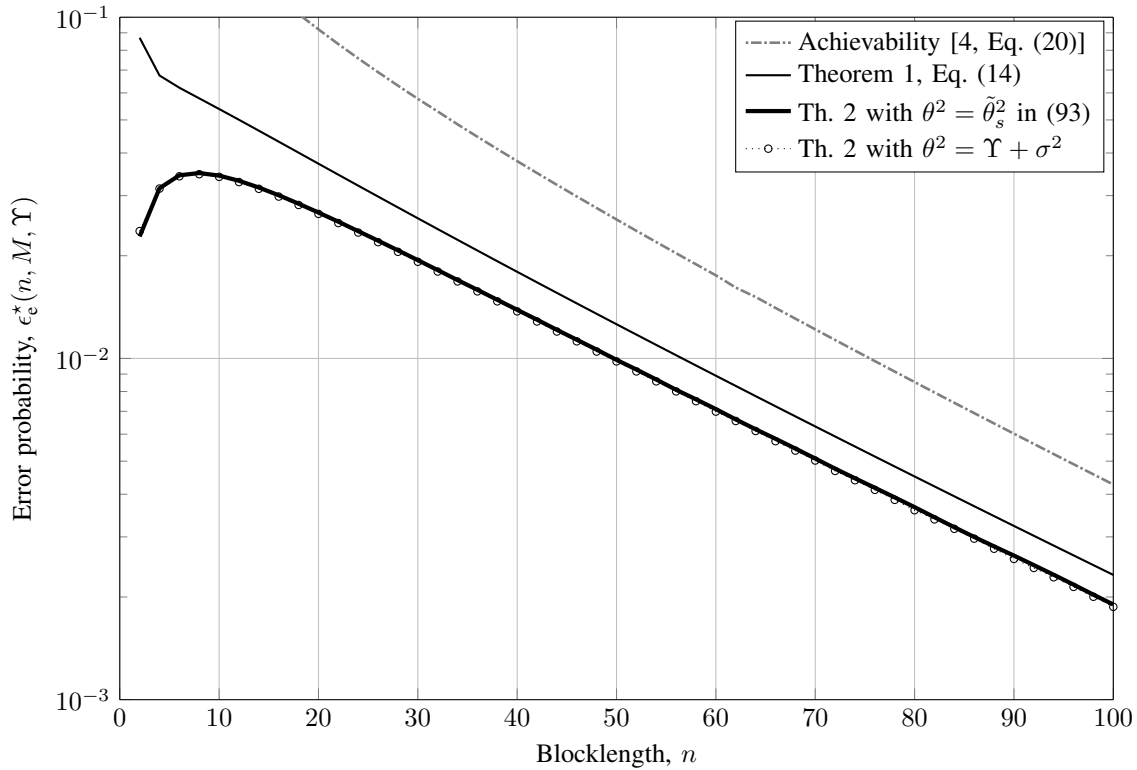%

\subsection{Comparison with previous results under the different power constraints}
\label{sec:numerical-2}

We consider the transmission of $M = 2^{nR}$ codewords over $n$ uses of an AWGN channel with $R=1.5$ bits/channel use and $\text{SNR}=10\log_{10}\frac{\Upsilon}{\sigma^2}=10$ dB. The channel capacity is $C = \frac{1}{2}\log_2\bigl(1+\frac{\Upsilon}{\sigma^2}\bigr) \approx 1.73$ bits/channel use under the three power constraints considered.

\begin{figure}[t]%
\vspace{-0.3mm}
\centering% This file was created by matlab2tikz v0.6.0 running on MATLAB 8.2.
%Copyright (c) 2008--2014, Nico Schlömer <nico.schloemer@gmail.com>
%All rights reserved.
%Minimal pgfplots version: 1.3
%  http://www.mathworks.com/matlabcentral/fileexchange/22022-matlab2tikz
\begin{tikzpicture}

\begin{axis}[%
width=.8\linewidth,
height=.55\linewidth,
scale only axis,
separate axis lines,
every outer x axis line/.append style={black},
every x tick label/.append style={font=\color{black}},
xmajorgrids,
xmin=0,
xmax=100,
xlabel={Blocklength, $n$},
every outer y axis line/.append style={black},
every y tick label/.append style={font=\color{black}},
ymode=log,
ymajorgrids,
ymin=1e-3,
ymax=0.1,
yminorticks=true,
ylabel={Error probability, $\epsilon_{\text{m}}^{\star}(n,M,\Upsilon)$},
legend style={at={(0.991,0.995)},anchor=north east,legend cell align=left,align=left,draw=black}
]
%\addplot [color=black,solid,line width=0.5pt,mark=*,mark options={solid,scale=0.5}]
\addplot [color=gray,dash pattern=on 1pt off 1pt on 4pt off 1pt,line width=1pt]
  table[row sep=crcr]{%
2	0.455955733219089\\
4	0.298266272463783\\
6	0.232825092838634\\
8	0.193415958578272\\
10	0.165736837942085\\
12	0.144633459590168\\
14	0.127777344897316\\
16	0.113874111297367\\
18	0.102138894188286\\
20	0.0921030051648228\\
22	0.0833850896930203\\
24	0.0757472429348077\\
26	0.0690036349818882\\
28	0.0630122543302112\\
30	0.0576601816961487\\
32	0.0529241114413876\\
34	0.0485335066867251\\
36	0.044626002880964\\
38	0.041084384667068\\
40	0.0378673698078911\\
42	0.0349387914965699\\
44	0.0322663461099882\\
46	0.0298244606161055\\
48	0.0275888781188258\\
50	0.0255393991987647\\
52	0.0236586933600312\\
54	0.0219295372861573\\
56	0.0203388956804966\\
58	0.0188741201465949\\
60	0.0175236198359466\\
62	0.0161306229294946\\
64	0.0151432866727508\\
66	0.0140636838968213\\
68	0.0130802102210177\\
70	0.0121702407840231\\
72	0.0113276444927851\\
74	0.0105471173388172\\
76	0.00982304560155835\\
78	0.00915212561085251\\
80	0.00852945524331934\\
82	0.00795123423288897\\
84	0.00741413901444186\\
86	0.00691509404755351\\
88	0.00645125400557414\\
90	0.00601998715636451\\
92	0.00561885985041757\\
94	0.00524562204452637\\
96	0.00489819379827264\\
98	0.00457465268664452\\
100	0.00427322207647964\\
};
\addlegendentry{Achievability \cite[Eq.~(20)]{Shannon59}};

%\addplot [color=gray,dash pattern=on 2pt off 1pt on 2pt off 1pt,line width=0.8pt]
%  table[row sep=crcr]{%
%2	0.087049668551521\\
%4	0.0674793315545587\\
%6	0.062130523253862\\
%8	0.0578319146837727\\
%10	0.0538268509396664\\
%12	0.0500468291688573\\
%14	0.0464952069049285\\
%16	0.0431758908824696\\
%18	0.040085759455498\\
%20	0.0372164965609607\\
%22	0.0345563522632975\\
%24	0.0320924509406799\\
%26	0.0298115013696256\\
%28	0.0277006657124931\\
%30	0.0257467333164426\\
%32	0.0239381426302476\\
%34	0.0222632964548267\\
%36	0.0207123649673819\\
%38	0.0192749961995984\\
%40	0.0179430155551334\\
%42	0.0167082769958404\\
%44	0.0155625662953534\\
%46	0.0144997482922194\\
%48	0.0135129859435369\\
%50	0.0125966182361117\\
%52	0.0117456919440054\\
%54	0.0109545208311098\\
%56	0.0102192120097656\\
%58	0.00953555459380961\\
%60	0.00889950073579302\\
%62	0.00830754822595631\\
%64	0.00775658986084426\\
%66	0.0072438892205652\\
%68	0.00676610664093838\\
%70	0.0063210799890924\\
%72	0.00590638452507471\\
%74	0.00551993601935682\\
%76	0.00515933857943288\\
%78	0.00482345309146015\\
%80	0.0045100976071068\\
%82	0.00421765853056049\\
%84	0.00394473173477974\\
%86	0.00368999621374883\\
%88	0.00345221138932129\\
%90	0.00323021406396209\\
%92	0.00302291512646593\\
%94	0.00282929609979286\\
%96	0.00264840560435701\\
%98	0.00247935579645121\\
%100	0.0023213188296439\\
%};
%\addlegendentry{Theorem~\ref{thm:shannon_lower_bound}, Eq.~\refE{shannon_lower_bound}};

\addplot [color=black,solid,line width=1.6pt]
  table[row sep=crcr]{%
2	0.0227771513207305\\
4	0.0315379363092323\\
6	0.0344147744112234\\
8	0.0349399399283127\\
10	0.0343427624661513\\
12	0.0331597680731154\\
14	0.0316880575946233\\
16	0.0300785587518243\\
18	0.0284146434301356\\
20	0.0267649208162544\\
22	0.0251667933961467\\
24	0.0236006674686066\\
26	0.0221399543879728\\
28	0.0207397922855153\\
30	0.0194117509291246\\
32	0.0181575407124722\\
34	0.0169683367623379\\
36	0.0158894542139695\\
38	0.0148535460669512\\
40	0.013885645616868\\
42	0.0129881678765352\\
44	0.0121276102118403\\
46	0.0113466357780985\\
48	0.0106138753770168\\
50	0.00990224068088397\\
52	0.00928279292319438\\
54	0.00867478448219586\\
56	0.00809238338130142\\
58	0.00759104829677458\\
60	0.00710475778067897\\
62	0.00662786081068722\\
64	0.00620143439840396\\
66	0.00581605513294296\\
68	0.00544574520395657\\
70	0.0050903590842115\\
72	0.00475394689436381\\
74	0.00445472354208347\\
76	0.00417832907126781\\
78	0.00391715108654566\\
80	0.0036668854765047\\
82	0.00342525869468506\\
84	0.00320211695908252\\
86	0.00299835129662343\\
88	0.00281397240606852\\
90	0.00264453620162069\\
92	0.00248011736075191\\
94	0.00232390118887678\\
96	0.00217431729234844\\
98	0.00203111518089692\\
100	0.00190169878648021\\
};
\addlegendentry{Th.~\ref{thm:maximal_lower_bound} with $\theta^2=\tilde\theta_s^2$ in \refE{theta_s-def}};

\addplot [color=black,dash pattern=on 4pt off 1pt,line width=0.8pt]
  table[row sep=crcr]{%
2	0.00943753459882599\\
4	0.018110498827246\\
6	0.0224901583167915\\
8	0.0244563905910394\\
10	0.0250875347273377\\
12	0.0249516600717901\\
14	0.0243657407279504\\
16	0.023516480847669\\
18	0.0225183615194793\\
20	0.0214437709105646\\
22	0.0203391026929083\\
24	0.0192344383412364\\
26	0.0181493931061522\\
28	0.0170964338505126\\
30	0.0160834862279196\\
32	0.0151153011256113\\
34	0.0141938778980303\\
36	0.013320193506208\\
38	0.0124940539637779\\
40	0.0117148109810981\\
42	0.0109806430743043\\
44	0.0102900245729463\\
46	0.00964108163093876\\
48	0.00903175751998009\\
50	0.00846012813352312\\
52	0.0079240069633093\\
54	0.00742164500985247\\
56	0.00695086237491236\\
58	0.00650977104241029\\
60	0.00609685972640129\\
62	0.00570997720198069\\
64	0.00534795832081581\\
66	0.00500899730341791\\
68	0.00469191281642828\\
70	0.00439500321318292\\
72	0.00411693397285068\\
74	0.0038568246228597\\
76	0.00361337068865556\\
78	0.00338541534773578\\
80	0.00317203808194201\\
82	0.00297232042025992\\
84	0.00278535926110741\\
86	0.0026102765499834\\
88	0.00244625297728952\\
90	0.00229299214927448\\
92	0.00214926705643916\\
94	0.0020147685029976\\
96	0.00188873890267636\\
98	0.00177082535917576\\
100	0.00166019901595387\\
};
\addlegendentry{Corollary~\ref{cor:shannon_lower_bound}, Eq.~\refE{cor_shannon_lower_bound} };

\addplot [color=black,solid,line width=0.8pt]
  table[row sep=crcr]{%
2	0.00105865976403765\\
4	0.00549184452442368\\
6	0.00934653603622666\\
8	0.0118971315596089\\
10	0.0134292048532865\\
12	0.0142508239155342\\
14	0.0145873299259398\\
16	0.0145940472729672\\
18	0.0143774737049785\\
20	0.0140113319470939\\
22	0.0135470232282006\\
24	0.0130207851187059\\
26	0.0124584372847311\\
28	0.0118783218420475\\
30	0.0112936613012644\\
32	0.0107139306552771\\
34	0.0101454484181667\\
36	0.00959288954416017\\
38	0.00905934989352526\\
40	0.00854702321251619\\
42	0.00805675575876155\\
44	0.00758925078573071\\
46	0.00714464410319122\\
48	0.00672268230988113\\
50	0.0063230095984452\\
52	0.00594489540558445\\
54	0.00558780723028283\\
56	0.0052507561494553\\
58	0.00493287934153157\\
60	0.00463352448416927\\
62	0.00435146268612587\\
64	0.00408617930093995\\
66	0.00383660127649306\\
68	0.0036021031616145\\
70	0.00338160772954037\\
72	0.00317429372004967\\
74	0.00297967145449723\\
76	0.00279688853993162\\
78	0.00262518711593696\\
80	0.00246397808326048\\
82	0.00231265694811438\\
84	0.00217061637840749\\
86	0.00203725566227697\\
88	0.00191200928053733\\
90	0.00179472129749857\\
92	0.00168447946040173\\
94	0.00158110047815896\\
96	0.0014840335435512\\
98	0.00139304722009306\\
100	0.00130752087721287\\
};
\addlegendentry{Theorem~\ref{thm:shannon_lower_bound} + Eq.~\refE{relations-1}};

\end{axis}
\end{tikzpicture}%
%
\caption{Upper and lower bounds to the channel coding error probability for an AWGN channel under a maximal power constraint. System parameters: $\text{SNR}=10$~dB, rate $R = 1.5$ bits/channel use.}\vspace{-2mm}\label{fig:maximal-AWGN-Pevsn-snr10dB}
\end{figure}
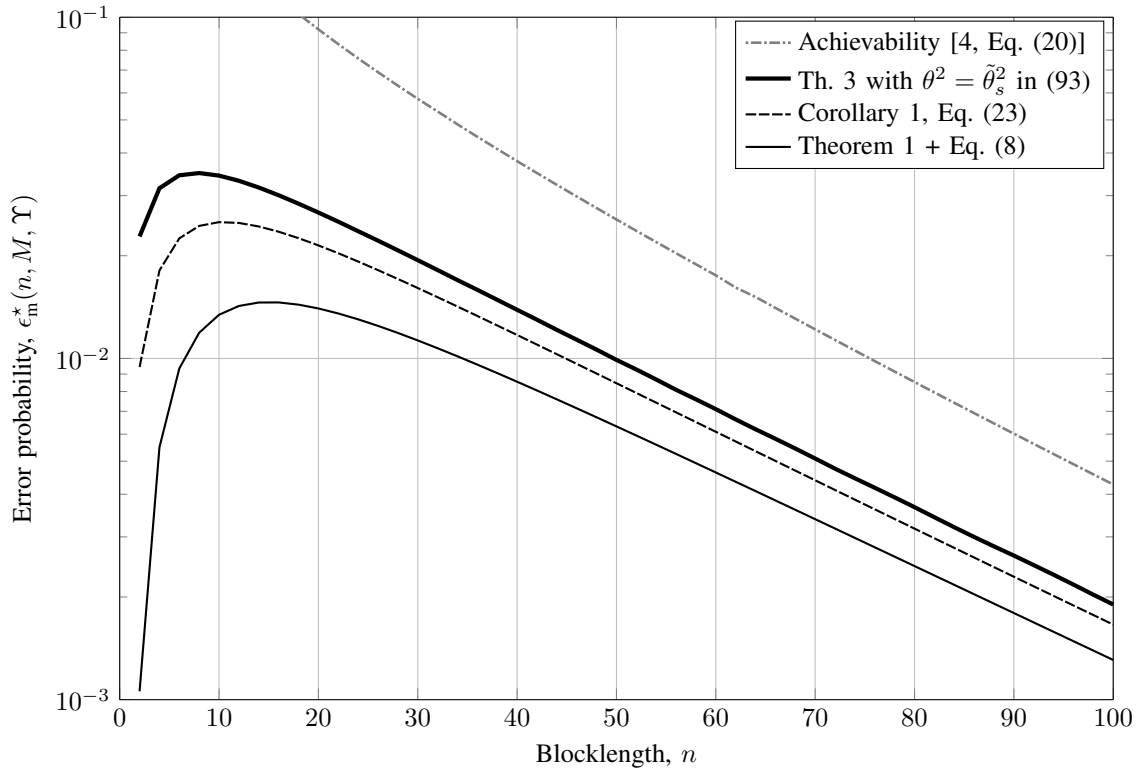%

\subsubsection{Equal power constraint}
In \refF{equal-AWGN-Pevsn-snr10dB}, we compare the lower bounds discussed in \refS{equal} for the AWGN channel under an equal power constraint.
For reference, we include the achievability part of \cite[Eq.~(20)]{Shannon59}, which was derived for an average power limitation and that, therefore, applies under equal, maximal and average power constraints. We observe that Shannon cone-packing bound from Theorem~\ref{thm:shannon_lower_bound} is the tightest lower bound in this setting. 
As discussed in \refS{equal-shannon}, by considering the optimal auxiliary (non-product) distribution, the meta-converse bound \refE{metaconverse} recovers the cone-packing bound and therefore, it coincides with the curve from Theorem~\ref{thm:shannon_lower_bound}.
The hypothesis testing bound with an auxiliary Gaussian distribution from Theorem~\ref{thm:PPV_lower_bound}  is slightly weaker.
Since the rate of the system $R=1.5$ bits/channel use is relatively close to channel capacity 
$C \approx 1.8$ bits/channel use, there is no much gain in this example by considering an auxiliary distribution equal to the exponent achieving output distribution with $\theta=\tilde\theta_s$ (see Section~\ref{sec:exponent-achieving}). Note however that both the capacity and exponent achieving output distributions are product distributions.

\subsubsection{Maximal power constraint} For the same system parameters as in the previous example, Fig.~\ref{fig:maximal-AWGN-Pevsn-snr10dB} compares different lower bounds derived under a maximal power constraint. In particular, we consider the combination of Theorem~\ref{thm:shannon_lower_bound} with \refE{relations-1}, the slightly sharper Corollary~\ref{cor:shannon_lower_bound} and the hypothesis testing bound from 
Theorem~\ref{thm:maximal_lower_bound}
with $\theta=\theta_s$ as defined in \refE{theta_s-def}.
In the figure, we can see that the tightest bound in this setting corresponds to the hypothesis testing bound from 
Theorem~\ref{thm:maximal_lower_bound}.
Applying the relation \refE{relations-1} to extend the cone-packing bound from Theorem~\ref{thm:shannon_lower_bound} to a maximal power constraint incurs in a certain loss, which can be slightly tightened as shown in Corollary~\ref{cor:shannon_lower_bound}.
In the figure we observe that the three curves present the same asymptotic slope as they feature the same error exponent.

\begin{figure}[t]%
\vspace{-0.3mm}
\centering% This file was created by matlab2tikz v0.6.0 running on MATLAB 8.2.
%Copyright (c) 2008--2014, Nico Schlömer <nico.schloemer@gmail.com>
%All rights reserved.
%Minimal pgfplots version: 1.3
%  http://www.mathworks.com/matlabcentral/fileexchange/22022-matlab2tikz
\begin{tikzpicture}

\begin{axis}[%
width=.8\linewidth,
height=.55\linewidth,
scale only axis,
separate axis lines,
every outer x axis line/.append style={black},
every x tick label/.append style={font=\color{black}},
xmajorgrids,
xmin=0,
xmax=100,
xlabel={Blocklength, $n$},
every outer y axis line/.append style={black},
every y tick label/.append style={font=\color{black}},
ymode=log,
ymajorgrids,
ymin=1e-5,
ymax=0.1,
yminorticks=true,
ylabel={Error probability, $\epsilon_{\text{a}}^{\star}(n,M,\Upsilon)$},
legend style={at={(0.991,0.995)},anchor=north east,legend cell align=left,align=left,draw=black}
]
%\addplot [color=black,solid,line width=0.5pt,mark=*,mark options={solid,scale=0.5}]
\addplot [color=gray,dash pattern=on 1pt off 1pt on 4pt off 1pt,line width=1pt]
  table[row sep=crcr]{%
2	0.455955733219089\\
4	0.298266272463783\\
6	0.232825092838634\\
8	0.193415958578272\\
10	0.165736837942085\\
12	0.144633459590168\\
14	0.127777344897316\\
16	0.113874111297367\\
18	0.102138894188286\\
20	0.0921030051648228\\
22	0.0833850896930203\\
24	0.0757472429348077\\
26	0.0690036349818882\\
28	0.0630122543302112\\
30	0.0576601816961487\\
32	0.0529241114413876\\
34	0.0485335066867251\\
36	0.044626002880964\\
38	0.041084384667068\\
40	0.0378673698078911\\
42	0.0349387914965699\\
44	0.0322663461099882\\
46	0.0298244606161055\\
48	0.0275888781188258\\
50	0.0255393991987647\\
52	0.0236586933600312\\
54	0.0219295372861573\\
56	0.0203388956804966\\
58	0.0188741201465949\\
60	0.0175236198359466\\
62	0.0161306229294946\\
64	0.0151432866727508\\
66	0.0140636838968213\\
68	0.0130802102210177\\
70	0.0121702407840231\\
72	0.0113276444927851\\
74	0.0105471173388172\\
76	0.00982304560155835\\
78	0.00915212561085251\\
80	0.00852945524331934\\
82	0.00795123423288897\\
84	0.00741413901444186\\
86	0.00691509404755351\\
88	0.00645125400557414\\
90	0.00601998715636451\\
92	0.00561885985041757\\
94	0.00524562204452637\\
96	0.00489819379827264\\
98	0.00457465268664452\\
100	0.00427322207647964\\
};
\addlegendentry{Achievability \cite[Eq.~(20)]{Shannon59}};

%\addplot [color=gray,dash pattern=on 2pt off 1pt on 2pt off 1pt,line width=0.8pt]
%  table[row sep=crcr]{%
%2	0.087049668551521\\
%4	0.0674793315545587\\
%6	0.062130523253862\\
%8	0.0578319146837727\\
%10	0.0538268509396664\\
%12	0.0500468291688573\\
%14	0.0464952069049285\\
%16	0.0431758908824696\\
%18	0.040085759455498\\
%20	0.0372164965609607\\
%22	0.0345563522632975\\
%24	0.0320924509406799\\
%26	0.0298115013696256\\
%28	0.0277006657124931\\
%30	0.0257467333164426\\
%32	0.0239381426302476\\
%34	0.0222632964548267\\
%36	0.0207123649673819\\
%38	0.0192749961995984\\
%40	0.0179430155551334\\
%42	0.0167082769958404\\
%44	0.0155625662953534\\
%46	0.0144997482922194\\
%48	0.0135129859435369\\
%50	0.0125966182361117\\
%52	0.0117456919440054\\
%54	0.0109545208311098\\
%56	0.0102192120097656\\
%58	0.00953555459380961\\
%60	0.00889950073579302\\
%62	0.00830754822595631\\
%64	0.00775658986084426\\
%66	0.0072438892205652\\
%68	0.00676610664093838\\
%70	0.0063210799890924\\
%72	0.00590638452507471\\
%74	0.00551993601935682\\
%76	0.00515933857943288\\
%78	0.00482345309146015\\
%80	0.0045100976071068\\
%82	0.00421765853056049\\
%84	0.00394473173477974\\
%86	0.00368999621374883\\
%88	0.00345221138932129\\
%90	0.00323021406396209\\
%92	0.00302291512646593\\
%94	0.00282929609979286\\
%96	0.00264840560435701\\
%98	0.00247935579645121\\
%100	0.0023213188296439\\
%};
%\addlegendentry{Theorem~\ref{thm:shannon_lower_bound}, Eq.~\refE{shannon_lower_bound}};

\addplot [color=black,solid,line width=1.6pt]
  table[row sep=crcr]{%
2	0.0227771513207305\\
4	0.0315379363092323\\
6	0.0344147744112234\\
8	0.0349399399283127\\
10	0.0343427624661513\\
12	0.0331597680731154\\
14	0.0316880575946233\\
16	0.0300785587518243\\
18	0.0284146434301356\\
20	0.0267649208162544\\
22	0.0251667933961467\\
24	0.0236006674686066\\
26	0.0221399543879728\\
28	0.0207397922855153\\
30	0.0194117509291246\\
32	0.0181575407124722\\
34	0.0169683367623379\\
36	0.0158894542139695\\
38	0.0148535460669512\\
40	0.013885645616868\\
42	0.0129881678765352\\
44	0.0121276102118403\\
46	0.0113466357780985\\
48	0.0106138753770168\\
50	0.00990224068088397\\
52	0.00928279292319438\\
54	0.00867478448219586\\
56	0.00809238338130142\\
58	0.00759104829677458\\
60	0.00710475778067897\\
62	0.00662786081068722\\
64	0.00620143439840396\\
66	0.00581605513294296\\
68	0.00544574520395657\\
70	0.0050903590842115\\
72	0.00475394689436381\\
74	0.00445472354208347\\
76	0.00417832907126781\\
78	0.00391715108654566\\
80	0.0036668854765047\\
82	0.00342525869468506\\
84	0.00320211695908252\\
86	0.00299835129662343\\
88	0.00281397240606852\\
90	0.00264453620162069\\
92	0.00248011736075191\\
94	0.00232390118887678\\
96	0.00217431729234844\\
98	0.00203111518089692\\
100	0.00190169878648021\\
};
\addlegendentry{Th.~\ref{thm:average_lower_bound} with $\theta^2=\tilde\theta_s^2$ in \refE{theta_s-def}};

\addplot [color=black,dash pattern=on 4pt off 1pt,line width=0.8pt]
  table[row sep=crcr]{%
2	1.86664110802483e-06\\
4	4.57719399211524e-06\\
6	1.44348277163094e-05\\
8	2.65011305184051e-05\\
10	3.767254730808e-05\\
12	4.6452894172052e-05\\
14	5.23900876265697e-05\\
16	5.63131198660072e-05\\
18	5.81228223844911e-05\\
20	5.83516422507151e-05\\
22	5.79783765722846e-05\\
24	5.64817740628744e-05\\
26	5.43142092767041e-05\\
28	5.21789399191925e-05\\
30	4.93278989488383e-05\\
32	4.67789635020707e-05\\
34	4.4261671278928e-05\\
36	4.15644077851092e-05\\
38	3.89272391822143e-05\\
40	3.62869140153513e-05\\
42	3.38670659032095e-05\\
44	3.13484006027321e-05\\
46	2.89143187737413e-05\\
48	2.7073319849001e-05\\
50	2.51756131190848e-05\\
52	2.35526171607146e-05\\
54	2.18144497560477e-05\\
56	2.02970090467377e-05\\
58	1.87804570859954e-05\\
60	1.7263817435345e-05\\
62	1.60169962758979e-05\\
64	1.49798710253434e-05\\
66	1.3772677184745e-05\\
68	1.27458917784906e-05\\
70	1.18279037118393e-05\\
72	1.0964331177172e-05\\
74	1.01602093782773e-05\\
76	9.44527485825061e-06\\
78	8.71674213034485e-06\\
80	8.01139189941525e-06\\
82	7.47369524415254e-06\\
84	6.89496771843005e-06\\
86	6.34348742748931e-06\\
88	5.85908427174932e-06\\
90	5.432850701587e-06\\
92	5.05561164882048e-06\\
94	4.72116329230668e-06\\
96	4.37305257936539e-06\\
98	4.02880371376403e-06\\
100	3.76601618722873e-06\\
};
\addlegendentry{Corollary~\ref{cor:shannon_lower_bound} + Eq.~\refE{relations-2} };

\addplot [color=black,solid,line width=0.8pt]
  table[row sep=crcr]{%
2	5.88492242582989e-09\\
4	2.00048774556915e-07\\
6	1.79458777406106e-06\\
8	5.4492486965326e-06\\
10	1.03295361232167e-05\\
12	1.53060228472409e-05\\
14	1.96491077801387e-05\\
16	2.333298815519e-05\\
18	2.59484369011611e-05\\
20	2.75703603202184e-05\\
22	2.8714969963395e-05\\
24	2.91685650121891e-05\\
26	2.89192825575518e-05\\
28	2.86695215456337e-05\\
30	2.78457912799364e-05\\
32	2.69412290190539e-05\\
34	2.60039683182106e-05\\
36	2.48547418600506e-05\\
38	2.36081163442849e-05\\
40	2.23275418373527e-05\\
42	2.11546015420811e-05\\
44	1.98051265404797e-05\\
46	1.84476547653244e-05\\
48	1.7431306587625e-05\\
50	1.63436211560608e-05\\
52	1.54564423069413e-05\\
54	1.44364472790998e-05\\
56	1.35028305887071e-05\\
58	1.26010440919879e-05\\
60	1.16633641003447e-05\\
62	1.08814091868864e-05\\
64	1.02311235121982e-05\\
66	9.44960145096982e-06\\
68	8.78451861542001e-06\\
70	8.21617320228426e-06\\
72	7.62946220618398e-06\\
74	7.10531426620864e-06\\
76	6.63743375707142e-06\\
78	6.1394898463463e-06\\
80	5.67847322519223e-06\\
82	5.30378309669509e-06\\
84	4.91813423317727e-06\\
86	4.53962921654836e-06\\
88	4.20546069963056e-06\\
90	3.90986262809232e-06\\
92	3.64792565628579e-06\\
94	3.41545943706349e-06\\
96	3.17978229883624e-06\\
98	2.92935926274168e-06\\
100	2.75044535205484e-06\\
};
\addlegendentry{Theorem~\ref{thm:shannon_lower_bound} + Eqs.~\refE{relations-1}-\refE{relations-2}};

\end{axis}
\end{tikzpicture}%
%
\caption{Upper and lower bounds to the channel coding error probability for an AWGN channel under an average power constraint. System parameters: $\text{SNR}=10$~dB, rate $R = 1.5$ bits/channel use.}\vspace{-2mm}\label{fig:average-AWGN-Pevsn-snr10dB}
\end{figure}
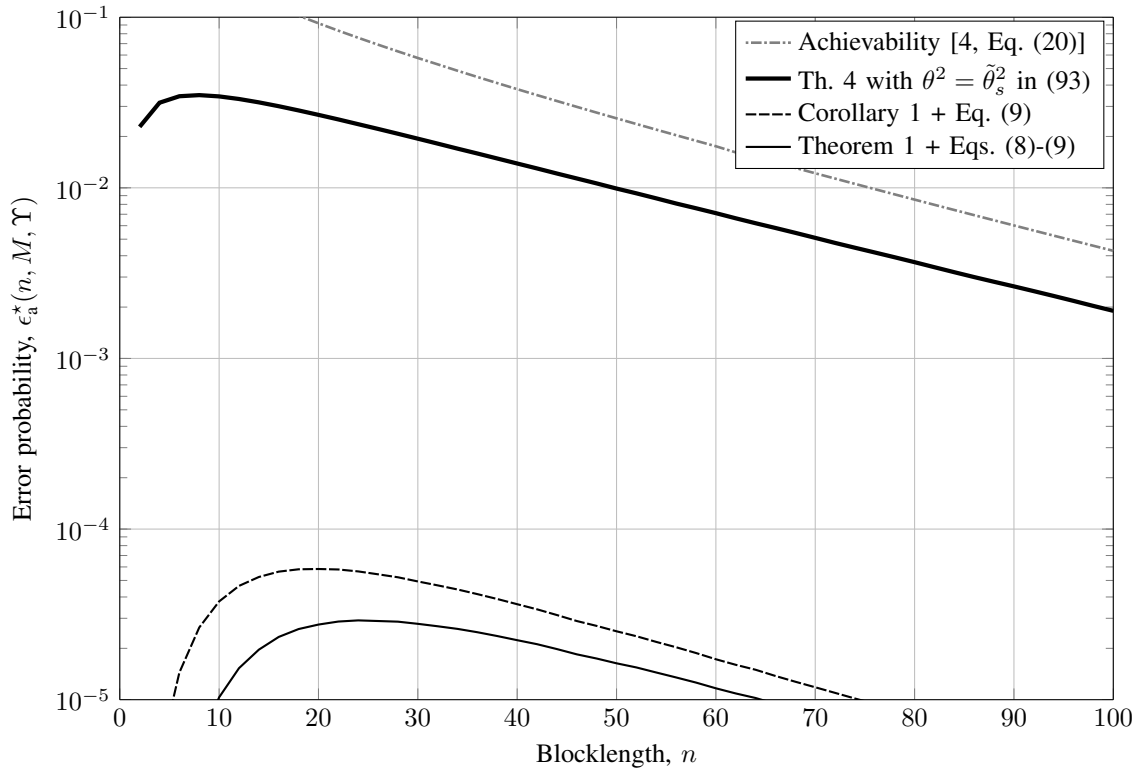%

\subsubsection{Average power constraint}
We compare now the bounds for an average power constraint. In particular, we consider the combination of Theorem~\ref{thm:shannon_lower_bound} with \refE{relations-1}-\refE{relations-2},
the combination of Corollary~\ref{cor:shannon_lower_bound} with \refE{relations-2} and the hypothesis testing bound from 
Theorem~\ref{thm:average_lower_bound}
with $\theta=\theta_s$ as defined in \refE{theta_s-def}.
For this set of system parameters, the condition in Corollary~\ref{cor:average_metaconverse_bound} is satisfied for all $n$, and the simplified bound \refE{cor_average_metaconverse_bound} can be used to evaluate Theorem~\ref{thm:average_lower_bound}.
The hypothesis testing bound  is again the tightest bound, as shown in Fig.~\ref{fig:average-AWGN-Pevsn-snr10dB}.
The application of \refE{relations-2} to obtain bounds for an average power constraint incurs in a large loss with respect to the corresponding direct hypothesis-testing bound.\footnote{All the bounds presented here hold under the average probability of error formalism. While the counterpart of \refE{relations-2} for maximal error probability is tighter (see~\cite[Lem.~65]{PolThesis}), the finite-length gap to Theorem~\ref{thm:average_lower_bound} is still significant.}

Since the condition in Corollary~\ref{cor:average_metaconverse_bound} is satisfied for all $n$, it follows that the bounds from Theorems~\ref{thm:PPV_lower_bound}, \ref{thm:maximal_lower_bound} and \ref{thm:average_lower_bound} coincide
in Figs.~\ref{fig:equal-AWGN-Pevsn-snr10dB}, \ref{fig:maximal-AWGN-Pevsn-snr10dB} and~\ref{fig:average-AWGN-Pevsn-snr10dB}.
While in the equal power constraint setting, Shannon cone-packing bound is still the tightest, under both maximal and average  power constraint the new hypothesis testing bounds yield tighter results.
Indeed, for an average power constraint the advantage of Theorem~\ref{thm:average_lower_bound} over previous results is significant in the finite blocklength regime, as shown in Fig.~\ref{fig:average-AWGN-Pevsn-snr10dB}.

%%%%%%%%%%%%%%%%%%%%%%%%%%%%%%%%%%%%%%%%%%%%%%%%%%%%%%%

\subsection{Exponent-achieving output distribution and numerical evaluation}

In the previous examples, the transmission rate was very close to channel capacity. Therefore, using the exponent achieving or the capacity achieving output distributions in the hypothesis-testing bounds did not result in significant differences. 
We consider now an power-constrained AWGN channel with $\text{SNR}=10\log_{10}\frac{\Upsilon}{\sigma^2}=5$~dB. The asymptotic capacity of this channel is $C\approx 1.03$ bits/channel use and its critical rate is $R_{\text{cr}}\approx 0.577$ bits/channel use.\footnote{The critical rate of a channel is defined as the point below which the sphere-packing exponent and the random-coding exponent start to diverge~\cite{Gall68}. For the power-constrained AWGN channel this point corresponds to the rate at which the maximum in \refE{Esp-def} is attained for $s=\frac{1}{2}.$}

\begin{figure}[t]%
\centering% This file was created by matlab2tikz v0.6.0 running on MATLAB 9.7.
%Copyright (c) 2008--2014, Nico Schlömer <nico.schloemer@gmail.com>
%All rights reserved.
%Minimal pgfplots version: 1.3
%  http://www.mathworks.com/matlabcentral/fileexchange/22022-matlab2tikz
\begin{tikzpicture}

\begin{axis}[%
width=.8\linewidth,
height=.55\linewidth,
scale only axis,
separate axis lines,
every outer x axis line/.append style={black},
every x tick label/.append style={font=\color{black}},
xmajorgrids,
xmin=0,
xmax=100,
xlabel={Blocklength, $n$},
every outer y axis line/.append style={black},
every y tick label/.append style={font=\color{black}},
ymode=log,
ymajorgrids,
yminorgrids,
ymin=1e-08,
ymax=1e-02,
yminorticks=true,
ylabel={Error probability, $\Pe$},
legend style={legend cell align=left,align=left,draw=black}
]

\addplot [color=gray,dash pattern=on 1pt off 1pt on 4pt off 1pt,line width=1pt]
  table[row sep=crcr]{%
2	nan\\
4	0.431559895221708\\
6	0.102686050383235\\
8	0.0602824237555683\\
10	0.039512337166487\\
12	0.0269900956958788\\
14	0.0188346633952803\\
16	0.0133306789681951\\
18	0.00952795806712819\\
20	0.00686034507678901\\
22	0.00494959062784987\\
24	0.00361497462083986\\
26	0.00264058757236993\\
28	0.0019352202211199\\
30	0.00142230808354369\\
32	0.00104792458607112\\
34	0.000741624312158378\\
36	0.000550235411261716\\
38	0.000410004688848877\\
40	0.000306641241394514\\
42	0.000230039031142246\\
44	0.000172999257287236\\
46	0.000130353511493014\\
48	9.83622199933107e-05\\
50	7.43009199219426e-05\\
52	5.61655297397154e-05\\
54	4.24764122392715e-05\\
56	3.21311685258774e-05\\
58	2.430755631695e-05\\
60	1.83878049931924e-05\\
62	1.39078116542179e-05\\
64	1.05171303925088e-05\\
66	7.95093353316535e-06\\
68	6.00915287865997e-06\\
70	4.54021184202002e-06\\
72	3.42921280018106e-06\\
74	2.58927780273558e-06\\
76	1.95443292732966e-06\\
78	1.47478194217658e-06\\
80	1.11250430668498e-06\\
82	8.3897686482857e-07\\
84	6.32531856839207e-07\\
86	4.76761661367018e-07\\
88	3.5926082974156e-07\\
90	2.70656770869996e-07\\
92	2.03861729521488e-07\\
94	1.53519631511049e-07\\
96	1.15585505055263e-07\\
98	8.70119900005422e-08\\
100	6.54903846078056e-08\\
};
\addlegendentry{Achievability \cite[Eq.~(20)]{Shannon59}};

\addplot [color=black,solid,line width=0.8pt]
  table[row sep=crcr]{%
2	0.00890838511366413\\
4	0.00685988295800842\\
6	0.00521460507103868\\
8	0.00386079869327319\\
10	0.00283077719729727\\
12	0.00206876897327971\\
14	0.00151103489329001\\
16	0.00110438686118638\\
18	0.000808148975541957\\
20	0.000592226008044074\\
22	0.000434652551629128\\
24	0.000319488199270983\\
26	0.000235183794471336\\
28	0.000173368384560031\\
30	0.000127971182647255\\
32	9.45805236831959e-05\\
34	6.99853879591279e-05\\
36	5.18438019421936e-05\\
38	3.844547094353e-05\\
40	2.85377723118094e-05\\
42	2.12030729757027e-05\\
44	1.57674374927624e-05\\
46	1.17350183981682e-05\\
48	8.74060488015689e-06\\
50	6.51515919914458e-06\\
52	4.85968170926315e-06\\
54	3.6273200793375e-06\\
56	2.7091429507065e-06\\
58	2.02461351949695e-06\\
60	1.51388842763851e-06\\
62	1.13262312643939e-06\\
64	8.47821445292154e-07\\
66	6.34938970976203e-07\\
68	4.75736152864268e-07\\
70	3.56615575525082e-07\\
72	2.67433008887238e-07\\
74	2.00640057677942e-07\\
76	1.50587206824516e-07\\
78	1.13064027361315e-07\\
80	8.49213326496523e-08\\
82	6.3806061118234e-08\\
84	4.79577940944939e-08\\
86	3.60576877826538e-08\\
88	2.71186657709214e-08\\
90	2.0401942312822e-08\\
92	1.53533787378033e-08\\
94	1.15573324932484e-08\\
96	8.70201308251009e-09\\
98	6.5540828451839e-09\\
100	4.93746923587669e-09\\
};
\addlegendentry{Theorem~\ref{thm:shannon_lower_bound}, Eq.~\refE{shannon_lower_bound}};

\addplot [color=black,solid,line width=1.5pt] 
  table[row sep=crcr]{%
2	0.00884141309330651\\
4	0.00632064753099055\\
6	0.00473330073379434\\
8	0.00349307632792853\\
10	0.00256097381353813\\
12	0.00187364706880733\\
14	0.00137072812739355\\
16	0.00100277353408155\\
18	0.000735076566104045\\
20	0.000538968825005736\\
22	0.000395739040731667\\
24	0.000291757159501106\\
26	0.000214847755404293\\
28	0.000158216058725371\\
30	0.000117050864864996\\
32	8.67417383009081e-05\\
34	6.40254126840559e-05\\
36	4.7648815574776e-05\\
38	3.53314897391691e-05\\
40	2.62121596369098e-05\\
42	1.95105127598246e-05\\
44	1.45221802314246e-05\\
46	1.08108380245602e-05\\
48	8.04604430683518e-06\\
50	6.02132631325491e-06\\
52	4.49964762588115e-06\\
54	3.35943780170881e-06\\
56	2.50190671893897e-06\\
58	1.87937009464855e-06\\
60	1.40541603706585e-06\\
62	1.04467394246245e-06\\
64	7.89253846918932e-07\\
66	5.87655405130676e-07\\
68	4.4377428807049e-07\\
70	3.30358557752205e-07\\
72	2.49884609808392e-07\\
74	1.85471372169221e-07\\
76	1.40785114664956e-07\\
78	1.05954765834932e-07\\
80	7.91937551109592e-08\\
82	5.99257646740885e-08\\
84	4.43120586601534e-08\\
86	3.37744973135506e-08\\
88	2.55536940480572e-08\\
90	1.8863415072811e-08\\
92	1.44004127810037e-08\\
94	1.09193391242394e-08\\
96	8.23255448912497e-09\\
98	6.12643492995898e-09\\
100	4.66571848116696e-09\\
};
\addlegendentry{Th.~\ref{thm:average_lower_bound} with $\theta^2=\tilde\theta_s^2$ in \refE{theta_s-def}};

\addplot [color=black,dash pattern=on 4pt off 1pt,line width=0.8pt]
  table[row sep=crcr]{%
2	0.00873114512722816\\
4	0.00562049677901826\\
6	0.00410264689071258\\
8	0.0030128576314681\\
10	0.00220942684063686\\
12	0.00161890499484507\\
14	0.0011863293988377\\
16	0.000869947029174999\\
18	0.00063860059575855\\
20	0.000469340476307014\\
22	0.000345380447907521\\
24	0.000254484867733144\\
26	0.000187746973434455\\
28	0.000138678658496021\\
30	0.000102553703738742\\
32	7.59234763999062e-05\\
34	5.62673148042902e-05\\
36	4.1741578011667e-05\\
38	3.09949225604076e-05\\
40	2.30357069668267e-05\\
42	1.71347428874215e-05\\
44	1.27556394062996e-05\\
46	9.50289806188811e-06\\
48	7.08470099005901e-06\\
50	5.28548866216743e-06\\
52	3.94574274266478e-06\\
54	2.94740339276676e-06\\
56	2.20295642296365e-06\\
58	1.64746148133423e-06\\
60	1.23268214592505e-06\\
62	9.22816133177633e-07\\
64	6.91165786564873e-07\\
66	5.17920658331717e-07\\
68	3.88267448901618e-07\\
70	2.9119210086671e-07\\
72	2.18481034908441e-07\\
74	1.6398804804058e-07\\
76	1.23133864781141e-07\\
78	9.24908480106061e-08\\
80	6.94973027390824e-08\\
82	5.22374757962529e-08\\
84	3.9277429983123e-08\\
86	2.9541504269093e-08\\
88	2.22259253253444e-08\\
90	1.6726859335779e-08\\
92	1.25915060225431e-08\\
94	9.4812951291263e-09\\
96	7.14125092173864e-09\\
98	5.38005096581011e-09\\
100	4.05418391037361e-09\\
};
\addlegendentry{Corollary~\ref{cor:shannon_lower_bound}, Eq.~\refE{cor_shannon_lower_bound} };

\addplot [color=black,dotted,mark=o,mark options={solid, scale=0.7}]
 table[row sep=crcr]{%
2	0.00459312212771501\\
4	0.00425711812277432\\
6	0.0033437000654932\\
8	0.0025040783785913\\
10	0.00184097802738326\\
12	0.00134285228041998\\
14	0.000976217207947288\\
16	0.000708873728695867\\
18	0.000514723623938078\\
20	0.000373877022737781\\
22	0.000271887815354174\\
24	0.000197842047050785\\
26	0.000144198170152418\\
28	0.000105168069459092\\
30	7.67795126520809e-05\\
32	5.6081182791142e-05\\
34	4.1036432492826e-05\\
36	3.00832908555392e-05\\
38	2.20612112839705e-05\\
40	1.61807400503264e-05\\
42	1.18785235960556e-05\\
44	8.73925104923781e-06\\
46	6.43004358967332e-06\\
48	4.73362134989876e-06\\
50	3.48707241671217e-06\\
52	2.56985004740022e-06\\
54	1.89346926739061e-06\\
56	1.39601661419716e-06\\
58	1.03319732206137e-06\\
60	7.63495390738233e-07\\
62	5.62558787702842e-07\\
64	4.17619535260074e-07\\
66	3.08434964531549e-07\\
68	2.28680487679005e-07\\
70	1.69232072581106e-07\\
72	1.25518277890098e-07\\
74	9.28784192013922e-08\\
76	6.90554944259602e-08\\
78	5.09241637505037e-08\\
80	3.80295059495516e-08\\
82	2.81709587558405e-08\\
84	2.09025294191399e-08\\
86	1.55832153248535e-08\\
88	1.15379622202682e-08\\
90	8.57731558461512e-09\\
92	6.40274443388352e-09\\
94	4.74995360703168e-09\\
96	3.51595876440344e-09\\
98	2.63475569407231e-09\\
100	1.96272479684448e-09\\
};
\addlegendentry{Th.~\ref{thm:average_lower_bound} with $\theta^2 = \Upsilon + \sigma^2$};

\addplot [color=black,dash pattern=on 4pt off 1pt,line width=0.8pt,mark=*,mark options={solid, scale=0.5}]
  table[row sep=crcr]{%
2	0.0009087863888233\\
4	3.31098391427129e-05\\
6	4.6854584430147e-06\\
8	1.72264094995995e-06\\
10	9.47683247569837e-07\\
12	6.06988602707494e-07\\
14	4.14865997966052e-07\\
16	2.91259428663852e-07\\
18	2.07266239201615e-07\\
20	1.47927465332571e-07\\
22	1.05821254117198e-07\\
24	7.60968346356401e-08\\
26	5.48673801187847e-08\\
28	3.94951123022475e-08\\
30	2.82797942318251e-08\\
32	2.04228162474648e-08\\
34	1.47212064312125e-08\\
36	1.06151903099734e-08\\
38	7.63808367598033e-09\\
40	5.50847548908945e-09\\
42	4.00589261161386e-09\\
44	2.89034120806877e-09\\
46	2.10071771345999e-09\\
48	1.51726593257205e-09\\
50	1.10006326694081e-09\\
52	7.96267941543932e-10\\
54	5.81571969223375e-10\\
56	4.18250007314234e-10\\
58	3.06553732650239e-10\\
60	2.24985024771521e-10\\
62	1.63617804436493e-10\\
64	1.1883925831523e-10\\
66	8.67079447918772e-11\\
68	6.34191499631634e-11\\
70	4.60535192582917e-11\\
72	3.38251604635074e-11\\
74	2.46715499914979e-11\\
76	1.81879489753285e-11\\
78	1.32568725318759e-11\\
80	9.75376186572067e-12\\
82	7.15671685547057e-12\\
84	5.25085140212175e-12\\
86	3.88484792955616e-12\\
88	2.82902364271044e-12\\
90	2.10251635607103e-12\\
92	1.53872397386377e-12\\
94	1.14043997284046e-12\\
96	8.39306650008909e-13\\
98	6.17891810723792e-13\\
100	4.54117575751713e-13\\
};
\addlegendentry{Corollary~\ref{cor:shannon_lower_bound} + Eq.~\refE{relations-2} };

\end{axis}
\end{tikzpicture}%
%
\caption{Upper and lower bounds to the channel coding error probability over an AWGN channel with $\text{SNR} = 5$~dB and $R= 0.58$ bits/channel use.}\label{fig:AWGN-Pevsn-R058-snr5dB}
\end{figure}
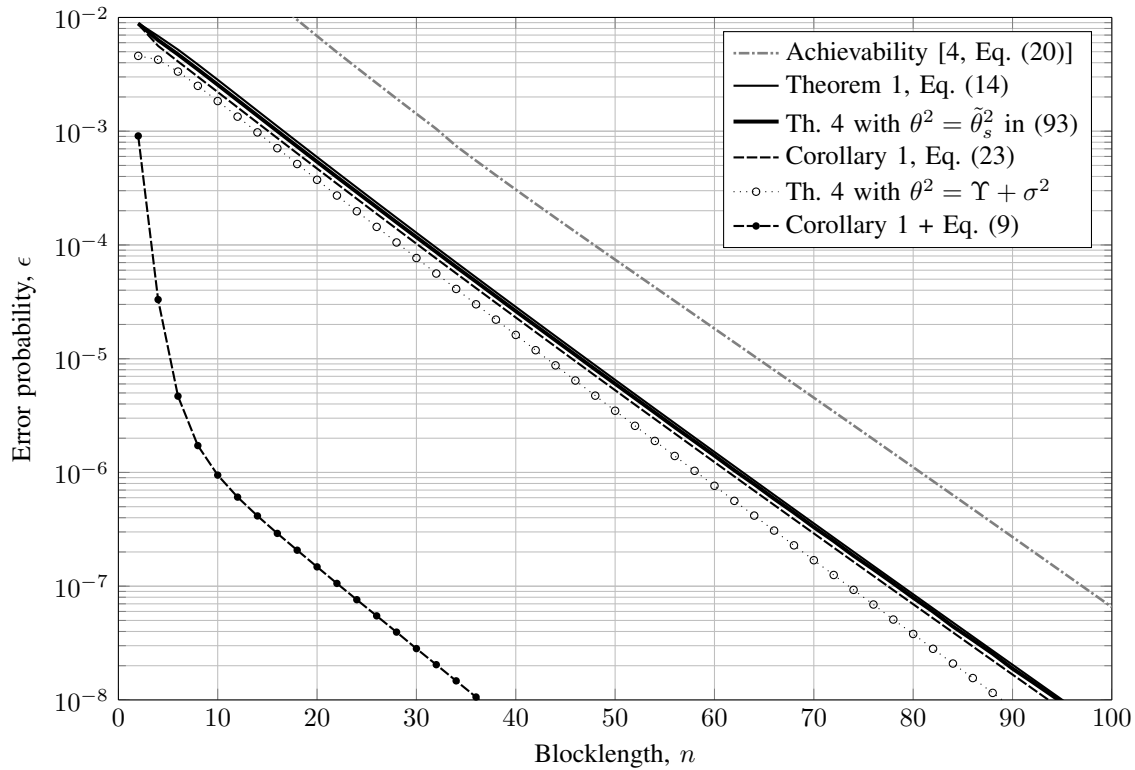%

\refFig{AWGN-Pevsn-R058-snr5dB} shows several of the bounds studied in the previous examples for $M = \lceil 2^{nR} \rceil$ messages with $R=0.58$ bits/channel use.
This transmission rate is slightly above the critical rate of the channel. We first note that the gap between the achievability part of \cite[Eq.~(20)]{Shannon59} and the converse bounds is larger than in the examples from Figures~\ref{fig:equal-AWGN-Pevsn-snr10dB}, \ref{fig:maximal-AWGN-Pevsn-snr10dB} and \ref{fig:average-AWGN-Pevsn-snr10dB}.
For an equal power constraint Shannon cone-packing bound Theorem~\ref{thm:shannon_lower_bound} is still the tightest bound. Nevertheless, the gap between Theorem~\ref{thm:shannon_lower_bound} and the curve of Theorem~\ref{thm:average_lower_bound} with $\theta^2=\tilde\theta_s^2$ in \refE{theta_s-def} for the value of $s$ optimizing \refE{HT-sp-exponent} is very small, and the later is valid under the three power constraints considered. In this example, Corollary~\ref{cor:shannon_lower_bound} and Corollary~\ref{cor:shannon_lower_bound} + Eq.~\refE{relations-2} yield weaker and much weaker bounds, respectively.
An error exponent analysis of these bounds shows that the asymptotic slope of the hypothesis-testing lower bound from Theorem~\ref{thm:average_lower_bound} with $\tilde\theta_s^2$ coincides with that of Shannon cone-packing lower bound from Theorem~\ref{thm:shannon_lower_bound}, hence both curves are parallel. In contrast, the curve with $\theta^2 = \Upsilon +\sigma^2$ presents a (slightly) larger error exponent and hence this bound will diverge as $n$ grows and become increasingly weaker. By using the value $\theta^2 = \tilde\theta_s^2$, we obtain not only the sphere-packing exponent but also tighter finite-length bounds, as shown in the figure.

\begin{figure}[t]%
\centering\input{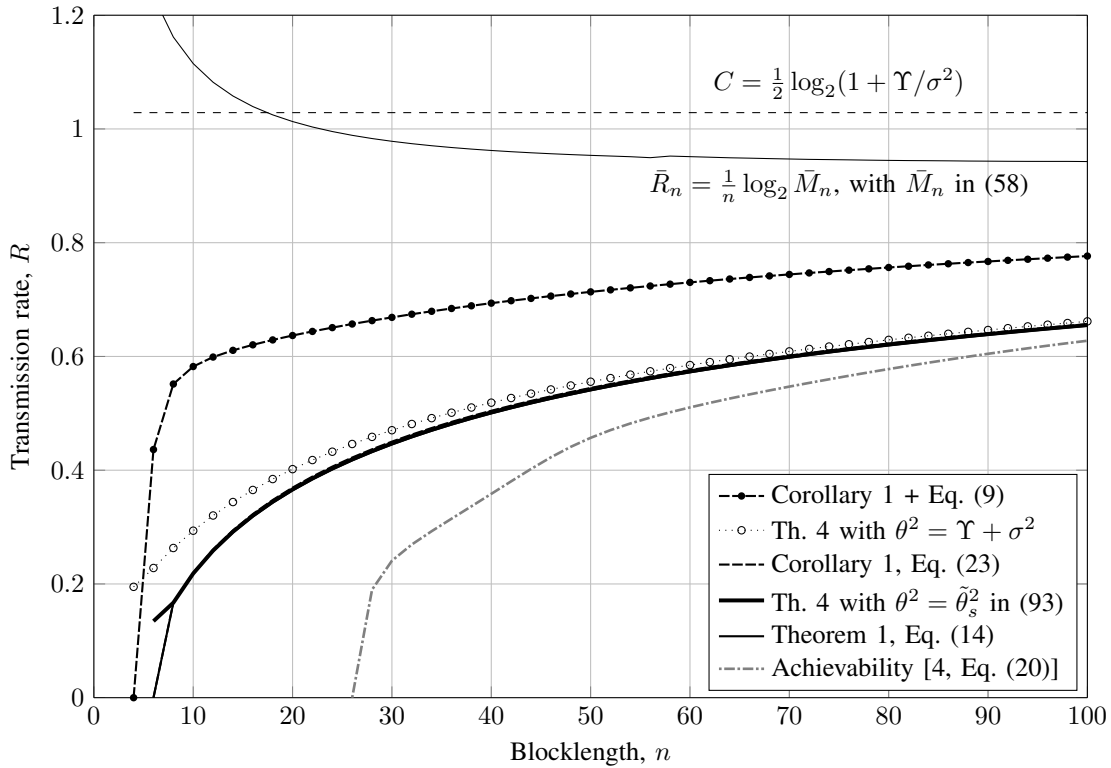}%
\caption{Bounds to the transmission rate in an AWGN channel with $\text{SNR} = 5$~dB and error probability $\Pe = 10^{-6}$.}\label{fig:AWGN-Rvsn-snr5dB}
\end{figure}%

The observations from \refF{AWGN-Pevsn-R058-snr5dB} are complemented in \refF{AWGN-Rvsn-snr5dB}. This figure analyzes the highest transmission rate versus the blocklength for a given error probability $\Pe=10^{-6}$.
The bounds included in this figure are precisely those from \refF{AWGN-Pevsn-R058-snr5dB}. For reference, we have also included the asymptotic channel capacity $C$ and the condition from Corollary~\ref{cor:average_metaconverse_bound} as an upper bound on the transmission rate of the system $R \leq \bar{R}_n = \frac{1}{n}\log_2 \bar{M}_n$.
In this plot the upper bounds from Theorem~\ref{thm:shannon_lower_bound}, Corollary~\ref{cor:shannon_lower_bound} and Theorem~\ref{thm:average_lower_bound} with $\theta^2=\tilde\theta_s^2$ are almost indistinguishable from each other. Note however that Theorem~\ref{thm:shannon_lower_bound} was derived for an equal power constraint, Corollary~\ref{cor:shannon_lower_bound} for a maximal power constraint and Theorem~\ref{thm:average_lower_bound} forr an average power constraint.
Comparing these bounds with the achievability bound \cite[Eq.~(20)]{Shannon59}, we observe that the behavior of the transmission rate approaching capacity is precisely characterized for blocklengths $n\geq 30$.
The upper bound from Theorem~\ref{thm:average_lower_bound} with $\theta^2=\Upsilon+\sigma^2$ is slightly looser than by considering the exponent achieving output distribution and Corollary~\ref{cor:shannon_lower_bound} + Eq.~\refE{relations-2} yields a much weaker bound for average power constraints.
The condition from Corollary~\ref{cor:average_metaconverse_bound} shows that the hypothesis testing bound \refE{cor_average_metaconverse_bound} can be used to evaluate Theorem~\ref{thm:average_lower_bound} in the range of values of $n$ considered, as $\bar{R}_n = \frac{1}{n}\log_2 \bar{M}_n$ is well-above of the Theorem~\ref{thm:average_lower_bound} curves.

\begin{figure}[t]%
\centering\input{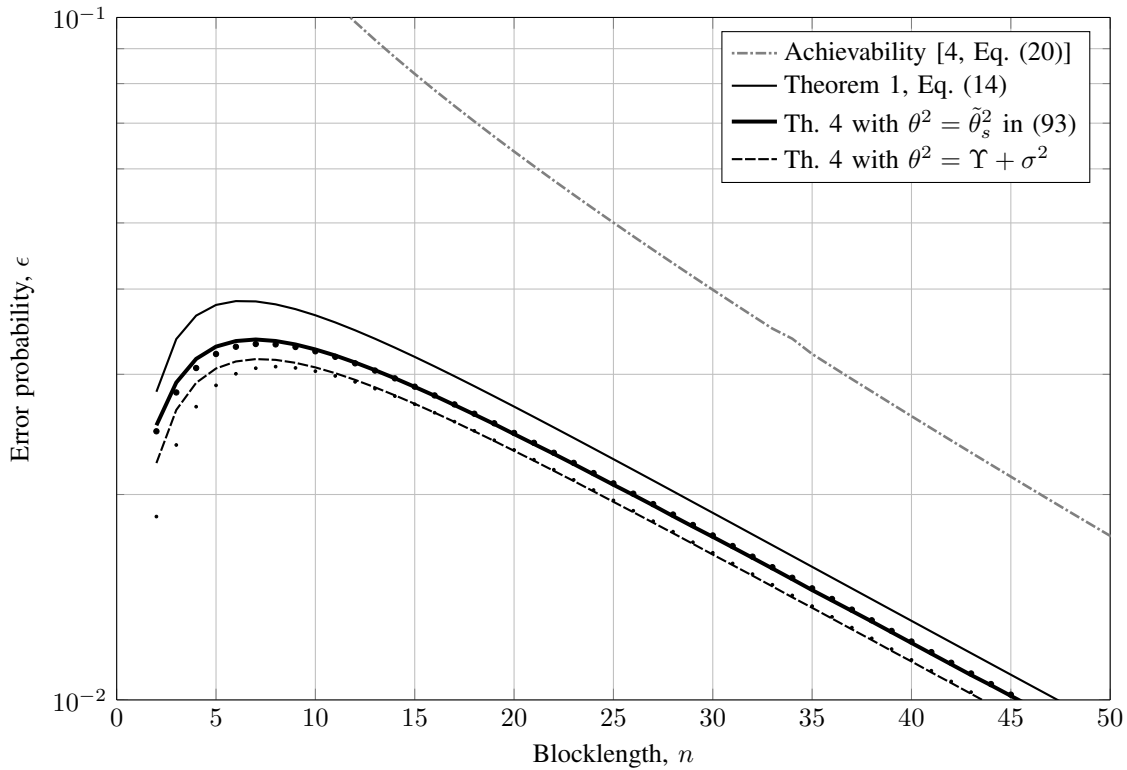}%
\caption{Lower bounds to the channel coding error probability over an AWGN channel with $\text{SNR} = 5$~dB and $R= 0.8$ bits/channel use. 
The bounds from Theorem~\ref{thm:average_lower_bound} have been evaluated using Theorem~\ref{thm:alpha-beta-opt} (lines) and using Theorem~\ref{thm:alpha-beta-sp-formulation} disregarding the small-o terms (markers $\bullet$). }\label{fig:AWGN-Pevsn-R08-snr5dB}
\end{figure}%

To evaluate the accuracy of the saddlepoint expansion introduced in Theorem~\ref{thm:alpha-beta-sp-formulation}, we show in  \refF{AWGN-Pevsn-R08-snr5dB} the exact hypothesis-testing bound computed according to Theorem~\ref{thm:alpha-beta-opt} (lines)  with the approximation that follows from disregarding the $ o\bigl(n^{-\frac{1}{2}}\bigr)$ terms in Theorem~\ref{thm:alpha-beta-sp-formulation} (markers $\bullet$). We can see that, for both the capacity-achieving variance $\theta^2$ and the exponent-achieving variance $\tilde\theta_s^2$, the approximation is accurate for blocklengths as short as $n=10$. This will also be true for larger values of $n$, for which numerical evaluation of the Marcum-$Q$ functions becomes infeasible using traditional methods.

If we compare the curves of the cone-packing bound from Theorem~\ref{thm:shannon_lower_bound} and
the hypothesis testing bounds from Theorems~\ref{thm:PPV_lower_bound},~\ref{thm:maximal_lower_bound} and
~\ref{thm:average_lower_bound} in Figs.~\ref{fig:equal-AWGN-Pevsn-snr10dB}-\ref{fig:AWGN-Pevsn-R08-snr5dB}, we only observe a small difference. Then, for practical purposes, it may be sufficient to use Theorem~\ref{thm:average_lower_bound}  as lower bound --since it was derived under an average power constraint, it applies for all equal, maximal and average power limitations-- and the achievability part of \cite[Eq. (20)]{Shannon59} as an upper bound --this bound was derived assuming an equal-power constraint, and since it is an achievability result, it applies for all equal, maximal and average power limitations. The saddlepoint approximation from Theorem~\ref{thm:alpha-beta-sp-formulation} is accurate for values of $n\geq 10$ and can be safely applied in the evaluation of Theorem~\ref{thm:average_lower_bound}.

\subsection{Constellation design for uncoded transmission  ($n=2$)}
\label{sec:constellations}

In the last example of this section, we consider the problem of transmitting $M\geq 2$ codewords over $n=2$ uses of an AWGN channel with $\text{SNR}=10$~dB. This problem corresponds to finding the best constellation for an uncoded quadrature communication system.

\begin{figure}[t]%
\centering% This file was created by matlab2tikz v0.6.0 running on MATLAB 8.2.
%Copyright (c) 2008--2014, Nico Schlömer <nico.schloemer@gmail.com>
%All rights reserved.
%Minimal pgfplots version: 1.3
%  http://www.mathworks.com/matlabcentral/fileexchange/22022-matlab2tikz
\begin{tikzpicture}

\begin{axis}[%
width=.8\linewidth,
height=.55\linewidth,
scale only axis,
separate axis lines,
every outer x axis line/.append style={black},
every x tick label/.append style={font=\color{black}},
xmajorgrids,
xmin=0,
xmax=70,
xlabel={Number of codewords, $M$},
every outer y axis line/.append style={black},
every y tick label/.append style={font=\color{black}},
ymajorgrids,
ymin=0,
ymax=0.8,
ylabel={Error probability, $\Pe$},
legend style={at={(0.995,0.01)},anchor=south east,legend cell align=left,align=left,draw=black}
]

\addplot [color=black,only marks,mark=o,mark options={solid},forget plot]
  table[row sep=crcr]{%
2	1e-05\\
4	0.00152\\
6	0.0257899999999991\\
8	0.0873799999999988\\
10	0.167910000000027\\
12	0.245830000000105\\
14	0.32047000000018\\
16	0.38029000000024\\
18	0.436580000000296\\
20	0.483980000000343\\
22	0.527320000000235\\
24	0.560290000000085\\
26	0.590819999999946\\
28	0.616129999999831\\
30	0.638529999999729\\
32	0.661719999999623\\
34	0.680749999999537\\
36	0.696979999999463\\
38	0.711419999999397\\
40	0.727059999999326\\
42	0.739949999999267\\
44	0.749539999999224\\
46	0.758069999999185\\
48	0.768829999999136\\
50	0.780339999999083\\
52	0.78551999999906\\
54	0.795339999999015\\
56	0.802799999998981\\
58	0.80755999999896\\
60	0.816629999998918\\
62	0.81852999999891\\
64	0.826069999998875\\
66	0.833179999998843\\
68	0.836559999998828\\
70	0.841649999998804\\
};
%\addlegendentry{$M$-PSK $\in\Lc_{\text{e}}$};

\addplot [color=black,only marks,mark=x,mark options={solid},forget plot]
  table[row sep=crcr]{%
2	0\\
4	0.00144\\
6	0.0262199999999991\\
8	0.0614700000000082\\
10	0.149280000000009\\
12	0.227110000000086\\
14	0.27110000000013\\
16	0.316840000000176\\
18	0.354900000000214\\
20	0.40078000000026\\
22	0.435750000000295\\
24	0.49028000000035\\
26	0.521490000000261\\
28	0.545270000000153\\
30	0.576940000000009\\
32	0.60525999999988\\
34	0.625859999999786\\
36	0.639799999999723\\
38	0.656189999999648\\
40	0.663349999999616\\
42	0.683079999999526\\
44	0.694929999999472\\
46	0.707649999999414\\
48	0.720679999999355\\
50	0.733729999999296\\
52	0.744949999999244\\
54	0.7546299999992\\
56	0.761869999999167\\
58	0.773229999999116\\
60	0.779369999999088\\
62	0.783999999999067\\
64	0.792839999999027\\
66	0.798489999999001\\
68	0.791349999999033\\
70	0.797209999999007\\
};
%\addlegendentry{$M$-APSK $\in\Lc_{\text{m}}$};

\addplot [color=black,only marks,mark=*,mark options={solid,scale=0.5},forget plot]
  table[row sep=crcr]{%
2	0\\
4	0.00144\\
6	0.0141399999999996\\
8	0.0439700000000028\\
10	0.100859999999994\\
12	0.135419999999995\\
14	0.163980000000023\\
16	0.20069000000006\\
18	0.234770000000094\\
20	0.274910000000134\\
22	0.309000000000168\\
24	0.3411100000002\\
26	0.361480000000221\\
28	0.385770000000245\\
30	0.403160000000262\\
32	0.422050000000281\\
34	0.442010000000301\\
36	0.458810000000318\\
38	0.47103000000033\\
40	0.487230000000347\\
42	0.499850000000359\\
44	0.511510000000307\\
46	0.522360000000258\\
48	0.5350000000002\\
50	0.542140000000167\\
52	0.550620000000129\\
54	0.565170000000063\\
56	0.570870000000037\\
58	0.583969999999977\\
60	0.585649999999969\\
62	0.595529999999925\\
64	0.601999999999895\\
66	0.607339999999871\\
68	0.618909999999818\\
70	0.623899999999795\\
};
%\addlegendentry{$M$-APSK $\in\Lc_{\text{a}}$};

%\addplot [color=blue,dotted,line width=1pt]
\addplot [color=gray,dash pattern=on 2pt off 1pt on 2pt off 1pt,line width=0.8pt]
  table[row sep=crcr]{%
2	3.87210821552205e-06\\
4	0.00156545802956606\\
6	0.0253602425788807\\
8	0.087049668551521\\
10	0.167064011083357\\
12	0.24719407720049\\
14	0.31980891614697\\
16	0.38311606620947\\
18	0.437589018234761\\
20	0.484374496045526\\
22	0.524688199022999\\
24	0.559616142687848\\
26	0.590069961608248\\
28	0.616796034652221\\
30	0.640400049128813\\
32	0.661372993179782\\
34	0.680114029405573\\
36	0.696949272512337\\
38	0.712146743692482\\
40	0.725928126093416\\
42	0.73847796608601\\
44	0.749950878448315\\
46	0.760477205949009\\
48	0.770167485336073\\
50	0.779115990604782\\
52	0.787403560649979\\
54	0.795099869433959\\
56	0.802265259588687\\
58	0.808952232229486\\
60	0.815206664471459\\
62	0.821068810020543\\
64	0.826574125953638\\
66	0.831753959445601\\
68	0.836636121024651\\
70	0.841245365399631\\
};
\addlegendentry{Th.~\ref{thm:shannon_lower_bound} (equal power constraint)};

%\addplot [color=black,dash pattern=on 1pt off 3pt on 3pt off 3pt,line width=1pt]
\addplot [color=black,solid,line width=0.8pt]
  table[row sep=crcr]{%
2	0\\
4	0.000554343446136664\\
6	0.00648235224246903\\
8	0.0231621008317434\\
10	0.0495886307485417\\
12	0.0821793869392142\\
14	0.117727001321701\\
16	0.154043597838796\\
18	0.189818720740142\\
20	0.224297908951651\\
22	0.257077396204988\\
24	0.287996590705995\\
26	0.317028716470207\\
28	0.344212104475585\\
30	0.369640254804997\\
32	0.393416452615887\\
34	0.415658600126905\\
36	0.436478099164263\\
38	0.455982319279488\\
40	0.474279834084087\\
42	0.491465277069596\\
44	0.507622331074684\\
46	0.522835103435229\\
48	0.53718074080763\\
50	0.550723082715317\\
52	0.563523864816942\\
54	0.575637220090718\\
56	0.587122442893334\\
58	0.598018361438232\\
60	0.608366982966327\\
62	0.618215355992831\\
64	0.627589081305372\\
66	0.63651572235913\\
68	0.64503669853054\\
70	0.653165967534063\\
};
\addlegendentry{Th.~\ref{thm:maximal_lower_bound} (maximal power constraint)};

\addplot [color=black,solid,line width=1.5pt]
  table[row sep=crcr]{%
2	0\\
4	0\\
6	0.00644504813743696\\
8	0.0235878130177086\\
10	0.0502550145675667\\
12	0.0829290000469514\\
14	0.118492324917788\\
16	0.154773547346803\\
18	0.19044874162717\\
20	0.224817044332309\\
22	0.257483115249551\\
24	0.288065266026151\\
26	0.315164428606167\\
28	0.339346930468133\\
30	0.36082522481176\\
32	0.380210437084461\\
34	0.397613250823361\\
36	0.413362102817421\\
38	0.427802222549644\\
40	0.4409841634428\\
42	0.453085401326318\\
44	0.464148931907218\\
46	0.474306985336001\\
48	0.483913019005225\\
50	0.492773341151806\\
52	0.501038624031549\\
54	0.508691663735015\\
56	0.515880571132588\\
58	0.522577030834834\\
60	0.528883382496607\\
62	0.534928881977286\\
64	0.540596537740422\\
66	0.545920699214883\\
68	0.551118860920067\\
70	0.556021727799477\\
};
\addlegendentry{Th.~\ref{thm:average_lower_bound} (average power constraint)};

%\addplot [color=gray,dashed,line width=1pt]
%  table[row sep=crcr]{%
%2	3.87032546086564e-06\\
%4	0.00027675968899457\\
%6	0.00257625744786184\\
%8	0.00943753459882599\\
%10	0.0218517818110924\\
%12	0.0392358372985184\\
%14	0.0603577292682198\\
%16	0.0839520918426264\\
%18	0.108962200409115\\
%20	0.1345864969168\\
%22	0.1602500544457\\
%24	0.185557012765763\\
%26	0.210245164980524\\
%28	0.234148990379265\\
%30	0.257171553982788\\
%32	0.279263751337099\\
%34	0.300409435363854\\
%36	0.320614723719897\\
%38	0.339900498307648\\
%40	0.358297094638003\\
%42	0.375840530402466\\
%44	0.392569909257836\\
%46	0.40852569064133\\
%48	0.423748370788649\\
%50	0.438277729115839\\
%52	0.452152365698414\\
%54	0.46540914834327\\
%56	0.47808322664358\\
%58	0.490207893184137\\
%60	0.501814555045044\\
%62	0.51293265301985\\
%64	0.523589881619107\\
%66	0.533812235209363\\
%68	0.54362391642218\\
%70	0.553047686489055\\
%};
%\addlegendentry{Sh'59 (maximal)};

\addplot [color=black,solid,line width=0.8pt,forget plot]
  table[row sep=crcr]{%
22.8148175140793	0\\
22.8148175140793	1\\
};
\node at (rel axis cs:0.25,0.93) {$M\leq \bar{M}$};

\end{axis}
\end{tikzpicture}%
%
\caption{Lower bounds to the channel coding error probability over an AWGN channel with $n=2$ and $\text{SNR}=10$~dB. Markers show the simulated error probability of a sequence of codes satisfying an equal ($\circ$), maximal ($\times$) and average ($\bullet$) power constraints. Vertical line corresponds to the boundary $M\leq \bar{M}$ from Corollary~\ref{cor:average_metaconverse_bound}. }\label{fig:AWGN-PevsM-n2-snr10dB}
\end{figure}
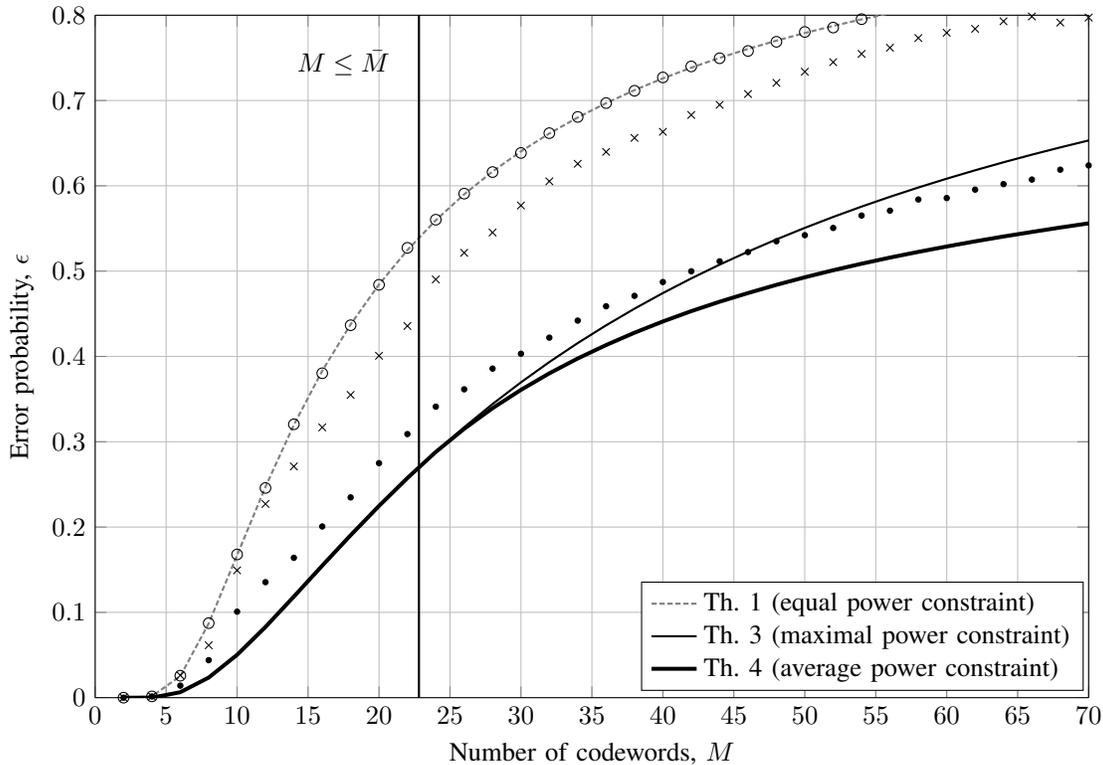%

Figure~\ref{fig:AWGN-PevsM-n2-snr10dB} depicts Shannon'59 lower bound from Theorem \ref{thm:shannon_lower_bound}, valid for equal power-constraints,
the bound from Theorem \ref{thm:maximal_lower_bound} for $\theta^2=\Upsilon+\sigma^2$, which is valid for maximal power-constraint,
and that from Theorem \ref{thm:average_lower_bound}, valid for average power-constraint. The vertical line shows the boundary of the region $M \leq \bar{M}$ defined in Corollary~\ref{cor:average_metaconverse_bound} where the bounds from Theorems \ref{thm:maximal_lower_bound} and \ref{thm:average_lower_bound} coincide. With markers, we show the simulated ML decoding error probability of a sequence of $M$-PSK (phase-shift keying) constellations satisfying the equal power constraint  ($\circ$) and that of a sequence of $M$-APSK (amplitude-phase-shift keying) constellations satisfying maximal ($\times$) and average ($\bullet$) power constraints.\footnote{The parameters of the $M$-APSK constellations (number of rings, number of points, amplitude and phase of each ring) have been optimized to minimize the error probability $\Pe$ for each value of $M$. To this end, the constellation parameters are randomly chosen around their best known values, and only the constellations with lower error probability are used in the next iteration of the stochastic optimization algorithm.}

Since the ML decoding regions of an $M$-PSK constellation are precisely $2$-dimensional cones, Shannon'59 lower bound coincides with the corresponding simulated probability ($\circ$). However, Shannon'59 lower bound does not apply to $M$-APSK constellations satisfying maximal ($\times$) and average ($\bullet$) power constraints, as discussed in \refS{maximal}.

We discuss now the results observed for codes satisfying maximal and average power constraints. We can see that while Theorem~\ref{thm:average_lower_bound} applies in both of these settings, this is not the case for Theorem~\ref{thm:maximal_lower_bound}, that in general only applies under maximal power constraint. As stated in Corollary~\ref{cor:average_metaconverse_bound}, the bounds from Theorems~\ref{thm:maximal_lower_bound} and \ref{thm:average_lower_bound} coincide for $M\leq\bar{M} \approx 22.8$.
Above this point, the two bounds diverge, and we can see from the figure that the average power constrained code ($\bullet$) violates the bound from Theorem~\ref{thm:maximal_lower_bound} for $M>45$.

Analyzing the constellations that violate Theorem \ref{thm:maximal_lower_bound}, we observe that they present several symbols concentrated at the origin of coordinates $(0,0)$.
As these symbols coincide, it is not possible to distinguish between them and they will often yield a decoding error. However, since the symbol $(0,0)$ does not require any energy for its transmission, the average power for the remaining constellation points is increased and this code yields an overall smaller error probability. This effect was also observed in \cite[Sec. 4.3.3]{PolThesis}, where a code with several codewords concentrated at the origin was used to study the asymptotics of the error probability in an average power constrained AWGN channel.
Interestingly, this structure is also suggested by the input distribution that follows from the derivation of Theorem~ \ref{thm:average_lower_bound}. 
As discussed in \refS{mc-implicit-distribution}, the lower bound~\refE{average_lower_bound} in Theorem~\ref{thm:average_lower_bound} corresponds to the value of the convex envelope $\underline{f}$
at the point $\bigl(\tfrac{1}{M},\Upsilon\bigr)$.
Whenever $M\leq\bar{M}$, this convex envelope corresponds to a convex combination of the functions $f\bigl(\bar{\beta},0\bigr)$ and $f\bigl(\beta_0,\gamma_0\bigr)$ with $\gamma_0>\Upsilon$.
Therefore, the input distribution induced by Theorem~\ref{thm:average_lower_bound} is composed by a mass point at the origin and by a uniform distribution over the spherical shell with squared radius $\gamma_0>\Upsilon$. While this distribution does not describe how the codewords of a good code are distributed over the space, it suggest that several codewords could be concentrated at $(0,0)$.

\section{Discussion}
\label{sec:discussion}

We studied the performance of block coding on an AWGN channel under different power limitations at the transmitter. In particular, we showed that the hypothesis-testing bound, \cite[Th. 41]{Pol10} which was originally derived under an equal power limitation, also holds under maximal power constraints (Theorem~\ref{thm:maximal_lower_bound}), and, for rates below a given threshold, under average power constraints (Corollary~\ref{cor:average_metaconverse_bound}). For rates close and above capacity, we proposed a new bound using the convex envelope of the error probability a certain binary hypothesis test (Theorem~\ref{thm:average_lower_bound}).

The performance bounds described above follow from the analysis of the meta-converse bound \cite[Th.~27]{Pol10}, which corresponds to the error probability a surrogate hypothesis test between the distribution induced by the channel and a certain auxiliary distribution. 
For the optimal auxiliary distribution and an equal power-constraint, Polyanskiy showed in~\cite[Sec. VI.F]{Pol13} that the meta-converse bound recovers Shannon cone-packing bound \cite[Eq. (20)]{Shannon59}. In this work, however, we chose the auxiliary distribution to be an i.i.d. Gaussian distribution with zero-mean and certain variance.
If the variance is chosen to be capacity achieving, the resulting bound has a sub-optimal error exponent~\cite{Nakiboglu19-SP,Nakiboglu19-Augustin}. 
Considering the variance of the exponent-achieving output distribution yields tighter finite-length bounds in general, which feature the sphere-packing exponent. Moreover, using the saddlepoint approximation from Theorem~\ref{thm:alpha-beta-sp-formulation}, it is possible to evaluate the bounds for the asymptotically optimal distribution without incurring in extra computational cost (Corollary~\ref{cor:alpha-beta-sp-exponent}). 

The numerical advantage of the new bounds compared to previous results in the literature is small for a maximal power constraint and significant for an average power constraint, as shown in Figures~\ref{fig:maximal-AWGN-Pevsn-snr10dB}-\ref{fig:AWGN-Rvsn-snr5dB}. Additionally, several of the theoretical contributions are of independent interest:
\begin{itemize}
\item We proposed a new geometric interpretation of \cite[Th. 41]{Pol10} which is analogous to the one in \cite{Shannon59}. The hypothesis testing bound \cite[Th. 41]{Pol10} can then be described as the probability of the noise moving a codeword $\x$ out of an $n$-dimensional sphere that roughly covers $1/M$-th of the output space. Interestingly, this sphere is not centered at the codeword $\x$ but at $\bigl(1+\frac{\sigma^2}{\Upsilon}\bigr)\x$.
\item This work addresses the optimization of the meta-converse bound over input distributions. While the results obtained are specific for an additive Gaussian noise channel, the techniques used can in principle be applied to more complicated channels, e.g., via the analysis of the saddlepoint expansion of the meta-converse~\cite{isit18}.
\item For an average power constraint and rates close to and above capacity, the input distribution that optimizes the meta-converse bound presents a mass point at the origin. This suggest that the optimal codes in this region must have several all-zeros codewords (as it occurs for the APSK constellations studied in \refS{constellations}) and motivates the fact that no strong-converse exists for an average power limitation at the transmitter~\cite[Th. 77]{PolThesis}.
\item  In Appendix~\ref{apx:f-beta-gamma}, we provide an exhaustive characterization of the error probability of a binary hypothesis test between two Gaussian distributions, which could be of interest in related problems.
\end{itemize}

In our derivations, we  did not impose any structure to the codebooks beyond the corresponding power limitation. Then, the results obtained are general and do not require the codes to belong to a certain family, to use a specific modulation, or to satisfy minimum distance constraints. Nevertheless, the study of lower bounds for structured codes remains an active area of research (see, e.g.,~\cite{Sas08}). Tight lower bounds for BPSK modulations (or general $M$-PSK modulations), can be obtained from the meta-converse bound \cite[Th. 27]{Pol10} using the results from \cite{isit18}. Evaluation of the meta-converse bound for general modulations is still an open problem due to the combinatorial nature of the optimization over input distributions.

\section*{Acknowledgment}
Fruitful discussions with Bar{\i}\c{s} Nakibo\u{g}lu, Tobias Koch and David Morales-Jimenez are gratefully acknowledged.

%%%%%%%%%%%%%%%%%%%%%%%%%%%%%%%%%%%%%%%%%%%%%%%%%%%%%%%%%%%%%%%%%%

\appendices

%%%%%%%%%%%%%%%%%%%%%%%%%%%%%%%%%%%%%%%%%%%%%%%%%%%%%%%%%%%%%%%%%%

\section{Analysis of $f(\beta,\gamma) = \alpha_{\beta} \bigl(\varphi_{\sqrt{\gamma},\sigma}^n, \varphi_{0,\theta}^n \bigr)$}
\label{apx:f-beta-gamma}

\subsection{Parametric computation of $f(\beta,\gamma)$}
\label{apx:alpha-beta-marcumQ}

\begin{proposition}
\label{prop:alpha-beta-marcumQ}
Let $\sigma,\theta>0$ and $n\geq 1$, be fixed parameters, and define $\delta \triangleq \theta^2-\sigma^2$.
The function trade-off between $\alpha$ and $\beta$ in $\alpha_{\beta} \bigl(\varphi_{\sqrt{\gamma},\sigma}^n, \varphi_{0,\theta}^n \bigr)$ admits the following parametric formulation as a function of the auxiliary parameter $t\geq 0$,
\begin{align}
\alpha(\gamma,t) &= Q_{\frac{n}{2}}\left(\sqrt{n\gamma}\frac{\sigma}{\delta},\frac{t}{\sigma} \right),\label{eqn:alpha-marcumQ}\\
\beta(\gamma,t) &= 1-Q_{\frac{n}{2}}\left(\sqrt{n\gamma}\frac{\theta}{\delta},\frac{t}{\theta} \right),
\label{eqn:beta-marcumQ}
\end{align}
where $Q_m(a,b)$ denotes the Marcum $Q$-function, defined in \refE{marcumQ-def}.

To compute $f(\beta,\gamma) = \alpha_{\beta} \bigl(\varphi_{\sqrt{\gamma},\sigma}^n, \varphi_{0,\theta}^n \bigr)$, let $t_{\star}$ satisfy $\beta(\gamma,t_{\star})=\beta$ according to \refE{beta-marcumQ}. Then, it holds that $f(\beta,\gamma) = \alpha(\gamma,t_{\star})$ according to \refE{alpha-marcumQ}.
\end{proposition}
\begin{IEEEproof}
The proof follows the lines of that of \cite[Th. 41]{Pol10}, and it is included here for completeness.\footnote{Note that the resulting trade-off \refE{alpha-marcumQ}-\refE{beta-marcumQ} is scale invariant provided that $\sigma^2$, $\theta^2$ and $\gamma$ are scaled by the same quantity. Therefore, Proposition~\ref{prop:alpha-beta-marcumQ} is not more general than \cite[Th. 41]{Pol10} by allowing $\sigma^2\neq 1$.}

Let $\sigma,\theta>0$ and $n\geq 1$, be fixed parameters. We define the log-likelihood ratio
\begin{align}
  j(\y) 
  &\triangleq \log \frac{\varphi_{\sqrt{\gamma},\sigma}^n(\y)}{\varphi_{0,\theta}^n (\y)}\\
  &=  n \log\frac{\theta}{\sigma}
  - \frac{1}{2} \sum_{i=1}^{n} \frac{\theta^2(y_i-\sqrt{\gamma})^2-\sigma^2 y_i^2}{\sigma^2 \theta^2}.
  \label{eqn:jrho-def}
\end{align}

According to the Neyman-Pearson lemma, the trade-off $\alpha_{\beta} \bigl(\varphi_{\sqrt{\gamma},\sigma}^n, \varphi_{0,\theta}^n \bigr)$ admits the parametric form 
\begin{align}
\alpha(t') = \Pr\bigl[  j(\Y_{0}) \leq t' \bigr],\label{eqn:alpha-jY}\\
\beta(t') = \Pr\bigl[  j(\Y_{1}) > t'\bigr],\label{eqn:beta-jY}
\end{align}
in terms of the auxiliary parameter $t'\in\RR$ and where $\Y_0 \sim \varphi_{\sqrt{\gamma},\sigma}^n$, $\Y_1 \sim \varphi_{0,\theta}^n$.

Using the change of variable $\z = (\y_0-\sqrt{\gamma})/\sigma$, we obtain that the distribution of the random variable $j(\Y_0)$, $\Y_0 \sim \varphi_{\sqrt{\gamma},\sigma}^n$ coincides with that of  $j_0(\Z)$, $\Z \sim \varphi_{0,1}^n$, where
\begin{align}
  j_{0}(\z) &\triangleq n \log\frac{\theta}{\sigma} +\frac{n}{2} \frac{\gamma}{\delta}-\frac{1}{2}\frac{\delta}{\theta^2} \sum_{i=1}^{n}\left(z_i-\frac{\sigma\sqrt{\gamma}}{\delta}\right)^2.\label{eqn:j0rho-def}
\end{align}
Analogously, if we define
\begin{align}
  j_{1}(\z) 
  &\triangleq n \log\frac{\theta}{\sigma} +\frac{n}{2} \frac{\gamma}{\delta}-\frac{1}{2}\frac{\delta}{\sigma^2} \sum_{i=1}^{n}\left(z_i-\frac{\theta\sqrt{\gamma}}{\delta}\right)^2,\label{eqn:j1rho-def}
\end{align}
it follows that the distributions of $j(\Y_{1})$, $\Y_1 \sim \varphi_{0,\theta}^n$, and that of $j_1(\Z)$, $\Z \sim \varphi_{0,1}^n$ coincide.

Then, we may rewrite \refE{alpha-jY}-\refE{beta-jY} as
\begin{align}
\alpha(t') = \Pr\bigl[  j_{0}(\Z) \leq t' \bigr],\label{eqn:alpha-jZ}\\
\beta(t')  = \Pr\bigl[  j_{1}(\Z) > t'\bigr],\label{eqn:beta-jZ}
\end{align}
where $\Z \sim \varphi_{0,1}^n$.
Given \refE{j0rho-def} and \refE{j1rho-def}, we conclude that $j_{0}(\Z)$ and $j_{1}(\Z)$ follow a (shifted and scaled) noncentral $\chi^2$ distribution with $n$ degrees of freedom and non-centrality parameters $n\gamma \sigma^2/\delta^2$ and $n\gamma \theta^2/\delta^2$, respectively.

Using \refE{j0rho-def} in \refE{alpha-jZ}, we obtain
\begin{align}
\alpha(t') &= \Pr\left[ n \log\frac{\theta}{\sigma} +\frac{n}{2} \frac{\gamma}{\delta}-\frac{1}{2}\frac{\delta}{\theta^2} \sum_{i=1}^{n}\left(Z_i-\frac{\sigma\sqrt{\gamma}}{\delta}\right)^2 \leq t' \right].\label{eqn:alpha-jZ-2}
\end{align}
We consider the change of variable $t'\leftrightarrow t$ such that 
\begin{align}
t' = n\log\frac{\theta}{\sigma} + \frac{n}{2}\frac{\gamma}{\delta} - \frac{\delta t^2}{ 2\sigma^2\theta^2}.
\label{eqn:change-of-variable-t}
\end{align}
Using \refE{change-of-variable-t} in \refE{alpha-jZ-2} and
making the dependence on the parameter $\gamma$ explicit,
we obtain
\begin{align}
\alpha(\gamma,t) &= \Pr\left[ \sum_{i=1}^{n}\left(Z_i-\frac{\sigma\sqrt{\gamma}}{\delta}\right)^2 \geq \left(\frac{t}{\sigma}\right)^2 \right].\label{eqn:alpha-jZ-3}
\end{align}
Proceeding analogously for \refE{beta-jZ} yields
\begin{align}
\beta(\gamma,t)  = \Pr\left[\sum_{i=1}^{n}\left(Z_i-\frac{\theta\sqrt{\gamma}}{\delta}\right)^2 < \left(\frac{t}{\theta}\right)^2 \right].\label{eqn:beta-jZ-3}
\end{align}
The cumulative density function of a non-central $\chi^2$ distribution with $n$ degrees of freedom and non-centrality parameter $\nu$ can be written in terms of the generalized Marcum $Q$-function $Q_m(a,b)$ as~\cite{nuttall75}
\begin{align}\label{eqn:chi2-cdf}
  F_{n,\nu}(x) = 1-Q_{\frac{n}{2}}\bigl(\sqrt{\nu},\sqrt{x}\bigr).
\end{align}
Noting that $F_{n,\nu}(x)$ is continuous, using \refE{chi2-cdf} in \refE{alpha-jZ-3} and \refE{beta-jZ-3}, we obtain the desired result.
\end{IEEEproof}

\subsection{Derivatives of $f(\beta,\gamma)$}
\label{apx:derivatives-f-beta-gamma}

Let $\sigma,\theta>0$ and $n\geq 1$, be fixed parameters, and define $\delta \triangleq \theta^2-\sigma^2$. To obtain the derivatives of $f(\beta,\gamma)$ with respect to $\beta$ and $\gamma$, we start from the parametric formulation from Proposition~\ref{prop:alpha-beta-marcumQ} and use the following auxiliary result.
For $a>0$ and $b>0$, the Marcum-$Q$ function is defined as
\begin{align}
Q_m(a,b) 
\triangleq \int_{b}^{\infty} \frac{t^m}{a^{m-1}}
 e^{-\frac{a^2+t^2}{2}} I_{m-1}(at) \diff t.
 \label{eqn:marcumQ-def}
\end{align}
\begin{proposition}
\label{prop:derivatives-marcumQ}
The derivatives of $Q_m(a,b)$ with respect to its parameters $a>0$ and $b>0$ are given by
\begin{align}
\frac{\partial{Q_m(a,b)}}{\partial a} &= 
\frac{b^m}{a^{m-1}}e^{-\frac{a^2+b^2}{2}} I_{m}(ab),\label{eqn:marcumQ-da}\\
\frac{\partial{Q_m(a,b)}}{\partial b} &= 
- \frac{b^m}{a^{m-1}}e^{-\frac{a^2+b^2}{2}} I_{m-1}(ab),\label{eqn:marcumQ-db}
\end{align}
where $I_{m}(\cdot)$ denotes the $m$-th order modified Bessel function of the first kind.
\end{proposition}
\begin{IEEEproof}
The derivative \refE{marcumQ-db} follows since the variable $b$ appears only in the lower limit of the definite integral in \refE{marcumQ-def}, then the derivative corresponds to the integrand evaluated at $t=b$.

To prove \refE{marcumQ-da}, let $n = m + \ell$ for some $\ell\in\ZZ^{+}$, and define
\begin{align}
  \tilde{Q}^{(n)}_m(a,b) \triangleq 1-e^{-\frac{a^2+b^2}{2}}\sum_{r=m}^{n} \Bigl(\frac{b}{a}\Bigr)^{r} I_{r}(ab),
\end{align}
with partial derivative
\begin{align}
\!&\frac{\partial{\tilde{Q}^{(n)}_m(a,b)}}{\partial a}
% \notag\\\!&
= e^{-\frac{a^2+b^2}{2}} \sum_{r=m}^{n}\Bigl(\frac{b}{a}\Bigr)^{r}\biggl(\Bigl(a\!+\!\frac{r}{a}\Bigr) I_{r}(ab) - b I'_{r}(ab)  \biggr).
\end{align}
Using the identity $I_{m}'(x) = \frac{m}{x} I_{m}(x) + I_{m+1}(x)$ \cite[Sec.~8.486]{Gradshteyn07} and canceling terms we obtain
\begin{align}
\!&\frac{\partial{\tilde{Q}^{(n)}_m(a,b)}}{\partial a}
% \notag\\\!&
= \frac{b^m}{a^{m-1}}e^{-\frac{a^2+b^2}{2}} I_{m}(ab)
- \frac{b^{n+1}}{a^{n}}e^{-\frac{a^2+b^2}{2}} I_{n+1}(ab).
\label{eqn:marcumQ-derivative-1}
\end{align}

We next show that the sequence \refE{marcumQ-derivative-1} presents uniform convergence to the right-hand side of
\refE{marcumQ-da}. Then, since the sequence of functions $\tilde{Q}^{(n)}_m(a,b)$ converges to $Q_m(a,b)$ as $n\to\infty$  \cite[eq. (4.63)]{Simon04}, the sequence \refE{marcumQ-derivative-1} must converge to ${\partial{Q_m(a,b)}}/{\partial a}$ \cite[Sec.~0.307]{Gradshteyn07} and the identity \refE{marcumQ-da} holds.

Indeed, using \cite[Sec.~8.431]{Gradshteyn07} it follows that, for $n\geq 2$
\begin{align}
 \biggl(\frac{b}{a}\biggr)^{n+1} I_{n+1}(ab) 
 &= \frac{(b^2/2)^{n+1}}{\Gamma\bigl(n+\frac{3}{2}\bigr)\Gamma\bigl(\frac{1}{2}\bigr)} \int_{-1}^{1} \bigl(1-t^2\bigr)^{n+\frac{1}{2}} e^{abt} \diff t\\
 &\leq \frac{(b^2/2)^{n+1} e^{ab}}{\Gamma\bigl(n+\frac{3}{2}\bigr)\Gamma\bigl(\frac{1}{2}\bigr)},
 \label{eqn:marcumQ-derivative-2}
\end{align}
where in the last step we used that $e^{abt} \leq e^{ab}$ for $t\in[-1,1]$ and that $\int_{-1}^{1} \bigl(1-t^2\bigr)^{n+\frac{1}{2}} \diff t < 1$ for $n\geq 2$.

Then, from  \refE{marcumQ-derivative-1} and \refE{marcumQ-derivative-2} we obtain
\begin{align}
\biggl|\frac{\partial{\tilde{Q}^{(n)}_m(a,b)}}{\partial a} - \frac{b^m}{a^{m-1}}e^{-\frac{a^2+b^2}{2}} I_{m}(ab)\biggr|
&= a e^{-\frac{a^2-2ab+b^2}{2}} \frac{(b^2/2)^{n+1}}{\Gamma\bigl(n+\frac{3}{2}\bigr)\Gamma\bigl(\frac{1}{2}\bigr)}
\end{align}
which is uniformly bounded for any $0 < a \leq \bar{a}$ with $\bar{a}<\infty$, any $b<\infty$ and $n$ sufficiently large, since the growth of $\Gamma\bigl(n+\frac{3}{2}\bigr)$ is asymptotically faster than that of $(b^2/2)^{n+1}$. 
\end{IEEEproof}
%\begin{remark}
%To the best of our knowledge, the form of the derivative in \refE{marcumQ-da} does not appear in the literature for non-integer values of $m$. 
%For integer values of $m$, \refE{marcumQ-da} can be easily
%obtained from \refE{marcumQ-def} by using the identities $Q_m(a,b) = 1 - Q_{1-m}(b,a)$ and $I_m(x) = I_{-m}(x)$.
%\end{remark}

Using the derivatives of the Marcum-$Q$ function \refE{marcumQ-da} and \refE{marcumQ-db}, we obtain that the derivatives of \refE{alpha-marcumQ} are
\begin{align}
\frac{\partial{\alpha(\gamma,t)}}{\partial \gamma} &= 
\frac{1}{2} \frac{\sigma \sqrt{{n}/{\gamma}}}{\delta} \frac{b^{\frac{n}{2}}}{a^{\frac{n}{2}-1}}e^{-\frac{a^2+b^2}{2}} I_{\frac{n}{2}}(ab),\label{eqn:da_dg}\\
\frac{\partial{\alpha(\gamma,t)}}{\partial t} &= -\frac{1}{\sigma}
\frac{b^{\frac{n}{2}}}{a^{\frac{n}{2}-1}}e^{-\frac{a^2+b^2}{2}} I_{\frac{n}{2}-1}(ab),\label{eqn:da_dt}
\end{align}
with $a = \sqrt{n \gamma}\frac{\sigma}{\delta}$ and $b=\frac{t}{\sigma}$. Proceeding analogously, for the derivatives of \refE{beta-marcumQ} we obtain
\begin{align}
\frac{\partial{\beta(\gamma,t)}}{\partial \gamma} &= 
- \frac{1}{2} \frac{\theta \sqrt{{n}/{\gamma}}}{\delta} \frac{\bar{b}^{\frac{n}{2}}}{\bar{a}^{\frac{n}{2}-1}}e^{-\frac{\bar{a}^2+\bar{b}^2}{2}} I_{\frac{n}{2}}(\bar{a}\bar{b}),
\label{eqn:db_dg}\\
\frac{\partial{\beta(\gamma,t)}}{\partial t} &= \frac{1}{\theta}
\frac{\bar{b}^{\frac{n}{2}}}{\bar{a}^{\frac{n}{2}-1}}e^{-\frac{\bar{a}^2+\bar{b}^2}{2}} I_{\frac{n}{2}-1}(\bar{a}\bar{b}),\label{eqn:db_dt}
\end{align}
where $\bar{a} = \sqrt{n\gamma}\frac{\theta}{\delta}$ and $\bar{b}=\frac{t}{\theta}$. Note that $ab=\bar{a}\bar{b}$, hence, $I_{\frac{n}{2}}(ab)=I_{\frac{n}{2}}(\bar{a}\bar{b})$ and $I_{\frac{n}{2}-1}(ab)=I_{\frac{n}{2}-1}(\bar{a}\bar{b})$.

We now proceed to obtain the derivatives of $f(\beta,\gamma)$ with respect to its parameters:

\subsubsection{Derivative $\partial f(\beta,\gamma)/\partial \gamma$ for fixed $\beta$}
Let $\beta \in[0,1]$ be fixed and let $t(\gamma)$ be such that $\beta\bigl(\gamma,t(\gamma)\bigr) = \beta$ from \refE{beta-marcumQ}. We apply the chain rule for total derivatives to write
\begin{align}
  \frac{\partial \beta\bigl(\gamma,t(\gamma)\bigr)}{\partial \gamma} &=  \biggl( \frac{\partial\beta(\gamma,t)}{\partial\gamma} + \frac{\partial \beta(\gamma,t)}{\partial t}  \frac{\partial t(\gamma)}{\partial \gamma} \biggr) \bigg|_{t=t(\gamma)}. \label{eqn:total-derivatives-beta}
\end{align}
As $\beta\bigl(\gamma,t(\gamma)\bigr) = \beta$ is fixed, then \refE{total-derivatives-beta} must be equal to $0$. Then, identifying \refE{total-derivatives-beta} to $0$ and solving for $\frac{\partial t(\gamma)}{\partial \gamma}$ yields
\begin{align}
\frac{\partial t(\gamma)}{\partial \gamma}
= - \frac{{\frac{\partial}{\partial\gamma} \beta(\gamma,t)}}{{\frac{\partial}{\partial t} \beta(\gamma,t)}}
= \frac{\theta^2}{2\delta} \sqrt{\frac{n}{\gamma}}
\frac{I_{\frac{n}{2}}\Bigl(\sqrt{n\gamma}\frac{t}{\delta}\Bigr)}
{I_{\frac{n}{2}-1}\Bigl(\sqrt{n\gamma}\frac{t}{\delta}\Bigr)}, \label{eqn:partial-t-g}
\end{align}
where $t=t(\gamma)$, and where we used \refE{db_dg} and \refE{db_dt}. Note that we obtained an expression for $\frac{\partial t(\gamma)}{\partial \gamma}$ without computing $t(\gamma)$ explicitly, as doing this would require to invert \refE{beta-marcumQ} which is not analytically tractable.

We apply now the chain rule for total derivatives to $\alpha\bigl(\gamma,t(\gamma)\bigr)$ to write
\begin{align}
\!\frac{\partial \alpha\bigl(\gamma,t(\gamma)\bigr)}{\partial \gamma} &=  \biggl( \frac{\partial \alpha(\gamma,t)}{\partial\gamma} + \frac{\partial \alpha(\gamma,t)}{\partial t}  \frac{\partial t(\gamma)}{\partial \gamma} \biggr) \bigg|_{t=t(\gamma)} \label{eqn:total-derivatives-alpha-1}
\end{align}
Note that, for fixed $\beta$, $\frac{\partial f(\beta,\gamma)}{\partial \gamma} = \frac{\partial \alpha(\gamma,t(\gamma))}{\partial \gamma}$. Hence, 
using \refE{da_dg}, \refE{da_dt} and \refE{partial-t-g} in \refE{total-derivatives-alpha-1} we finally obtain
\begin{align}
\frac{\partial f(\beta,\gamma)}{\partial \gamma}
&= - \frac{n}{2\delta} \biggl(\frac{t\delta}{\sigma^2\sqrt{n \gamma}}\biggr)^{\frac{n}{2}} e^{-\frac{1}{2}\left( \frac{n\gamma\sigma^2}{\delta^2} + \frac{t^2}{\sigma^2}\right)} I_{\frac{n}{2}}\biggl(\sqrt{n\gamma}\frac{t}{\delta}\biggr), \label{eqn:partial-f-g}
\end{align}
where $t$ satisfies $\beta(\gamma,t) = \beta$ with $\beta(\gamma,t)$ given in \refE{beta-marcumQ}. 

\subsubsection{Derivative $\partial f(\beta,\gamma)/\partial \beta$ for fixed $\gamma$}

In this case we use \refE{da_dt} and \refE{db_dt} to obtain
\begin{align}
  \frac{\partial f(\beta,\gamma)}{\partial \beta}
  &= \frac
  {\frac{\partial}{\partial t}\alpha(\gamma,t)}
  {\frac{\partial}{\partial t}  \beta(\gamma,t)}
  = -\frac{\theta^n}{\sigma^n} e^{\frac{1}{2}\left(\frac{n\gamma}{\delta} - t^2\left(\frac{1}{\sigma^{2}}-\frac{1}{\theta^{2}}\right)\right)}
  \label{eqn:partial-f-b}
\end{align}
where $t$ satisfies $\beta(\gamma,t) = \beta$ with $\beta(\gamma,t)$ given in \refE{beta-marcumQ}.

\subsubsection{Derivative $\partial^2 f(\beta,\gamma)/(\partial\beta \partial\gamma)$}

Taking the derivative of \refE{partial-f-b} with respect to $\gamma$ yields
\begin{align}
  \frac{\partial^2 f(\beta,\gamma)}{\partial \beta \partial \gamma}
  &= -\frac{\theta^n}{\sigma^n} e^{\frac{1}{2}\left(\frac{n\gamma}{\delta} - t^2\left(\frac{1}{\sigma^{2}}-\frac{1}{\theta^{2}}\right)\right)}
  \left( \frac{n}{2\delta}- t \left(\frac{1}{\sigma^{2}}-\frac{1}{\theta^{2}}\right)
\frac{\partial t(\gamma)}{\partial \gamma}  \right)
  \label{eqn:partial-f-b-g}
\end{align}
where $t$ satisfies $\beta(\gamma,t) = \beta$ with $\beta(\gamma,t)$ given in \refE{beta-marcumQ}, and where $\frac{\partial t(\gamma)}{\partial \gamma}$ is given in \refE{partial-t-g}.

\subsubsection{Derivative ${\partial^2 f(\beta,\gamma)}/{(\partial \beta)^2}$}

Taking the derivative of \refE{partial-f-b} with respect to $\beta$ yields
\begin{align}
  \frac{\partial^2 f(\beta,\gamma)}{(\partial \beta)^2}
  &= t \frac{\theta^n}{\sigma^n} e^{\frac{1}{2}\left(\frac{n\gamma}{\delta} - t^2\left(\frac{1}{\sigma^{2}}-\frac{1}{\theta^{2}}\right)\right)}
   \left(\frac{1}{\sigma^{2}}-\frac{1}{\theta^{2}}\right)
\frac{\partial t}{\partial \beta} 
  \label{eqn:partial-f-b2}
\end{align}
where $t$ satisfies $\beta(\gamma,t) = \beta$ with $\beta(\gamma,t)$ given in \refE{beta-marcumQ}, and where the term $\frac{\partial t}{\partial \beta}$ can be obtained from \refE{db_dt},
\begin{align}
\frac{\partial t}{\partial\beta} &=
\left(\frac{\partial{\beta(\gamma,t)}}{\partial t}\right)^{-1}
= \frac{\delta}{\sqrt{n\gamma}}
\left(\frac{\theta^2\sqrt{n\gamma}}{t\delta}\right)^{\frac{n}{2}}
e^{\frac{1}{2}\left( \frac{n\gamma\theta^2}{\delta^2} + \frac{t^2}{\theta^2}\right)}
\left(I_{\frac{n}{2}-1}\biggl(\sqrt{n\gamma}\frac{t}{\delta}\biggr)\right)^{-1}.\label{eqn:partial-t-b}
\end{align}

\subsubsection{Derivative ${\partial^2 f(\beta,\gamma)}/{(\partial \gamma)^2}$} 

Taking the derivative of \refE{partial-f-g} with respect to $\gamma$, straightforward but tedious algebra yields
\begin{align}
  \frac{\partial^2 f(\beta,\gamma)}{(\partial \gamma)^2}
  &= - \frac{n}{ 4 \delta} \Bigl(\frac{t\delta}{\sigma^2\sqrt{n\gamma}}\Bigr)^{\frac{n}{2}} e^{-\frac{1}{2}\left( \frac{n\gamma\sigma^2}{\delta^2} + \frac{t^2}{\sigma^2}\right)} I_{\frac{n}{2}}\biggl(\sqrt{n\gamma}\frac{t}{\delta}\biggr)\notag\\
  &\quad \times\left( \frac{n}{\delta} - \frac{n}{\gamma} + \sqrt{\frac{n}{\gamma}} \frac{t}{\delta}
\left(
  \frac{I_{\frac{n}{2}-1}\Bigl(\sqrt{n\gamma}\frac{t}{\delta}\Bigr)}{I_{\frac{n}{2}}\Bigl(\sqrt{n\gamma}\frac{t}{\delta}\Bigr)}
   - \frac{\theta^2}{\sigma^2} 
  \frac{I_{\frac{n}{2}}\Bigl(\sqrt{n\gamma}\frac{t}{\delta}\Bigr)}{I_{\frac{n}{2}-1}\Bigl(\sqrt{n\gamma}\frac{t}{\delta}\Bigr)}
   \right)
\right), \label{eqn:partial-f-g2}
\end{align}
where $t$ satisfies $\beta(\gamma,t) = \beta$  with $\beta(\gamma,t)$ given in \refE{beta-marcumQ}.
Here, we used the identity 
$I_{m}'(x) = I_{m-1}(x) - \frac{m}{x} I_{m}(x)$
\cite[Sec.~8.486]{Gradshteyn07}.

\subsection{Derivatives of $f(\beta,\gamma)$ at $\gamma=0$}
\label{apx:derivatives-f-beta-gamma-0}
The function $f(\beta,0)$ can be evaluated by setting $\gamma=0$ and using \refE{alpha-marcumQ}-\refE{beta-marcumQ}. However, the preceding expressions for the derivatives of $f(\beta,\gamma)$ often yield an indeterminacy in this case. This can be avoided by taking the limit as $\gamma \to 0$ and using that~\cite[Sec.~8.445]{Gradshteyn07}
\begin{align}
 I_{m}(x) = \frac{\bigl(\frac{x}{2}\bigr)^m}{\Gamma\bigl(m+1\bigr)}
            + o(x^m),  \label{eqn:Imx-asympt}
\end{align}
where $\Gamma(\cdot)$ denotes the gamma function and $o\bigl(g(x)\bigr)$ summarizes the terms that approach zero faster than $g(x)$, \textit{i.e.}, $\lim_{x\to 0}\frac{o(g(x))}{g(x)}=0$.
For example, using \refE{Imx-asympt} and $\frac{\Gamma(m+1)}
{\Gamma(m)} = m$ we obtain from \refE{partial-t-g} that
\begin{align}
\frac{\partial t(\gamma)}{\partial \gamma}\biggr|_{\gamma=0}
&= \frac{t}{2} \frac{\theta^2}{\delta^2}. \label{eqn:partial-t-g-0}
\end{align}

Proceeding analogously for the derivatives of $f(\beta,\gamma)$, it follows that
\begin{align}
\frac{\partial f(\beta,\gamma)}{\partial \gamma}\biggr|_{\gamma=0}
&= - \frac{1}{\delta} \frac{t_0^n}{\sigma^n} \frac{e^{-\frac{1}{2} \frac{t_0^2}{\sigma^2}}}{\Gamma\bigl(\tfrac{n}{2}\bigr)  2^{\frac{n}{2}}}, \label{eqn:partial-f-g-0}\\
\frac{\partial f(\beta,\gamma)}{\partial \beta}\biggr|_{\gamma=0}
&= -\frac{\theta^n}{\sigma^n} e^{-\frac{1}{2} t_0^2\left(\frac{1}{\sigma^{2}}-\frac{1}{\theta^{2}}\right)},
  \label{eqn:partial-f-b-0}\\
\frac{\partial^2 f(\beta,\gamma)}{\partial\beta\partial\gamma}\biggr|_{\gamma=0}
&= -\frac{\theta^n}{\sigma^n} 
  \left( \frac{n}{2\delta}- \frac{t_0^2}{2\delta\sigma^{2}}\right)
  e^{-\frac{1}{2} t_0^2\left(\frac{1}{\sigma^{2}}-\frac{1}{\theta^{2}}\right)},
  \label{eqn:partial-f-b-g-0}\\
  \frac{\partial^2 f(\beta,\gamma)}{(\partial \beta)^2}\biggr|_{\gamma=0}
  &= \frac{\theta^n}{\sigma^n}
\biggl(\frac{\theta\sqrt{2}}{t_0}\biggr)^{n-2}
\frac{\delta}{\sigma^2}
\Gamma\bigl(\tfrac{n}{2}\bigr) 
e^{-\frac{1}{2}t_0^2\frac{\delta - \sigma^2}{\theta^2\sigma^2}}, \label{eqn:partial-f-b2-0}\\
  \frac{\partial^2 f(\beta,\gamma)}{(\partial \gamma)^2}\biggr|_{\gamma=0}
  &= - \frac{n}{ 4 \delta} \frac{t_0^n}{\sigma^n 2^{\frac{n}{2}}}
  \left(\frac{n}{\delta} +  \left( \frac{n}{n+2} - \frac{\theta^2}{\sigma^2}\right) \frac{t_0^2}{\delta^2}\right)
       \frac{e^{-\frac{1}{2}\frac{t_0^2}{\sigma^2}}}{\Gamma(\frac{n}{2}+1)},
\label{eqn:partial-f-g2-0}
\end{align}
where in all cases $t_0$ satisfies $\beta(0,t_0) = \beta$ with $\beta(\gamma,t)$ given in \refE{beta-marcumQ}.
To obtain \refE{partial-f-g2-0} from \refE{partial-f-g2} we used \refE{Imx-asympt} and the expansions
\begin{align}
\frac{I_{m-1}(x)}{I_{m}(x)} = \frac{2m}{x} + \frac{x}{2(m+1)} + o(x),\qquad
\frac{I_{m}(x)}{I_{m-1}(x)} = \frac{x}{2m} + o(x).
\end{align}

%%%%%%%%%%%%%%%%%%%%%%%%%%%%%%%%%%%%%%%%%%%%%%%%%%%%%%%%%%%%

\section{Proof of Lemma~\ref{lem:envelope_equals_f}}
\label{apx:envelope_equals_f}

We characterize the region where $f(\beta,\gamma)$ and its convex envelope $\underline{f}(\beta,\gamma)$ coincide using  the following result.

\begin{proposition}\label{prop:convex-envelope}
Suppose $g$ is differentiable with gradient $\nabla g$. Let $\Ac$ denote the domain of $g$, and let  $a_0\in\Ac$. If the inequality
\begin{align}
  g(\bar a) \geq g(a_0) + \nabla g(a_0)^T (\bar a - a_0),
  \label{eqn:prop-convex-envelope}
\end{align}
is satisfied for all $\bar a \in \Ac$, then, $g(a_0) = g^{**}(a_0)$ holds.
\end{proposition}
\begin{IEEEproof}
As $g^{**}$ is the lower convex envelope of $g$, then $g(a_0) \geq g^{**}(a_0)$ trivially. It remains to show that \refE{prop-convex-envelope} implies $g(a_0) \leq g^{**}(a_0)$. Fenchel's inequality~\cite[Sec.~3.3.2]{Boyd04} yields
\begin{align}
  g^{**}(a_0) \geq \langle a_0,b \rangle - g^{*}(b),
  \label{eqn:fenchel-ineq-0}
\end{align}
for any $b$ in the domain of $g^{*}$.

Setting $b = \nabla g(a_0)$ and using \refE{LF-transform} 
in \refE{fenchel-ineq-0}, we obtain
\begin{align}
  g^{**}(a_0)  &\geq \nabla g(a_0)^T a_0 - \max_{\bar{a}\in\Ac} \bigl\{
  \nabla g(a_0)^T\bar{a} - g(\bar a)  \bigr\}
  \label{eqn:fenchel-ineq-1}\\
  &= \min_{\bar{a}\in\Ac} \bigl\{
  \nabla g(a_0)^T(a_0-\bar{a}) + g(\bar a)  \bigr\}
  \label{eqn:fenchel-ineq-2}\\
  &\geq \min_{\bar{a}\in\Ac} \bigl\{ g(a_0) \bigr\},
  \label{eqn:fenchel-ineq-3}
\end{align}
where in the last step we used \refE{prop-convex-envelope} to lower bound $g(\bar a)$. Since the objective of \refE{fenchel-ineq-3} does not depend on $\bar{a}$, we conclude from \refE{fenchel-ineq-1}-\refE{fenchel-ineq-3} that $g(a_0) \leq g^{**}(a_0)$ and the result follows.
\end{IEEEproof}

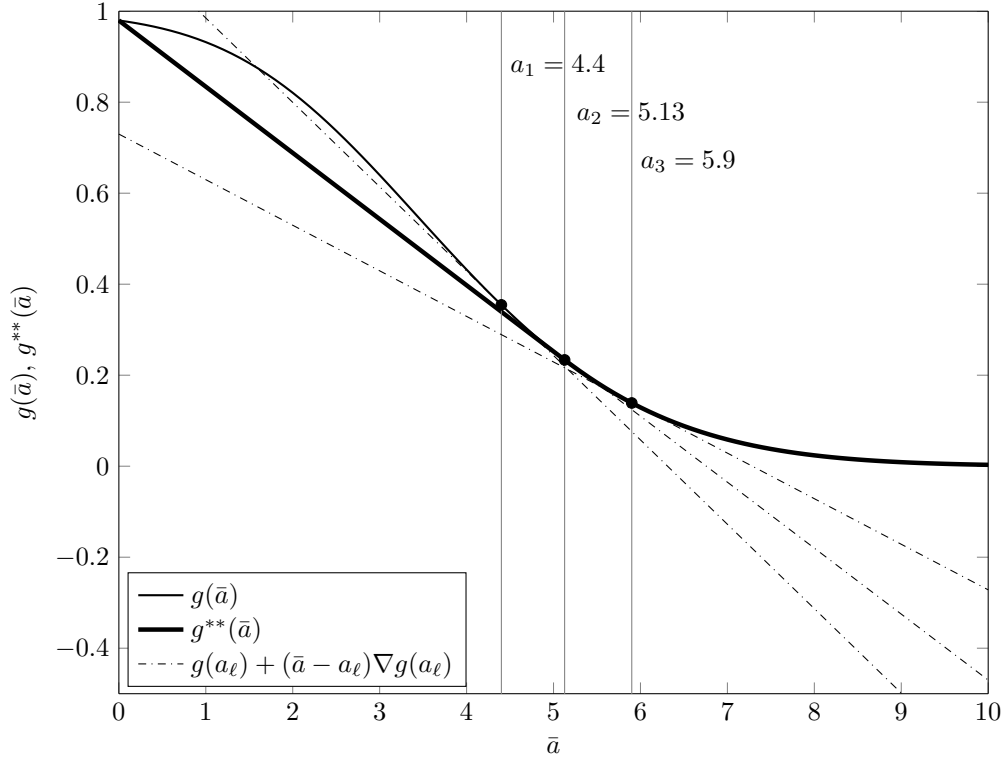
\begin{figure}[t]%
\centering% This file was created by matlab2tikz v0.6.0 running on MATLAB 8.2.
%Copyright (c) 2008--2014, Nico Schlömer <nico.schloemer@gmail.com>
%All rights reserved.
%Minimal pgfplots version: 1.3
%  http://www.mathworks.com/matlabcentral/fileexchange/22022-matlab2tikz
\begin{tikzpicture}

\begin{axis}[%
width=.7\linewidth,
height=.55\linewidth,
scale only axis,
separate axis lines,
every outer x axis line/.append style={black},
every x tick label/.append style={font=\color{black}},
xmin=0,
xmax=10,
xlabel={$\bar{a}$},
every outer y axis line/.append style={black},
every y tick label/.append style={font=\color{black}},
ytick={-0.4,-0.2,0,0.2,0.4,0.6,0.8,1},
ymin=-0.5,
ymax=1,
ylabel={$g(\bar{a}),\, g^{**}(\bar{a})$},
legend style={at={(0.01,0.01)},anchor=south west,legend cell align=left,align=left,draw=black}
]
\addplot [color=black,solid,line width=0.8pt]
  table[row sep=crcr]{%
0	0.97959777901405\\
0.1	0.976800959471999\\
0.2	0.97366198633429\\
0.3	0.970148410881538\\
0.4	0.966223033808878\\
0.5	0.961854045477556\\
0.6	0.957001569771807\\
0.7	0.951629747817022\\
0.8	0.94570242374794\\
0.9	0.939183303084772\\
1	0.932036105691001\\
1.1	0.924230837829455\\
1.2	0.915734843868026\\
1.3	0.906524769819627\\
1.4	0.896573787482287\\
1.5	0.885865334411872\\
1.6	0.874387237130194\\
1.7	0.86213191274658\\
1.8	0.84909652950083\\
1.9	0.835283128255816\\
2	0.820715175831309\\
2.1	0.805396803676944\\
2.2	0.78935989898761\\
2.3	0.772628804772277\\
2.4	0.755250198898824\\
2.5	0.737250739024797\\
2.6	0.718689562180382\\
2.7	0.699617404798655\\
2.8	0.680080265643311\\
2.9	0.660136872542212\\
3	0.639856487927015\\
3.1	0.619302513186311\\
3.2	0.598529830948357\\
3.3	0.577602476373171\\
3.4	0.556584174944848\\
3.5	0.535541356978797\\
3.6	0.514547297551659\\
3.7	0.493647252637399\\
3.8	0.472898279888807\\
3.9	0.452354576584325\\
4	0.43207832723504\\
4.1	0.412121325985752\\
4.2	0.392513700732596\\
4.3	0.373295283836609\\
4.4	0.354527412701067\\
4.5	0.336225217505878\\
4.6	0.318409032040712\\
4.7	0.301134890187879\\
4.8	0.284397324030113\\
4.9	0.268221071961423\\
5	0.252630910363518\\
5.1	0.237616376151686\\
5.2	0.223211571694168\\
5.3	0.209392711336536\\
5.4	0.196183353057605\\
5.5	0.18356696489523\\
5.6	0.171539062893571\\
5.7	0.160108288708049\\
5.8	0.149244446672944\\
5.9	0.138946326536343\\
6	0.129209746141278\\
6.1	0.120006353990628\\
6.2	0.11132260526048\\
6.3	0.10314535947915\\
6.4	0.0954649432457501\\
6.5	0.0882534851190706\\
6.6	0.0814932955673795\\
6.7	0.0751660960276739\\
6.8	0.0692531950356181\\
6.9	0.0637356527501478\\
7	0.0585944328445525\\
7.1	0.0538105410221596\\
7.2	0.0493651496900609\\
7.3	0.0452397085836716\\
7.4	0.041416041374263\\
7.5	0.0378764285075942\\
7.6	0.0346036767119847\\
7.7	0.031583500381131\\
7.8	0.0288005424694332\\
7.9	0.0262363700983372\\
8	0.0238762581282023\\
8.1	0.0217069869020641\\
8.2	0.0197196993254652\\
8.3	0.0178960724395026\\
8.4	0.0162243503381146\\
8.5	0.0146990509466517\\
8.6	0.0133031490700497\\
8.7	0.0120294782333805\\
8.8	0.0108688662721098\\
8.9	0.00981048191363134\\
9	0.00884922089339409\\
9.1	0.00797387353847451\\
9.2	0.00718054235312602\\
9.3	0.00645981677027991\\
9.4	0.00580721241113407\\
9.5	0.0052163224150486\\
9.6	0.00468110216599182\\
9.7	0.00419862675748235\\
9.8	0.00376229096399404\\
9.9	0.00336850275803135\\
10	0.00301430413982268\\
};
\addlegendentry{$g(\bar{a})$};

\addplot [color=black,solid,line width=1.7pt]
  table[row sep=crcr]{%
0	0.97959777901405\\
5.1276	0.2336\\
5.2	0.223211571694168\\
5.3	0.209392711336536\\
5.4	0.196183353057605\\
5.5	0.18356696489523\\
5.6	0.171539062893571\\
5.7	0.160108288708049\\
5.8	0.149244446672944\\
5.9	0.138946326536343\\
6	0.129209746141278\\
6.1	0.120006353990628\\
6.2	0.11132260526048\\
6.3	0.10314535947915\\
6.4	0.0954649432457501\\
6.5	0.0882534851190706\\
6.6	0.0814932955673795\\
6.7	0.0751660960276739\\
6.8	0.0692531950356181\\
6.9	0.0637356527501478\\
7	0.0585944328445525\\
7.1	0.0538105410221596\\
7.2	0.0493651496900609\\
7.3	0.0452397085836716\\
7.4	0.041416041374263\\
7.5	0.0378764285075942\\
7.6	0.0346036767119847\\
7.7	0.031583500381131\\
7.8	0.0288005424694332\\
7.9	0.0262363700983372\\
8	0.0238762581282023\\
8.1	0.0217069869020641\\
8.2	0.0197196993254652\\
8.3	0.0178960724395026\\
8.4	0.0162243503381146\\
8.5	0.0146990509466517\\
8.6	0.0133031490700497\\
8.7	0.0120294782333805\\
8.8	0.0108688662721098\\
8.9	0.00981048191363134\\
9	0.00884922089339409\\
9.1	0.00797387353847451\\
9.2	0.00718054235312602\\
9.3	0.00645981677027991\\
9.4	0.00580721241113407\\
9.5	0.0052163224150486\\
9.6	0.00468110216599182\\
9.7	0.00419862675748235\\
9.8	0.00376229096399404\\
9.9	0.00336850275803135\\
10	0.00301430413982268\\
};
\addlegendentry{$g^{**}(\bar{a})$};

\addplot [color=black,dash dot]
  table[row sep=crcr]{%
0	1.17050222820002\\
10	-0.683985988843051\\
};
\addlegendentry{$g(a_{\ell}) + (\bar{a}-a_{\ell})\nabla g(a_{\ell})$};

\addplot [color=black,only marks,mark=*,mark options={solid},forget plot]
  table[row sep=crcr]{%
4.4	0.354527412701067\\
};
\addplot [color=black!50!white,solid,forget plot]
  table[row sep=crcr]{%
4.4	-1\\
4.4	1\\
};
\addplot [color=black,dash dot,forget plot]
  table[row sep=crcr]{%
0	0.97959777901405\\
10	-0.47\\
};
\addplot [color=black,only marks,mark=*,mark options={solid},forget plot]
  table[row sep=crcr]{%
5.1276	0.2336\\
};
\addplot [color=black!50!white,solid,forget plot]
  table[row sep=crcr]{%
5.1276	-1\\
5.1276	1\\
};
\addplot [color=black,dash dot,forget plot]
  table[row sep=crcr]{%
0	0.729898252692595\\
10	-0.271715181470543\\
};
\addplot [color=black,only marks,mark=*,mark options={solid},forget plot]
  table[row sep=crcr]{%
5.9	0.138946326536343\\
};
\addplot [color=black!50!white,solid,forget plot]
  table[row sep=crcr]{%
5.9	-1\\
5.9	1\\
};
\node at (rel axis cs:0.54-0.035,0.92) {$a_1 = 4.4$};
\node at (rel axis cs:0.624-0.035,0.85) {$a_2 = 5.13$};
\node at (rel axis cs:0.69-0.035,0.78) {$a_3 = 5.9$};
\end{axis}
\end{tikzpicture}%
%
\caption{Example of Proposition~\ref{prop:convex-envelope} for the one-dimensional function $g(a) = f(\beta,a)$ with $\beta=0.001$, $n=6$, $\sigma^2=1$ and $\theta^2=3$, which is defined for $a\geq 0$.}\label{fig:convAnalysis}
\end{figure}%

\refFig{convAnalysis} shows an example of Proposition~\ref{prop:convex-envelope} for a certain one-dimensional function $g$. When $a_0=a_1$, the figure shows that \refE{first-order-condition} is violated as the dash-dotted line is above $g(\bar{a})$ (thin solid line) for small values of $\bar{a}$. Then, the (one-dimensional) convex envelope  $g^{**}$ (thick solid line) is strictly smaller than the function $g$ at the point $a_0=a_1$. In contrast, for $a_0=a_2$ \refE{first-order-condition} is satisfied for all values of $\bar{a}\geq 0$. Therefore $g$ coincides with its convex envelope $g^{**}$ at $a_0 = a_2$. This is also true for any $a_0 > a_2$ (e.g., for $a_0 = a_3$), and therefore $g$ and its convex envelope $g^{**}$ coincide for any $a_0 \geq a_2$.

We apply Proposition~\ref{prop:convex-envelope} to the function  $f(\beta,\gamma)$. We recall that $f(\beta,\gamma)$ is differentiable for $\beta\in[0,1]$ and $\gamma\geq 0$ with derivatives given in Appendix~\ref{apx:f-beta-gamma}. We define the gradients
\begin{align}
\nabla_{\beta} f(b,g) &\triangleq \frac{\partial f(\beta,\gamma)}{\partial \beta}\Big|_{\beta=b,\gamma=g},\\
\nabla_{\gamma} f(b,g) &\triangleq \frac{\partial f(\beta,\gamma)}{\partial \gamma}\Big|_{\beta=b,\gamma=g}.
\end{align}

According to Proposition~\ref{prop:convex-envelope}, the function $f(\beta_0,\gamma_0)$ and its convex envelope $\underline{f}(\beta_0,\gamma_0)$ coincide if 
\begin{align}
f(\bar\beta,\bar\gamma) \;\geq\; f(\beta_0,\gamma_0) &+ (\bar\beta-\beta_0)\nabla_{\beta} f(\beta_0,\gamma_0)
%\notag\\ &
+ (\bar\gamma-\gamma_0)\nabla_{\gamma} f(\beta_0,\gamma_0).
\label{eqn:first-order-condition}
\end{align}
is satisfied for all $\beta\in[0,1]$ and $\gamma\geq 0$.
This condition implies that the first-order Taylor approximation of $f(\beta,\gamma)$ at $(\beta_0,\gamma_0)$ is a global under-estimator of the original function $f$.

The derivatives of $f(\beta,\gamma)$, given in Appendix~\ref{apx:f-beta-gamma}, imply that the function is decreasing in both parameters, convex with respect to $\beta\in[0,1]$, and jointly convex with respect to $(\beta,\gamma)$ except for a neighborhood near the axis $\gamma=0$. Using these properties, it can be shown that the condition \refE{first-order-condition} only needs to be verified along the axis $\bar\gamma=0$. For example, for the one-dimensional function $g$ in \refF{convAnalysis}, we can see that if the first-order condition is satisfied at $\bar{a}=0$, it is also satisfied for any $\bar{a}\geq 0$.

Then, we conclude that $f(\beta_0,\gamma_0) = \underline{f}(\beta_0,\gamma_0)$ if \refE{first-order-condition} holds for every $\bar\beta\in[0,1]$ and $\bar\gamma= 0$, i.e., if
\begin{align}
f(\beta_0,\gamma_0) - f(\bar\beta,0) &\geq  (\beta_0-\bar\beta)\nabla_{\beta} f(\beta_0,\gamma_0)
%\notag\\&\qquad\;\;\;
+ \gamma_0 \nabla_{\gamma} f(\beta_0,\gamma_0).
\label{eqn:first-order-condition-1}
\end{align}

Let $\theta \geq \sigma > 0$, $n\geq 1$. Let $t_0$ be the value such that $\beta(\gamma_0,t_0) = \beta_0$ and let $\bar{t}$ satisfy $\beta(0,\bar{t}) = \bar\beta$, for $\beta(\gamma,t)$ defined in \refE{beta-marcumQ}.
Using \refE{alpha-marcumQ} and the derivatives \refE{partial-f-g} and \refE{partial-f-b} from Appendix~\ref{apx:f-beta-gamma}, we obtain the identities
\begin{align}
\!f(\beta_0,\gamma_0) - f(\bar\beta,0) &= Q_{\frac{n}{2}}\!\left(\!\sqrt{n\gamma_0}\frac{\sigma}{\delta},\frac{t_0}{\sigma}\!\right) - Q_{\frac{n}{2}}\!\left(\!0,\frac{\bar{t}}{\sigma}\!\right)\!,\!\label{eqn:first-order-term-1}\\
\nabla_{\beta} f(\beta_0,\gamma_0) &= -\frac{\theta^n}{\sigma^n} e^{\frac{1}{2}\left(\frac{n\gamma_0}{\delta} - t_0^2\left(\frac{1}{\sigma^{2}}-\frac{1}{\theta^{2}}\right)\right)},\label{eqn:first-order-term-2}\\
\nabla_{\gamma} f(\beta_0,\gamma_0) &=  - \frac{n}{2\delta} \biggl(\frac{t_0\delta}{\sigma^2\sqrt{n \gamma_0}}\biggr)^{\frac{n}{2}} 
 e^{-\frac{1}{2}\left( n\gamma_0\frac{\sigma^2}{\delta^2} + \frac{t_0^2}{\sigma^2}\right)} I_{\frac{n}{2}}\biggl(\sqrt{n\gamma_0}\frac{t_0}{\delta}\biggr).\label{eqn:first-order-term-3}
\end{align}

As $\beta(\gamma_0,t_0) = \beta_0$ and $\beta(0,\bar{t}) = \bar\beta$, using \refE{beta-marcumQ}, it follows that
\begin{align}
\beta_0-\bar\beta =  Q_{\frac{n}{2}}\left(0,\frac{\bar{t}}{\theta} \right) - Q_{\frac{n}{2}}\left(\sqrt{n\gamma_0}\frac{\theta}{\delta},\frac{t_0}{\theta} \right).\label{eqn:first-order-term-4}
\end{align}
Then, substituting \refE{first-order-term-1} and \refE{first-order-term-4} in \refE{first-order-condition-1}, reorganizing terms, it yields
\begin{align}
&Q_{\frac{n}{2}}\left(\sqrt{n\gamma_0}\frac{\sigma}{\delta},\frac{t_0}{\sigma} \right)
+ \nabla_{\beta} f(\beta_0,\gamma_0)
Q_{\frac{n}{2}}\left(\sqrt{n\gamma_0}\frac{\theta}{\delta},\frac{t_0}{\theta} \right)
%\notag\\&
- \gamma_0 \nabla_{\gamma} f(\beta_0,\gamma_0)
\,\geq\, h(\bar{t}),
\label{eqn:first-order-condition-2}
\end{align}
where $h(t)$ is given by
\begin{align}
h(t) \triangleq Q_{\frac{n}{2}}\left(0,\frac{t}{\sigma} \right) +\nabla_{\beta} f(\beta_0,\gamma_0)
Q_{\frac{n}{2}}\left(0,\frac{t}{\theta}\right).
\label{eqn:first-order-condition-ht}
\end{align}
The interval $\bar\beta\in[0,1]$ corresponds to $\bar{t} \geq 0$. We maximize \refE{first-order-condition-ht} over $t = \bar{t} \geq 0$ and we only verify the condition \refE{first-order-condition-2} for this maximum value.

To this end, we find the derivative of \refE{first-order-condition-ht} with respect to $t$, we identify the resulting expression with zero and solve for $t$.
Using \refE{marcumQ-db} and \refE{Imx-asympt} it follows that
\begin{align}
\frac{\partial}{\partial b} Q_{m}\left(0,b \right)
= -\frac{b^{2m-1}}{2^{m-1}}\frac{e^{-\frac{b^2}{2}}}{\Gamma(m)}\label{eqn:marcumQ-db-a0}
\end{align}
and therefore
\begin{align}
\frac{\partial}{\partial t} h(t) 
&=
- \frac{1}{\sigma} \frac{(t/\sigma)^{n-1}}{2^{\frac{n}{2}-1}} \frac{e^{-\frac{t^2}{2\sigma^2}}}{\Gamma(n/2)}
- \frac{\nabla_{\beta} f(\beta_0,\gamma_0)}{\theta} \frac{(t/\theta)^{n-1}}{2^{\frac{n}{2}-1}} \frac{e^{-\frac{t^2}{2\theta^2}}}{\Gamma(n/2)}\\
&=
- \frac{t^{n-1}}{\sigma^n 2^{\frac{n}{2}-1}} \frac{e^{-\frac{t^2}{2\sigma^2}}}{\Gamma(n/2)}
+ \frac{t^{n-1}}{\sigma^n 2^{\frac{n}{2}-1}} \frac{e^{-\frac{t^2}{2\theta^2}}}{\Gamma(n/2)} e^{\frac{1}{2}\left(\frac{n\gamma_0}{\delta} - t_0^2\left(\frac{1}{\sigma^{2}}-\frac{1}{\theta^{2}}\right)\right)}
\label{eqn:first-order-condition-dht}
\end{align}
where in the second step we used \refE{first-order-term-2}. Identifying \refE{first-order-condition-dht} with zero, we obtain the roots $t=0$ and (after some algebra)
\begin{align}
t^2 = t_0^2 - n\gamma \frac{\sigma^2\theta^2}{\delta^2}.
\label{eqn:first-order-condition-dht0}
\end{align}
By evaluating the second derivative of \refE{first-order-condition-ht}, it can be verified that \refE{first-order-condition-dht0} indeed corresponds to a maximum of $h(t)$.

Therefore, we conclude that the right-hand side of   \refE{first-order-condition-2} is maximized for
\begin{align}
\bar{t}_{\star} = \sqrt{\bigl(t_0^2- n\gamma {\sigma^2\theta^2}/{\delta^2}\bigr)_{+}}
\label{eqn:first-order-opt-bart}
\end{align}
where the threshold $(a)_{+} = \max(0,a)$ follows from the constraint $\bar{t} \geq 0$.
Using \refE{first-order-term-2}, \refE{first-order-term-3} and \refE{first-order-opt-bart} in \refE{first-order-condition-2} we obtain the desired characterization for the region of interest.
For the statement of the result in Lemma~\ref{lem:envelope_equals_f}, we select the smallest $t_0$ that fulfills \refE{first-order-condition-2} (which satisfies the condition with equality) and use the simpler notation $(\beta,\gamma)$ instead of $(\beta_0,\gamma_0)$.

\begin{figure}[t]
\centering\input{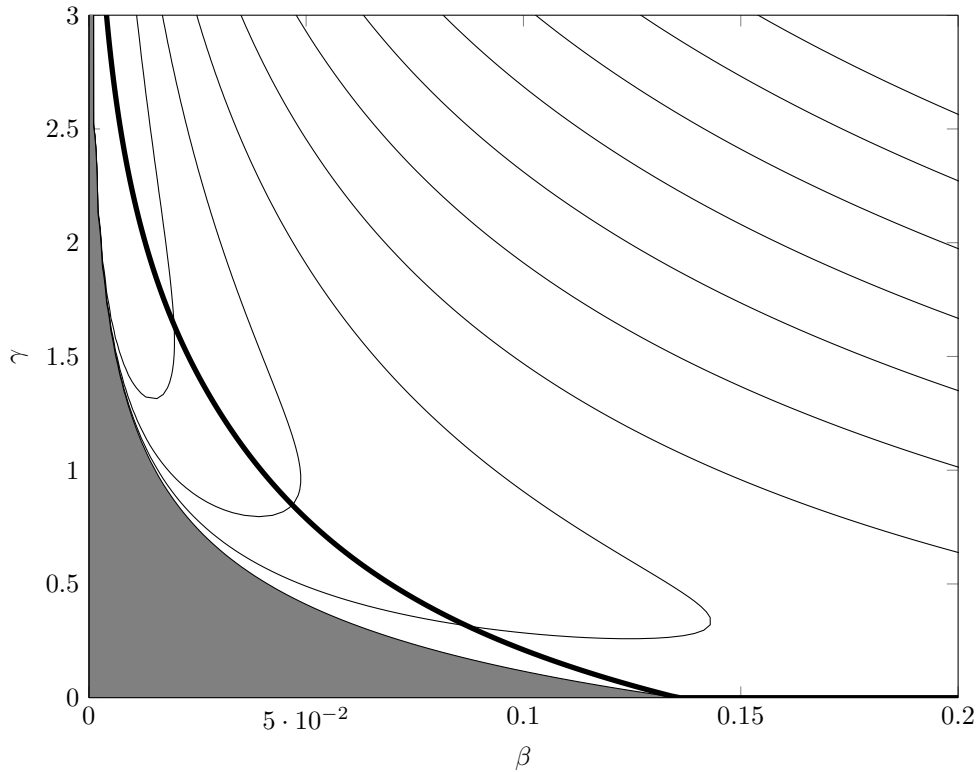}%
\caption{Level curves of $\det \nabla^2 f(\beta,\gamma)$ for $n=6$, $\sigma^2=1$, $\theta^2=2$. The region where $\det \nabla^2 f(\beta,\gamma)<0$ is shaded in gray. The bold line corresponds to the points where $\beta = 1- Q_{\frac{n}{2}}\bigl( \sqrt{n\gamma}{\theta}/{\delta},\, t_0/{\theta} \bigr)$ as described in Lemma~\ref{lem:envelope_equals_f}.}\label{fig:detHessian}%
\end{figure}%

We emphasize that the condition for Lemma~\ref{lem:envelope_equals_f} derived in this appendix does not correspond to the region where $f(\beta,\gamma)$ is locally convex, but it precisely characterizes the region where $f(\beta,\gamma) = \underline{f}(\beta,\gamma)$. \refFig{detHessian} shows the difference between these two regions for a given set of parameters: the shaded area shows the points where $f(\beta,\gamma)$ is locally non-convex, while the bold line corresponds to the lower-boundary of the region where $f(\beta,\gamma) = \underline{f}(\beta,\gamma)$.

\bibliographystyle{IEEEtran}
\bibliography{bib/references}

\end{document}